\documentclass [letter, 12pt]{article}
\usepackage{a4wide, amsmath, amsfonts, amssymb, geometry}

\usepackage{bm}
\usepackage{fancyhdr, dsfont}

\usepackage[pdftex]{graphicx} 
\usepackage{graphicx}
\usepackage{float}

\restylefloat{figure}
\usepackage{epstopdf}

\def\vev#1{\langle #1 \rangle}

\newcommand{\nwc}{\newcommand}
\nwc{\ba}  {\begin{array}}
\nwc{\ea}  {\end{array}}
\nwc{\bdm} {\begin{displaymath}}
\nwc{\edm} {\end{displaymath}}

\nwc{\bea} {\begin{equation}\ba{lcl}}
\nwc{\eea} {\ea\end{equation}}

\nwc{\bda} {\bdm\ba{lcl}} 
\nwc{\eda} {\ea\edm}

\nwc{\bc}  {\begin{center}}
\nwc{\ec}  {\end{center}}

\nwc{\ds}  {\displaystyle}
\nwc{\bmat}{\left(\ba}
\nwc{\emat}{\ea\right)}
\nwc{\nn}  {\nonumber}
\nwc{\nnn} {\nonumber \vspace{.2cm} \\ }
\nwc{\ra}  {\rightarrow}
\nwc{\lra} {\longrightarrow}

\nwc{\p} {\partial}

\def\beq{\begin{equation}}
\def\eeq{\end{equation}}

\newcommand{\vecb}{\left(\begin{array}{c}}
\newcommand{\vece}{\end{array}\right)}
\newcommand{\ccb}{\left(\begin{array}{cc}}
\newcommand{\cce}{\end{array}\right)}
\newcommand{\cccb}{\left(\begin{array}{ccc}}
\newcommand{\ccce}{\end{array}\right)}
\newcommand{\ccccb}{\left(\begin{array}{cccc}}
\newcommand{\cccce}{\end{array}\right)}
\newcommand{\cccccb}{\left(\begin{array}{ccccc}}
\newcommand{\ccccce}{\end{array}\right)}

\def\cc[#1]#2{C_{#1}^{\phantom{#1}#2}}

\newcommand{\ve}{\vec}
\newcommand{\pa}{\partial}

\newcommand{\al}{\alpha}
\newcommand{\be}{\beta}
\newcommand{\ga}{\gamma}
\newcommand{\de}{\delta}
\newcommand{\ep}{\epsilon}
\newcommand{\vep}{\varepsilon}

\newcommand{\io}{\iota}

\newcommand{\si}{\sigma}
\newcommand{\la}{\lambda}
\newcommand{\ka}{\kappa}
\newcommand{\om}{\omega}

\newcommand{\Ga}{\Gamma}

\newcommand{\Om}{\Omega}

\newcommand{\mto}{\rightarrow}
\newcommand{\te}{\textrm}
\newcommand{\eq}{ \ \ = \ \ }
\newcommand{\co}{\ , \ \ \ \ \ \ }
\newcommand{\dd}{\mathrm{d}}

\newcommand{\half}{\tfrac{1}{2}}

\newcommand{\dal}{\dot{\alpha}}
\newcommand{\dbe}{\dot{\beta}}
\newcommand{\dga}{\dot{\gamma}}
\newcommand{\dde}{\dot{\delta}}
\newcommand{\dep}{\dot{\epsilon}}

\newcommand{\dka}{\dot{\kappa}}

\newcommand{\di}{\dot{\io}}

\newcommand{\CC}{\mathbb C}
\newcommand{\NN}{\mathbb N}
\newcommand{\ZZ}{\mathbb Z}

\newcommand{\gab}{\bar{\gamma}}

\newcommand{\tspin}{\Theta^{\vec{a}}_{\vec{b}}}
\newcommand{\spin}{^{\vec{a}}_{\vec{b}}}

\newcommand{\tspinb}{\Theta^{\vec{a}}_{\vec{b}} \left[ \begin{smallmatrix}}
\newcommand{\tspine}{\end{smallmatrix} \right]}
\def\teta#1{\tspinb #1 \tspine}
\def\vev#1{\langle #1 \rangle}
\def\vevs#1{\langle #1 \rangle \spin}

\newcommand{\ttt}{\bigl[\Theta^{\vec{a}}_{\vec{b}}(\ve{0})\bigr]^3}
\newcommand{\tttt}{\bigl[\Theta^{\vec{a}}_{\vec{b}}(\ve{0})\bigr]^4}
\newcommand{\ttttt}{\bigl[\Theta^{\vec{a}}_{\vec{b}}(\ve{0})\bigr]^5}

\allowdisplaybreaks

\begin{document}

\title{\textbf{Higher~Loop~Spin Field Correlators}\\ 
\textbf{in Various Dimensions}\\[0.5cm]}
\author{D. H\"artl, O. Schlotterer}
\date{}
\smallskip
\maketitle
\centerline{\it Max--Planck--Institut f\"ur Physik}
\centerline{\it Werner--Heisenberg--Institut}
\centerline{\it 80805 M\"unchen, Germany}

\medskip\bigskip\vskip2cm \abstract{\noindent We compute higher-point superstring correlators involving spin fields in
  various even space-time dimensions $D$ at tree-level and to arbitrary loop order. This generalizes previous work in
  $D=4$ space-time dimensions. The main focus are $D=6,8$ and $D=10$ superstring compactifications for which correlation
  functions with four and more spin fields are computed. More precisely, we present every non-vanishing six-point
  function.  A number of results can even be derived for arbitrary $D$.  A closed formula for the correlators
  $\langle\psi\ldots\psi SS \rangle$ with any number of fermions $\psi$ and two spin fields $S$ in $D$ space-time
  dimensions is given for arbitrary genus.
  
  Moreover, in $D=6$ and for arbitrary genus, we find a general formula for the correlators $\langle S \dot S \ldots S
  \dot S\rangle$. The latter serve as basic building blocks to construct higher-point fermionic correlation functions.
  In $D=8$ we can profit from the $SO(8)$-triality to derive further tree-level correlators with a large numbers of spin
  fields.}

\vskip3.5cm
\begin{flushright}
{\small  MPP--2010--2}
\end{flushright}


\thispagestyle{empty}

\newpage
\setcounter{tocdepth}{2}
\tableofcontents

\numberwithin{equation}{section}

\newpage
\section{Introduction}
\label{sec:Introduction}

Multi-leg superstring amplitudes are of both considerable theoretical interest in the framework of a full--fledged
superstring theory \cite{Stieberger:2006bh,Stieberger:2007jv,Stieberger:2007am,Medina:2006uf} and of phenomenological
relevance. Both tree- and higher-loop superstring amplitudes in diverse dimensions are important to test various aspects
of duality symmetries relating different string vacua, see e.g.\ \cite{Vafa:1997pm}. In addition, in $D=4$ dimensions
parton scattering may become relevant in describing corrections to jet cross sections from low string scale or large
extra dimension physics \cite{LHC1,LHC2,LHC3}.  Hence, it is essential to develop a detailed account of building blocks
necessary to compute amplitudes in these scenarios.

\medskip In the Ramond--Neveu--Schwarz (RNS) formalism of superstring theory, one of the main obstacles in computing
amplitudes is the interacting nature of the Neveu--Schwarz fermion (NS) $\psi^\mu$ and the Ramond (R) spin field $S_\al$.
Since these fields enter the vertex operators the underlying super conformal field theory (SCFT) correlators are hard to
obtain.  Dimensional reduction from $D=10$ to lower dimensions $D<10$ organizes the RNS fields $\psi^\mu,S_\al,S^{\dbe}$
into vector, scalar and spinor representations of the Lorentz group $SO(1,D-1)$. The calculation of various correlation
functions involving these fields at arbitrary genus $g$ in $D=6,8$ and $D=10$ is the main purpose of this paper.


\medskip In $D=4$ space-time dimensions, the problem of computing $SO(1,3)$ covariant correlation functions involving
vectors $\psi^\mu$ and spinors $S_\al, S^{\dbe}$ has been solved in complete generality at tree-level \cite{tree} and
for large classes of correlators on higher genus \cite{loop4D}.  One motivation for this paper is to extend this work to
$D=6,8$ and $D=10$ space-time dimensions -- both at tree-level and at higher genus. As we will explain later, many CFT
results for the correlators involving the fields $\psi^\mu$ and $S_\al, S^{\dbe}$ can be obtained for arbitrary even
dimension $D=2n$.

\medskip In even dimensions $D=2n$, the covariant RNS fields can be represented by $n$ independent copies of an $SO(2)$
spin system $\{\Psi_i^\pm,s_i^\pm\}$, $i=1,\dots,n$ \cite{LE1}. The fields $\Psi_i^\pm$ have conformal weight $h=1/2$
and carry Ramond charge $\pm 1$, whereas the spin fields $s_i^\pm$ have weight $1/8$ and Ramond charge $\pm 1/2$. Their
interacting CFT can be understood by bosonization. We refer to \cite{AG2,AG3} for a discussion of the subtleties arising
for spin systems on higher genus. Individual spin systems with all their correlation functions are well-understood on
arbitrary genus \cite{AS1,AS3,AS2}. Therefore all the following results have been calculated by making use of $n$
$SO(2)$ spin systems for $D=2n$.


\medskip Due to the group structure of $SO(1,D-1)$ the technical challenges in handling several spin fields become more
and more involved with increasing dimension $D$.  In the language of spin models, the combinatorical possibilities how
to combine the Ramond charges $\pm \frac{1}{2}, \, \pm 1$ rapidly grows with the space-time dimension $2n$ and the number of $SO(1,D-1)$ spin
fields involved in the correlator in question. For this reason, the generality of correlators given in this paper for
different numbers of space-time dimensions $D$ decreases from $D=6$ towards $D=10$, and only those cases with at most
two spin fields can be treated in a universal way for arbitrary $D$.

\medskip In non-compactified $D=10$ superstring theory, the GSO projected fermion vertex in its canonical
$(-1/2)$-ghost picture introduces only left-handed spin fields $S_\al$ of $SO(1,9)$. But picture changing in $D=10$ as
well as compactifying $S_\al$ to lower dimensional spinors introduce spin fields $S^{\dbe}$ with right-handed chirality
from the corresponding $SO(1,2n-1)$ point of view. It is therefore essential to consider any chirality configuration in
spin field correlations, regardless of $D$.

\medskip For amplitudes involving gluons $g$ and/or (anti-)gauginos $\chi,\bar \chi$ the necessary classes of RNS open
string correlators to be computed are the same for every $D$.  More precisely, for the tree-level $N$-~gluon amplitude
${\cal M}(g_1 \ldots g_N)$ the correlator $\langle \psi_1\ldots\psi_{2N-2} \rangle$ is required. Furthermore, for the
amplitude ${\cal M}(g_1 \ldots g_N\,\chi\,\bar\chi)$ one needs the correlator $\vev{ \psi_1\ldots\psi_{2N-1}\,S\,S}$,
while ${\cal M}(g_1 \ldots g_N\,\chi\,\bar\chi\,\chi\,\bar\chi)$ involves the correlator
$\vev{\psi_1\ldots\psi_{2N}\,S\,S\,S\,S}$. At higher genus $g \neq 0$, at most $2g$ additional fields $\psi^\mu$ may
enter the correlation functions due to the picture changing operators necessary to cancel the total superghost
background charge of $2g-2$.  In $D=4,6,8$ and $D=10$, color-stripped tree-level amplitudes of $N$ gluons and only two
massless fermions are universal whether the fermions are adjoint gauginos or chiral quarks and leptons located at
D-brane intersections \cite{LHC1,LHC2}. However, this statement generically fails at higher genus or in the presence of
further fermion pairs.  A similar observation is made for two fermion disk amplitudes involving massive higher spin
excitations of gluons, quarks or gauginos \cite{LHC3,HS} -- independent on $D$.

\medskip An other important aspect of multi-leg correlators involving many spin fields arises when studying disk
couplings of brane and bulk fields in superstring compactifications in the presence of NS and R fluxes \cite{Becker}.
These interactions are determined by computing the relevant disk amplitude involving open and closed string states. The
latter can be reduced to disk amplitude of pure open strings \cite{Stieberger:2009hq}.

\medskip 
It is instructive to give the schematic form of an $N$-point multi-loop amplitude ${\cal M}_g(\Phi_1,..., \Phi_N)$ of
open string states $\Phi_{i}$ at genus $g$ (for more details see \cite{VV1,Olaf3,HP}):
\begin{align}
  {\cal M}_g(\Phi_1,...,\Phi_N) \eq &\int \frac{\dd^{N_g} \Om}{\det \Om}  \int \frac{ \prod_{i=1}^N \dd z^i}{{\cal V}_{\te{CKG}}^g} \; \sum_{I} {\cal K}_I(\Phi_i) \, C_{X}^I(z_i,\Phi_i,\Om) \notag \\
  & \times \ \sum_{(\ve{a},\ve{b})} {\cal Z}^{(\ve{a},\ve{b})}(\Om) \, C^{I;(\ve{a},\ve{b})}_{\psi,S}(z_i,\Phi_i,\Om) \,
  C^{(\ve{a},\ve{b})}_{\te{ghost}}(z_i, \Phi_i,\Om) \, C^{(\ve{a},\ve{b})}_{\te{int}}(z_i, \Phi_i,\Om)\,.
\label{amp}
\end{align}
The only information about the SCFT of the internal geometry comes from the genus $g$ partition function ${\cal
  Z}^{(\ve{a},\ve{b})}$ and the internal CFT correlators $C_{\te{int}}$\footnote{The latter may be built in and
  described by some character valued partition function or elliptic genus \cite{EG1,EG2,EG3,EG4,EG5,EG6}}. The internal
part of the spin fields and their higher loop interactions depend on the compactification details. They are not further
discussed in this work. The superghost contribution $C_{\te{ghost}}$, on the other hand, is model independent and
well-understood \cite{ghost}.  The space-time kinematics in the amplitude is determined by the correlation function
built from the bosonic string coordinates $X^\mu$ and the correlators involving the RNS fields $\psi^\mu,S_\al,
S^{\dbe}$. The latter give rise to non-trivial $SO(1,D-1)$ Lorentz structures in the amplitude. In these lines, the
index $I$ in \eqref{amp} refers to a set of kinematical terms ${\cal K}_I$, which originate from contractions of the
space-time fields $X^\mu, \psi^\mu, S_\al, S^{\dbe}$, i.e. from correlators with non-trivial Lorentz structure. We
denote the dependence on $(z_i,\Om)$ associated with the contraction ${\cal K}_I$ by $C_{X}^I$ and $C_{\psi,S}^I$.
 
\medskip Further steps towards computing the amplitude are, first, the sum over spin structures $(\ve{a},\ve{b})$ of the
partition function ${\cal Z}^{(\ve{a},\ve{b})}$ as well as three $(\ve{a},\ve{b})$ dependent correlation functions and,
second, the integrals $\int \frac{ \prod_{i=1}^N \dd z^i}{{\cal V}_{\te{CKG}}^g}$ over world-sheet positions and $\int
\frac{\dd^{N_g} \Om}{\det \Om}$ over the genus $g$ moduli space. The former can be performed by means of generalized
Riemann identities \cite{Olaf3,fay,mum,igu,Sloop1,Sloop2} whereas the latter still lacks a unified treatment and waits
for investigation in future work.
 
\medskip The organization of this work is as follows: We start in Section \ref{sec_review} by reviewing the RNS CFT in
$D$ dimensions and the genus $g$ correlation functions of the underlying $SO(2)$ spin models. In Section \ref{sec_uni},
correlation functions with two spin fields and arbitrary numbers of NS fermions are generalized from $D=4$ (see
\cite{tree,loop4D}) to general $D$. From Section \ref{sec_d6} to \ref{sec_d10} we discuss correlators with four and more
spin fields for the cases $D=6,8,10$. The $D=6$--dimensional case in Section \ref{sec_d6} still has a sufficiently simple
group structure that it admits a general solution for $2N$-point correlation function $\langle \prod_{i=1}^N
S_{\al_i}(x_i) S^{\dbe_i}(y_i) \rangle$. This is no longer possible in $D=8$ dimensions, but as explained in Section
\ref{sec_d8}, the triality symmetry of the $SO(8)$ Lorentz group simplifies the computation of many tree-level
correlators with large numbers of spin fields. Both these simplifying features are absent in the non-compactified case
$D=10$. So Section \ref{sec_d10} is a collection of correlators with a finite number of fields which can be computed
with reasonable effort. In all these dimension $D=6,8,10$, we have computed every six-point function (with every possible
combination of $\psi^\mu,S_\al,S^{\dbe}$) on arbitrary genus. Two appendices provide technical details of manipulating
gamma matrices.

\section{Review}
\label{sec_review}

\subsection{The RNS CFT in even dimensions $D$}
\label{review1}

Let us first of all introduce the interacting CFT of the Neveu--Schwarz fermion $\psi^\mu$ with the Ramond spin fields
$S_\al, S^{\dbe}$ in even space-time dimension $D$ \cite{FMS,Cohn,LE1}. For this purpose, the Dirac notation
$(\Ga^\mu)_A \, \! ^B$ for gamma matrices and spin fields $S_A \equiv S_\al \oplus S^{\dbe}$ is most convenient. See
Appendix \ref{appA} for details of the decomposition into chiral halves.

\medskip
The singular behavior of the NS fermions as well as the interaction between fermions and spin fields are insensitive to the number of dimensions we are working in:
\begin{subequations}
\label{ope}
\begin{align}
\psi^{\mu}(z)\;\psi^{\nu}(w) \  &= \ \frac{\eta^{\mu \nu}}{z-w}\ +\ \dots\,,
\label{ope1} \\
\psi^{\mu}(z) \; S_{A}(w) \ &= \ \frac{(\Ga^{\mu})_A \, \! ^B}{\sqrt{2}}\;(z-w)^{-1/2} \; S_B(w)\ +\ \dots  \label{ope2}\,.
\end{align}
\end{subequations}
On the other hand, conformal weights and therefore the mutual short distance behavior of the spin fields do depend on
the dimensionality. The most singular term of the $D$-dimensional spin field self-OPE depends on the relative chirality
of the spin fields. This is why we display two contributions\footnote{Here we are redefining the spin fields by a factor
  of $i$ relative to the convention of \cite{tree} in order to avoid proliferation of minus signs. For comparison with
  this reference, correlators with $2n$ spin fields need to be multiplied by $(-1)^n$.}:
\begin{equation}
S_A(z) \; S_B(w) \eq (z-w)^{-D/8} \, {\cal C}_{AB} \ + \ \frac{ (\Ga^\mu \, {\cal C})_{AB} }{\sqrt{2}} \; (z-w)^{-D/8 \, +  \, 1/2} \, \psi_\mu(w) \ + \ \dots\,.  \label{ope20}
\end{equation}
Depending on the chiral structure of the charge conjugation matrix ${\cal C}$, we obtain different scenarios:
\begin{itemize}
\item $D \ = \ 0 \ \te{mod} \ 4$ OPEs 
\begin{subequations}
\label{ope4}
\begin{align}
  S_{\al}(z)\;S_{\be}(w) \ &= \ (z-w)^{-D/8}\, C_{\al \be}\ +\ \dots \ , \label{ope43} \\
 S_{\al}(z)\;S^{\dbe}(w) \ &= \ \frac{(\ga^\mu \, C)_\al \, \!^{\dbe}}{\sqrt{2}}\;(z-w)^{-D/8 \, + \, 1/2} \,\psi_{\mu}(w)\ +\ \dots \ , \label{ope44}
\end{align}
\end{subequations}
\item $D \ = \ 2 \ \te{mod} \ 4$ OPEs 
\begin{subequations}
\label{ope6}
\begin{align}
  S_{\al}(z)\;S^{\dbe}(w) \ &= \ (z-w)^{-D/8}\,C_{\al} \, ^{\dbe}\ + \ \dots\ , \label{ope63} \\
 S_{\al}(z)\;S_{\be}(w) \ &= \ \frac{(\ga^{\mu} \, C)_{\al \be}}{\sqrt{2}}\;(z-w)^{-D/8 + 1/2}\, \psi_{\mu}(w)\ + \ \dots\,.  \label{ope64}
\end{align}
\end{subequations}
\end{itemize}
It is still possible to factorize fermions like in four dimensions \cite{tree} by multiplying the OPE (\ref{ope20}) with $( {\cal C}^{-1} \Ga^\nu)^{BA}$. Using $\te{Tr} \bigl\{ \Ga^\mu \Ga^\nu \bigr\} = -2^{D/2} \eta^{\mu \nu}$, we conclude
\begin{equation}
\psi^\mu(w) \eq - \, 2^{(1-D)/2} \, ({\cal C}^{-1} \, \Ga^{\mu})^{ BA } \; \lim_{z \mto w} (z-w)^{D/8-1/2} \, S_{A}(z) \, S_{B}(w) \ .
\label{factorD}
\end{equation}
In particular, the chiral trace relation $\te{Tr} \bigl\{ \ga^\mu \gab^\nu \bigr\} = -2^{(D-2)/2} \eta^{\mu \nu}$ admits to invert the two subcases (\ref{ope44}) and (\ref{ope64}):
\begin{equation}
\psi^\mu(w) \eq - \, 2^{(3-D)/2}  \, \lim_{z \mto w} (z-w)^{D/8 - 1/2} \ \times \ \left\{ \begin{array}{cl} \displaystyle
 (C^{-1} \, \gab^{\mu}) _{\dbe} \, \! ^{\al} \,  S_{\al}(z) \, S^{\dbe}(w) &\ \ : \, D \ = \ 0 \ \te{mod} \ 4\,, \\ [.4cm] \displaystyle
 (C^{-1} \, \gab^{\mu})^{ \be \al } \, S_{\al}(z) \, S_{\be}(w) &\ \ : \, D \ = \ 2 \ \te{mod} \ 4\,. \end{array} \right.
\label{factorize}
\end{equation}
However, one must admit that this factorization technique, which was essential in solving the $D=4$ case \cite{tree,
  loop4D} loses its efficiency for computing unknown correlators with increasing dimension because the spinor
representations become more and more complex in higher dimensions. This can be seen best by comparing the dimension $D$
of the vector representation of $SO(1,D-1)$ with the number $2^{(D-2)/2}$ of Weyl spinor components -- exponential
growth of the spinor clearly dominates over linear growth of the vector.


\subsection{Spin system correlators and theta functions}

Correlations functions of the NS fermions $\psi^\mu$ and the R spin fields $S_\al$ in $D=2n$ dimensions can be assembled
out of the $SO(2)$ spin system as discussed in \cite{AS1, AS3, AS2}, one needs $n$ independent copies of the $SO(2)$
spin system $\{\Psi_i^\pm,s_i^\pm\}$, $i=1,\dots,n$. The fields $\Psi_i^\pm$ are conformal fields of dimension $h=1/2$
and carry Ramond charge $\pm 1$, whereas the spin fields $s_i^\pm$ have $h=1/8$ and Ramond charge $\pm 1/2$. The
$\Psi_i^{\pm}$ from the spin system are the Cartan--Weyl representation of the $SO(1,D-1)$ vector $\psi^\mu$
\begin{subequations}
\label{reppsi}
\begin{align}
  \psi^{2i-1}(z) & \ \ \equiv \ \ \frac{1}{\sqrt{2}} \; \bigl(\Psi^+_i(z)+\Psi^-_i(z) \bigr)\,, \label{reps1} \\
  \psi^{2i}(z) & \ \ \equiv \ \  \frac{1}{\sqrt{2} \, i} \; \bigl(\Psi^+_i(z)-\Psi^-_i(z) \bigr)\,, \label{reps2}
\end{align}
\end{subequations}
whereas the R spin fields can be written as
\begin{equation}
  \label{reps}
  S_A(z) \eq \bigotimes_{i=1}^n s_i^\pm(z)\,.
\end{equation}
Since each of the $n$ Ramond charges can be chosen independently, there are $2^n=2^{D/2}$ such operators. This coincides
with the number of components of a Dirac spinor in $D$ dimensions. We take the convention that operators with an even
number of $s^-$ operators are left-handed, whereas those with an odd number are right-handed. Correlation functions of $SO(1,D-1)$ covariant RNS fields factorize into correlators of a single spin system.

\medskip Generalized theta functions \cite{fay,mum,igu} are the natural objects to express correlation functions at
non-zero genus. They assure the required periodicity along the homology cycles of the $g$-loop string world-sheet. They
can be derived from
\begin{equation}
\Theta( \ve{x} \, | \, \Om) \, \equiv \, \sum_{\ve{n} \in \ZZ^g} \exp \left[ 2\pi i \, \left( \tfrac{1}{2} \; \ve{n} \, \Om \, \ve{n} \ + \ \ve{n} \, \ve{x} \right) \right]
\label{defT1}
\end{equation}
by shifting the first argument according to some spin structure $(\ve{a},\ve{b})$:
\begin{align}
\Theta^{\ve{a}}_{\ve{b}}
 ( \ve{x} \, | \, \Om) \,  &\equiv \, \exp \left[ 2\pi i \, \left( \tfrac{1}{8} \; \ve{a} \, \Om \, \ve{a} \ + \ \tfrac{1}{2} \; \ve{a} \, \ve{x}  \ + \ \tfrac{1}{4} \; \ve{a}\, \ve{b} \right) \right] \, \Theta  \left( \ve{x} \, + \, \tfrac{\ve{b}}{2}  \, + \, \tfrac{\Om \, \ve{a}}{2} \, | \, \Om \right) \notag \\
    & = \, \sum_{\ve{n} \in \ZZ^g} \exp \left[ \pi i \, \left(  \ve{n} \, + \, \tfrac{\ve{a}}{2} \right) \, \Om \, \left(  \ve{n} \, + \, \tfrac{\ve{a}}{2} \right) \ + \ 2\pi i \, \left(  \ve{n} \, + \, \tfrac{\ve{a}}{2} \right) \, \left( \ve{x} \, + \, \tfrac{\ve{b}}{2} \right) \right]\,.
\label{defT2}
\end{align}
In our situations, the $g$-dimensional vectors $\ve{a},\ve{b}$ with entries zero or one characterize the periodicity of the fermion fields along the $2g$ homology cycles of the Riemann surface. The second argument of $\Theta$ is the $g \times g$ period matrix $\Om$.

\medskip We parametrize the two-dimensional string world-sheet by a complex coordinate $z$. The Abel map $z \mapsto
\smallint^z_{p} \ve{\om}$ lifts $z$ to the Jacobian variety of the world-sheet $\CC^g / (\ZZ^g + \Om \ZZ^g)$. These
integrals are then natural arguments for the theta function. The periodicity properties of the theta function under
transport of $z$ around a homology cycle are summarized in Appendix A of \cite{loop4D}.

\medskip
An important expression constructed out of the generalized theta functions is the prime form $E$,
\begin{equation}
E(z,w) \, \equiv \, \frac{\Theta^{\ve{a_0}}_{\ve{b}_0} \left( \int^z_w \ve{\om} \, | \, \Om \right) }{ h^{\ve{a}_0}_{\ve{b}_0}(z)\,h^{\ve{a}_0}_{\ve{b}_0}(w)}\,,
\end{equation}
where $(\ve{a_0},\ve{b_0})$ is an arbitrary odd spin structure such that $E(z,w)=-E(w,z)$. The half differentials
$h^{\ve{a}_0}_{\ve{b}_0}$ in the denominator are defined by
\begin{equation}
h^{\ve{a}_0}_{\ve{b}_0}(z) \, \equiv \, \sqrt{ \sum_{j=1}^g \, \om_j(z) \, \pa_j \Theta^{\ve{a}_0}_{\ve{b}_0}\left( \ve{0} \, | \, \Om \right)  } \,.
\end{equation}
They assure that $E$ is independent of the choice of $(\vec{a}_0,\vec{b}_0)$ as long as it is odd. Given the leading behaviour $E(z,w) \sim z-w \, + \, {\cal O} \bigl( (z-w)^3 \bigr)$, singularities in correlation functions are caused by appropriate powers of prime forms.

\medskip
Using these expressions, the correlator of an arbitrary number of fermions $\Psi^\pm$ and spin fields $s^\pm$ of an $SO(2)$ spin model can be derived for every
loop order $g$ \cite{AS2}
\begin{align}
&\left \langle \prod_{i=1}^{N_1} s^{+}(y_i) \, \prod_{j=1}^{N_2} s^{-}(z_j) \, \prod_{k=1}^{N_3} \Psi^{-}(u_k) \, \prod_{l=1}^{N_4} \Psi^{+}(v_l) \right \rangle^{\ve{a}}_{\ve{b}} \eq \left( \frac{\prod_{r<s}^{N_1} E(y_r,y_s) \, \prod_{r<s}^{N_2} E(z_r,z_s) }{ \prod_{i=1}^{N_1} \prod_{j=1}^{N_2} E(z_j,y_i) } \right)^{1/4}   \notag \\
& \ \ \ \times \,  \left( \frac{ \prod_{r<s}^{N_3} E(u_r,u_s) \, \prod_{r<s}^{N_4} E(v_r,v_s)  }{ \prod_{k=1}^{N_3} \prod_{l=1}^{N_4} E(v_l,u_k) }\right) \, \left( \frac{ \prod_{j=1}^{N_2} \prod_{k=1}^{N_3} E(u_k,z_j) \,  \prod_{i=1}^{N_1} \prod_{l=1}^{N_4} E(v_l,y_i) }{  \prod_{i=1}^{N_1} \prod_{k=1}^{N_3} E(u_k,y_i) \, \prod_{j=1}^{N_2} \prod_{l=1}^{N_4} E(v_l,z_j)} \right)^{1/2}    \notag \\
& \ \ \ \times \, \left( \tspin(\ve{0}) \right)^{-1} \ \tspin \left( \tfrac{1}{2}  \sum_{i=1}^{N_1} \smallint^{y_i}_{p} \ve{\om} \, - \, \tfrac{1}{2}  \sum_{j=1}^{N_2} \smallint^{z_j}_{p} \ve{\om} \, - \, \sum_{k=1}^{N_3} \smallint^{u_k}_{p} \ve{\om} \, + \,  \sum_{l=1}^{N_4} \smallint^{v_l}_{p} \ve{\om}\right)\,.
\label{at2}
\end{align}
Due to Ramond charge conservation $\half(N_1-N_2)-N_3+N_4 = 0$, the arbitrary reference point $p$ appearing in the Abel map
drops out. In the following we make use of the shorthands $E_{ij}\equiv E(z_i,z_j)$ and, from section \ref{sec_d6} on,
 \begin{equation}
   \tspin \left( \, \frac{1}{2} \, \left[  \smallint^{z_i}_{z_l} \ve{\om} \, + \,  \smallint^{z_j}_{z_m} \ve{\om} \, + \dots  + \,  \smallint^{z_k}_{z_n} \ve{\om} \right] \, \right) \, \equiv \, \tspin \left[ \begin{smallmatrix} i &j &\ldots &k \\ l &m &\ldots &n \end{smallmatrix} \right]\,.
\end{equation}
Note in particular that the factor $\frac{1}{2}$ in the argument of $\tspin$ -- which is ubiquitous in presence of spin fields -- will always be implicit.

\medskip Considerable simplifications occur for $g=0$ or $g=1$, i.e.\ scattering at tree-level or one loop. For $g=1$
the period matrix $\Om$ reduces to the modular parameter $\tau$ of the torus and the theta functions become the standard
ones:
\begin{equation}
  \theta_1 \equiv \Theta^1_1\,,\quad \theta_2 \equiv \Theta^1_0\,,\quad \theta_3 \equiv \Theta^0_0\,,\quad \theta_4 \equiv \Theta^0_1\,.
\end{equation}
On a $g=0$ world-sheet, the spin structure dependent theta functions trivialize, $\Theta^{\ve{a}}_{\ve{b}} \to 1$, and
the prime form reduces to $E(z,w)\to z-w$.

\subsection{The algorithm}

With the background on spin systems in mind, we can now calculate RNS correlation functions $\vevs{\psi^{\mu_1}\dots\psi^{\mu_\ell}\,S_{\al_1}\dots
  S_{\al_m}\,S^{\dbe_1} \dots S^{\dbe_n}}$ for specific choices of $\mu_i,\al_i$ and $\dbe_i$ by organizing the RNS
fields into their spin system content via \eqref{reppsi}, \eqref{reps} and applying \eqref{at2} for the $n$ individual $SO(2)$ correlators. The final
goal is to express the results in a covariant form, i.e.\ in terms of Clebsch--Gordan coefficients. These are built from
gamma matrices and the charge conjugation matrix and hence carry all the vector and spinor indices. They can be viewed
as $SO(1,2n-1)$ covariant Ramond charge conserving delta functions, schematically $C_{\al \be} \sim \de(\al + \be)$ and
$(\ga^\mu C)_{\al}{}^{\dbe} \sim \de(\mu + \al + \dbe)$ where $\mu,\al,\be,\dbe$ are treated as $n$ component Ramond
charge vectors such as $\mu \equiv (0,\pm1,0, \ldots,0)$ and $\al \equiv (\pm \frac{1}{2}, \ldots, \pm \frac{1}{2})$.

\medskip As a starting point we make an ansatz for the correlation function with a minimal set of Clebsch--Gordan
coefficients. Each of these index terms is accompanied by a $z$-dependent coefficient consisting of prime forms $E$ and
theta functions $\Theta^{\ve{a}}_{\ve{b}}$. The results obtained for special choices of $\mu_i,\al_i$ and $\dbe_i$ have
to be matched with this ansatz. It is most economic to first look at configurations $(\mu_i,\al_i,\dbe_i)$ where only
one tensor is non-zero. Then the loop-level result \eqref{at2} yields the coefficient for the respective index term. In
some cases, however, it is not possible to make all Clebsch Gordan coefficients vanish except for one, then more than one index term contributes for every choice of
$(\mu_i,\al_i,\dbe_i)$. Then, it can be helpful to switch to different Lorenz tensors which are (anti-)symmetric in
some vector- or spinor indices, see Appendix \ref{appA.2}. In other cases, Fay's trisecant identities \cite{fay,loop4D} have to be used
to determine the unknown coefficients. Sign issues can be resolved by calculating certain limits $z_i \rightarrow z_j$
at tree-level using the RNS OPEs \eqref{ope4} or \eqref{ope6}.

\medskip Let us illustrate this procedure with an easy example, the correlation function $\vevs{\psi^\mu \psi^\nu
  \psi^\la S_\al S_{\be}}$ in $D=6$ dimensions. A convenient ansatz in terms of four Clebsch--Gordan coefficients is
\begin{align}
\vevs{\psi^\mu(z_1) \, \psi^\nu(z_2) \, \psi^\la(z_3) \,& S_\al(z_4) \, S_{\be}(z_5)} \,=\, F_1(z)  \, (\ga^{\mu \nu \la} \, C)_{\al \be}\notag\\
&+\ F_2(z) \, \eta^{\mu \nu} \, (\ga^\la \, C)_{\al \be} \ + \ F_3(z) \, \eta^{\mu \la} \, (\ga^\nu \, C)_{\al \be} \ + \ F_4(z) \, \eta^{ \nu \la } \, (\ga^\mu \, C)_{\al \be}\,.
\label{example1}
\end{align}
The task is now to determine $F_1,F_2,F_3,F_4$ by making several choices for $\mu,\nu,\la,\al,\be$.

The coefficient $F_1$ can easily be obtained by setting $\mu=0,\,\nu=2,\,\la=4$. As the metric $\eta$ is diagonal all
the other index terms vanish for this configuration. Then by means of \eqref{reppsi} the NS fermions become
$(\psi^\mu,\psi^\nu,\psi^\la)=\frac{1}{\sqrt{2}}\,(\Psi_1^++\Psi_1^-,\Psi_2^++\Psi_2^-,\Psi_3^++\Psi_3^-)$ and we choose
for the spin fields $S_\al=S_\be=(s_1^+,s_2^-,s_3^-)$.  Hence we have to calculate
\begin{equation}
  \frac{1}{2\,\sqrt{2}}\,\vevs{(\Psi_1^++\Psi_1^-)(z_1)\,s_1^+(z_4)\,s_1^+(z_5)}\,\vevs{(\Psi_2^++\Psi_2^-)(z_2)\,s_2^-(z_4)\,s_2^-(z_5)}\,\vevs{(\Psi_3^++\Psi_3^-)(z_3)\,s_3^-(z_4)\,s_3^-(z_5)}\,.
\end{equation}
Due to Ramond charge conservations $\Psi_1^+(z_1),\,\Psi_2^-(z_2)$ and $\Psi_3^-(z_3)$ drop out and by using \eqref{at2} we
obtain the coefficient $F_1$ up to a sign
\begin{equation}
  F_1 \eq \pm \, \frac{\teta{1&1\\4&5}\,\teta{2&2\\4&5}\,\teta{3&3\\4&5}\, E_{45}^{3/4}}{ 2\,\sqrt{2} \, \ttt \, (E_{14}\,E_{15}\,E_{24}\,E_{25}\,E_{34}\,E_{35})^{1/2}}\,.
\end{equation}
The coefficient $F_2$ can be determined in a similar way by setting $\mu=\nu=0,\,\la=2$ and $S_\al=(s_1^+,s_2^+,s_3^+)$,
$S_\be=(s_1^-,s_2^+,s_3^-)$. No other tensor than $\eta^{\mu\nu}\,(\ga^\la\,C)_{\al\be}$ contributes as the
metric is diagonal and $\ga^{\mu\nu\la}$ totally antisymmetric. One finds that the results consists of two terms due to
the two in-equivalent fermion configurations $\Psi_1^+(z_1)\,\Psi_1^-(z_2)$ and $\Psi_1^-(z_1)\,\Psi_1^+(z_2)$ in the
first spin system. 
\begin{equation}
  F_2\eq \pm\,\frac{ \teta{3 &3 \\ 4 &5} \, \teta{4\\ 5} \, \bigl(
E_{14} \, E_{25} \,  \teta{1 &1 &4 \\ 2 &2 &5} \ + \ E_{15} \, E_{24} \,  \teta{1 &1 &5 \\ 2 &2 &4} \bigr)
  }{2\,\sqrt{2} \, \ttt \,  E_{12} \, (E_{14}\,E_{15}\,E_{24}\,E_{25}\,E_{34}\,E_{35})^{1/2} \, E_{45}^{1/4}}\,.
\end{equation}
The remaining $z_i$ functions $F_3$ and $F_4$ follow from $F_2$ by permutation in the vector indices and the $(1,2,3)$ labels.

The signs of the individual coefficients are easily fixed by requiring $\frac{\eta^{\mu\nu}}{z_{12}} \langle
\psi^\la(z_3) S_\al(z_4) S_{\be}(z_5)\rangle \spin$ to emerge in the $z_1 \rightarrow z_2$ limit or $\frac{(\ga^\rho
  C)_{\al \be}}{z_{45}^{1/4}} \, \langle \psi^\mu(z_1) \psi^\nu(z_2) \, \psi^\la(z_3) \psi_\rho(z_5) \rangle$ in the
$z_4 \rightarrow z_5$ limit for instance.

\section{Universal correlators: Two spin fields}
\label{sec_uni}

Correlators of type $\langle \psi^{\mu_1} ... \psi^{\mu_n} S_\al S_{\be} \rangle$ and $\langle \psi^{\mu_1} \dots
\psi^{\mu_n} S_\al S^{\dbe} \rangle$ have been calculated in full generality for $D=4$ dimensions in
\cite{tree,loop4D}. It turns out that their structure is almost unchanged in higher dimensions. The only thing we have
to pay attention to is the relative chirality of spin fields in non-zero correlations, but this subtlety can be bypassed in Dirac spinor notation, see our final result (\ref{OmegaD}) and (\ref{omegaD}) for general even $D$.

\subsection{The four-dimensional result}

In four dimensions, correlators $\langle \psi^n S S \rangle$ with an odd number $n \in 2\NN-1$ of fermions require spin fields of opposite chirality for a non-vanishing result. In \cite{loop4D}, it was shown by induction that these $2n+1$-point functions are given as follows:

\bigskip
\framebox{\begin{minipage}{6.3in}
\begin{align}
& \ \ \Om_{(n,D=4)}^{\mu_{1} ... \mu_{2n-1}}\,_{\al}{}^{\dbe}(z_{i})  \ \ := \ \
\langle \psi^{\mu_{1}}(z_{1}) \, \psi^{\mu_{2}}(z_{2}) \, ... \, \psi^{\mu_{2n-1}}(z_{2n-1}) \, S_{\al}(z_{A}) \, S^{\dbe}(z_{B}) \rangle \spin \ \Bigl. \Bigr|_{D=4} \notag \\
& \ \  = \ \ \frac{\left[ \tspin \left( \tfrac{1}{2} \smallint^{z_A} _{z_B} \ve{\om} \right) \right]^{2-n}}{\sqrt{2} \, \tspin ( \ve{0} ) \, \tspin ( \ve{0} )  \, \prod_{i=1}^{2n-1} (E_{iA} \, E_{iB})^{1/2} } \, \sum_{\ell = 0}^{n-1} \, \biggl( \frac{E_{AB}}{2 \, \tspin \left( \tfrac{1}{2} \smallint^{z_A} _{z_B} \ve{\om} \right)} \biggr)^{\ell} \notag \\
& \ \ \ \ \times \sum_{\rho \in S_{2n-1}/{\cal P}_{n,\ell}} \! \! \!  \te{sgn}(\rho) \, \bigl(\si^{\mu_{\rho(1)}} \, \bar{\si}^{\mu_{\rho(2)}} \, ... \, \bar{\si}^{\mu_{\rho(2\ell)}} \, \si^{\mu_{\rho(2\ell+1)}} \, \vep \bigr)_{\al} \, \!^{\dbe} \, \prod_{k=1}^{2\ell+1} \tspin \left( \tfrac{1}{2} \smallint^{z_A} _{z_{\rho(k)}} \ve{\om} \, + \, \tfrac{1}{2} \smallint^{z_B} _{z_{\rho(k)}} \ve{\om} \right) \notag \\
& \ \ \ \ \times \  \prod_{j=1}^{n-\ell-1} \frac{\eta^{\mu_{\rho(2\ell+2j)} \mu_{\rho(2\ell+2j+1)}}}{E_{\rho(2\ell+2j),\rho(2\ell+2j+1)} } \; E_{\rho(2\ell+2j),A} \, E_{\rho(2\ell+2j+1),B}  \, \tspin \left( \smallint^{z_{\rho(2\ell+2j)}}_{z_{\rho(2\ell+2j+1)}} \ve{\om} \, + \, \tfrac{1}{2} \smallint^{z_A} _{z_B} \ve{\om} \right)\,.
\label{Omega}
\end{align}
\end{minipage}}

\bigskip
\noindent
Their relatives with even number of NS fermions and two alike spin fields read
\bigskip

\framebox{\begin{minipage}{6.3in}
\begin{align}
& \ \ \om_{(n,D=4)}^{\mu_{1} ... \mu_{2n-2}}\,_{\al \be}(z_{i})  \ \ := \ \
\langle \psi^{\mu_{1}}(z_{1}) \, \psi^{\mu_{2}}(z_{2}) \, ... \, \psi^{\mu_{2n-2}}(z_{2n-2}) \, S_{\al}(z_{A}) \, S_{\be}(z_{B}) \rangle \spin \ \Bigl. \Bigr|_{D=4} \notag \\
& \ \  = \ \ \frac{ \left[ \tspin \left( \tfrac{1}{2} \smallint^{z_A} _{z_B} \ve{\om} \right) \right]^{3-n}}{ \tspin ( \ve{0} ) \, \tspin ( \ve{0} ) \, E_{AB}^{1/2}  \, \prod_{i=1}^{2n-2} (E_{iA} \, E_{iB})^{1/2} } \, \sum_{\ell = 0}^{n-1} \, \biggl( \frac{E_{AB}}{2 \, \tspin \left( \tfrac{1}{2} \smallint^{z_A} _{z_B} \ve{\om} \right)} \biggr)^{\ell} \notag \\
& \ \ \ \ \times \sum_{\rho \in S_{2n-2}/{\cal Q}_{n,\ell}} \! \! \!  \te{sgn}(\rho) \, \bigl(\si^{\mu_{\rho(1)}} \, \bar{\si}^{\mu_{\rho(2)}} \, ... \, \bar{\si}^{\mu_{\rho(2\ell)} } \, \vep \bigr)_{\al \be} \, \prod_{k=1}^{2\ell} \tspin \left( \tfrac{1}{2} \smallint^{z_A} _{z_{\rho(k)}} \ve{\om} \, + \, \tfrac{1}{2} \smallint^{z_B} _{z_{\rho(k)}} \ve{\om} \right) \notag \\
& \ \ \ \ \times \  \prod_{j=1}^{n-\ell-1} \frac{\eta^{\mu_{\rho(2\ell+2j-1)} \mu_{\rho(2\ell+2j)}}}{E_{\rho(2\ell+2j-1),\rho(2\ell+2j)} } \; E_{\rho(2\ell+2j-1),A} \, E_{\rho(2\ell+2j),B}  \, \tspin \left( \smallint^{z_{\rho(2\ell+2j-1)}}_{z_{\rho(2\ell+2j)}} \ve{\om} \, + \, \tfrac{1}{2} \smallint^{z_A} _{z_B} \ve{\om} \right) \ .
\label{omega}
\end{align}
\end{minipage}}

\bigskip
\noindent
We are using the notation from  Wess \& Bagger for the four-dimensional Clebsch--Gordan coefficients. They fit into our $D$-dimensional framework via $\si^\mu_{\al \dbe} \equiv \ga^\mu_{\al \dbe} \bigl. \bigr|_{D=4}$ and $\vep_{\al \be} \equiv C_{\al \be} \bigl. \bigr|_{D=4}$.

\medskip
The summation ranges $\rho \in S_{2n-1}/{\cal P}_{n,\ell}$ and $\rho \in S_{2n-2}/{\cal Q}_{n,\ell}$ certainly require some explanation. The conventions are taken from \cite{tree} where a more exhaustive presentation can be found. Formally, we define
\begin{subequations}
\begin{align}
S_{2n-1}/ {\cal P}_{n,\ell} \ \equiv \ \Bigl\{ &\rho \in S_{2n-1} : \ \rho(1) < \rho(2) < ... < \rho(2\ell + 1) \ , \Bigr. \notag \\
&\rho(2\ell + 2j) < \rho(2\ell + 2j + 1) \ \forall \ j = 1,2,...,n-\ell - 1 \ , \notag \\
\Bigl. &\rho(2\ell + 3) < \rho(2\ell + 5) < ... < \rho(2n -1) \Bigr\} 
\label{npt,4}\ , \\
S_{2n-2}/ {\cal Q}_{n,\ell} \ \equiv \ \Bigl\{ &\rho \in S_{2n-2} : \ \rho(1) < \rho(2) < ... < \rho(2\ell) \ , \Bigr. \notag \\
&\rho(2\ell + 2j-1) < \rho(2\ell + 2j) \ \forall \ j = 1,2,...,n-\ell - 1 \ , \notag \\
\Bigl. &\rho(2\ell + 2) < \rho(2\ell + 4) < ... < \rho(2n -2) \Bigr\} \ .
\label{npt,5}
\end{align}
\end{subequations}
In other words, the sums over $S_{2n-1}/{\cal P}_{n,\ell}$- and $S_{2n-2}/{\cal Q}_{n,\ell}$ in (\ref{Omega}) and
(\ref{omega}) run over those permutations $\rho$ of $(1,2,...,2n-1)$ or $(1,2,...,2n-2)$ which satisfy the following
constraints:
\begin{itemize}
\item Only ordered $\si$ products are summed over: The indices $\mu_{\rho(i)}$ attached to a chain of $\si$ matrices are
  increasingly ordered, e.g. whenever the product $\si^{\mu_{\rho(i)}} \bar{\si}^{\mu_{\rho(j)}} \si^{\mu_{\rho(k)}}$
  appears, the sub-indices satisfy $\rho(i) < \rho(j) < \rho(k)$.
\item On each metric $\eta^{\mu_{\rho(i)} \mu_{\rho(j)}}$ the first index is the ``lower'' one, i.e.\ $\rho(i) < \rho(j)$.
\item Products of several $\eta$'s are not double counted. So once we get $\eta^{\mu_{\rho(i)} \mu_{\rho(j)}}
  \eta^{\mu_{\rho(k)} \mu_{\rho(l)}}$, the term $\eta^{\mu_{\rho(k)} \mu_{\rho(l)}} \eta^{\mu_{\rho(i)} \mu_{\rho(j)}}$
  does not appear.
\end{itemize}
These restrictions on the occurring $S_{2n-1}$ (or $S_{2n-2}$) elements are abbreviated by a quotient ${\cal
  P}_{n,\ell}$ and ${\cal Q}_{n,\ell}$. The subgroups removed from $S_{2n-1}$ ($S_{2n-2}$) are $S_{2\ell+1} \times
S_{n-\ell - 1} \times (S_{2})^{n-\ell-1}$ and $S_{2\ell} \times S_{n-\ell - 1} \times (S_{2})^{n-\ell-1}$ respectively,
therefore the number of terms in (\ref{Omega}) and (\ref{omega}) at fixed $(n,\ell)$ is given by
\begin{subequations}
\begin{align}
\bigl| S_{2n-1} / {\cal P}_{n,\ell} \bigr| \ \ &= \ \ \frac{(2n\, - \, 1)!}{(2\ell \, + \, 1)! \, (n \, - \, \ell \, - \, 1)! \, 2^{n-\ell-1}}\ , \label{npt,7a} \\
\bigl| S_{2n-2} / {\cal Q}_{n,\ell} \bigr| \ \ &= \ \ \frac{(2n\, - \, 2)!}{(2\ell )! \, (n \, - \, \ell \, - \, 1)! \, 2^{n-\ell-1}}\ .
\label{npt,7}
\end{align}
\end{subequations}
To have some easy examples, let us explicitly evaluate the sums over $S_{2n-1} / {\cal P}_{n,\ell}$ and $S_{2n-2} / {\cal Q}_{n,\ell}$ occurring in the five- and six-point functions $\langle \psi^\mu \psi^\nu \psi^\la S_\al S^{\dbe} \rangle$ and $\langle \psi^\mu \psi^\nu \psi^\la \psi^\rho S_\al S_{\be} \rangle$. The formula (\ref{Omega}), applied to $n=2$, schematically tell us that (up to a $z$ dependent pre-factor)
\begin{align}
\langle \psi^{\mu_1} \,  &\psi^{\mu_2} \, \psi^{\mu_3} \, S_\al \, S^{\dbe} \rangle \ \ \sim \ \ \sum_{\ell = 0}^{1}   \sum_{\rho \in S_{3}/{\cal P}_{2,\ell}} \! \! \!  \te{sgn}(\rho) \, \bigl(\si^{\mu_{\rho(1)}} \, ... \, \si^{\mu_{\rho(2\ell+1)}} \, \vep \bigr)_{\al} \, \!^{\dbe} \, f^{\rho}_{\ell}(z_i) \notag \\
&\sim \ \ \underbrace{(\si^{\mu_3} \, \vep )_{\al} \, \! ^{\dbe} \,\eta^{\mu_1 \mu_2} \,  f^{(312)}_{\ell=0} \ - \ (\si^{\mu_2} \, \vep)_{\al} \, \! ^{\dbe} \, \eta^{\mu_1 \mu_3} \,  f^{(213)}_{\ell=0} \ + \ (\si^{\mu_1}\, \vep)_{\al} \, \! ^{\dbe} \, \eta^{\mu_2 \mu_3} \,  f^{(123)}_{\ell=0}}_{\rho \in S_3 / {\cal P}_{2,0}} \notag \\
& \ \ \ \ \ \ \ \ \ \ + \  \underbrace{ (\si^{\mu_1} \, \bar \si^{\mu_2} \, \si^{\mu_3} \, \vep)_{\al} \, \! ^{\dbe} \, f^{(123)}_{\ell=1} }_{\rho \in S_3 / {\cal P}_{2,1}}\,, \label{yy1} 
\end{align}
with $z_{ij}$ dependencies
\begin{align}
f_{\ell=0}^{(312)} \eq &\frac{E_{14} \, E_{25}}{E_{12}} \; \tspin \left( \tfrac{1}{2} \smallint^{z_4} _{z_{3}} \ve{\om} \, + \, \tfrac{1}{2} \smallint^{z_5} _{z_{3}} \ve{\om} \right) \, \tspin \left(  \smallint^{z_1} _{z_2} \ve{\om} \, + \, \tfrac{1}{2} \smallint^{z_4} _{z_{5}} \ve{\om} \right)\,, \notag \\
f_{\ell=0}^{(213)} \eq &\frac{E_{14} \, E_{35}}{E_{13}} \; \tspin \left( \tfrac{1}{2} \smallint^{z_4} _{z_{2}} \ve{\om} \, + \, \tfrac{1}{2} \smallint^{z_5} _{z_{2}} \ve{\om} \right) \, \tspin \left(  \smallint^{z_1} _{z_3} \ve{\om} \, + \, \tfrac{1}{2} \smallint^{z_4} _{z_{5}} \ve{\om} \right)\,, \notag \\
 f_{\ell=0}^{(123)} \eq &\frac{E_{24} \, E_{35}}{E_{23}} \; \tspin \left( \tfrac{1}{2} \smallint^{z_4} _{z_{1}} \ve{\om} \, + \, \tfrac{1}{2} \smallint^{z_5} _{z_{1}} \ve{\om} \right) \, \tspin \left(  \smallint^{z_2} _{z_3} \ve{\om} \, + \, \tfrac{1}{2} \smallint^{z_4} _{z_{5}} \ve{\om} \right)\,, \notag \\
 f_{\ell=1}^{(123)} \eq &\frac{ E_{45}\, \tspin \left( \tfrac{1}{2} \smallint^{z_4} _{z_{1}} \ve{\om} \, + \, \tfrac{1}{2} \smallint^{z_5} _{z_{1}} \ve{\om} \right) \, \tspin
 \left( \tfrac{1}{2} \smallint^{z_4} _{z_{2}} \ve{\om} \, + \, \tfrac{1}{2} \smallint^{z_5} _{z_{2}} \ve{\om} \right) \,
  \tspin \left( \tfrac{1}{2} \smallint^{z_4} _{z_{3}} \ve{\om} \, + \, \tfrac{1}{2} \smallint^{z_5} _{z_{3}} \ve{\om} \right) }{2 \, \tspin \left( \tfrac{1}{2} \smallint^{z_4} _{z_{5}} \ve{\om} \right)} \,. \label{yy2} 
\end{align}
Equation (\ref{omega}) for $n=3$ is expanded as
\begin{align}
\langle &\psi^{\mu_1} \, \psi^{\mu_2} \, \psi^{\mu_3} \, \psi^{\mu_4} \, S_\al \, S_{\be} \rangle \ \ \sim \ \ 
\sum_{\ell = 0}^{2}   \sum_{\rho \in S_{4}/{\cal Q}_{3,\ell}} \! \! \!  \te{sgn}(\rho) \, \bigl(\si^{\mu_{\rho(1)}} \, \bar{\si}^{\mu_{\rho(2)}} \, ... \, \bar{\si}^{\mu_{\rho(2\ell)} } \, \vep \bigr)_{\al \be} \, g_{\ell}^{\rho} \notag \\
&\sim \ \ \vep_{\al \be} \, \eta^{\mu_1 \mu_2} \, \eta^{\mu_3 \mu_4}\, g_{\ell = 0}^{(1234)} \ - \ \vep_{\al \be} \, \eta^{\mu_1 \mu_3} \, \eta^{\mu_2 \mu_4}\, g_{\ell = 0}^{(1324)} \ + \ \vep_{\al \be} \, \eta^{\mu_2 \mu_3} \, \eta^{\mu_1 \mu_4}\, g_{\ell = 0}^{(2314)} \notag \\
& \ \ \ \ + \ (\si^{\mu_1} \, \bar \si^{\mu_2} \, \vep)_{\al \be} \, \eta^{\mu_3 \mu_4}\, g_{\ell = 1}^{(1234)} \ - \ (\si^{\mu_1} \, \bar \si^{\mu_3} \, \vep)_{\al \be} \, \eta^{\mu_2 \mu_4}\, g_{\ell = 1}^{(1324)} \ + \ (\si^{\mu_1} \, \bar \si^{\mu_4} \, \vep)_{\al \be} \, \eta^{\mu_2 \mu_3}\, g_{\ell = 1}^{(1423)} \notag \\
& \ \ \ \ + \ (\si^{\mu_2} \, \bar \si^{\mu_3} \, \vep)_{\al \be} \, \eta^{\mu_1 \mu_4}\, g_{\ell = 1}^{(2314)} \ - \ (\si^{\mu_2} \, \bar \si^{\mu_4} \, \vep)_{\al \be} \, \eta^{\mu_1 \mu_3}\, g_{\ell = 1}^{(2413)} \ + \ (\si^{\mu_3} \, \bar \si^{\mu_4} \, \vep)_{\al \be} \, \eta^{\mu_1 \mu_2}\, g_{\ell = 1}^{(3412)} \notag \\
& \ \ \ \ + \ (\si^{\mu_1} \, \bar \si^{\mu_2} \, \si^{\mu_3} \, \bar \si^{\mu_4} \, \vep)_{\al \be} \,  g_{\ell = 2}^{(1234)} \,.
\end{align}
The associated world-sheet functions $g_\ell^{\rho}$ can be found in equation (4.6) of \cite{loop4D}.


\subsection{Generalization to higher even dimension $D$}
\label{sec:highpt}

The generalization of correlation functions $\langle \psi^n S S \rangle$ to even space-time dimensions $D>4$ only
requires minor modifications, basically multiplication with the minimal spin system correlator
\beq
\langle s^+ (z_A) \, s^-(z_B) \rangle \spin \eq \frac{1}{\tspin(\ve{0}) \, E_{AB}^{1/4}} \; \tspin \left(\tfrac{1}{2} \smallint^{z_A} _{z_B} \ve{\om} \right)  \ ,
\label{mini}
\eeq
The first higher dimensional analogue of the $D=4$ results $\Om_{(n,D=4)}^{\mu_{1}
  ... \mu_{2n-1}}\,_{\al} \, \! ^{\dbe}$ and $\om_{(n,D=4)}^{\mu_{1} ... \mu_{2n-2}}\,_{\al \be}$ with the same
chirality structures occurs in $D=8$ dimensions:
\begin{align}
& \ \ \Om_{(n,D=8)}^{\mu_{1} ... \mu_{2n-1}}\,_{\al} \, \! ^{\dbe}(z_{i})  \ \ := \ \
\langle \psi^{\mu_{1}}(z_{1}) \, \psi^{\mu_{2}}(z_{2}) \, ... \, \psi^{\mu_{2n-1}}(z_{2n-1}) \, S_{\al}(z_{A}) \, S^{\dbe}(z_{B}) \rangle \spin \ \Bigl. \Bigr|_{D=8} \notag \\
& \ \  = \ \ \frac{\left[ \tspin \left( \tfrac{1}{2} \smallint^{z_A} _{z_B} \ve{\om} \right) \right]^{4-n}}{\sqrt{2} \,  \bigl[ \, \tspin ( \ve{0} ) \, \bigr]^4  \, E_{AB}^{1/2} \, \prod_{i=1}^{2n-1} (E_{iA} \, E_{iB})^{1/2} } \, \sum_{\ell = 0}^{n-1} \, \biggl( \frac{E_{AB}}{2 \, \tspin \left( \tfrac{1}{2} \smallint^{z_A} _{z_B} \ve{\om} \right)} \biggr)^{\ell} \notag \\
& \ \ \ \ \times \sum_{\rho \in S_{2n-1}/{\cal P}_{n,\ell}} \! \! \!  \te{sgn}(\rho) \, \bigl(\ga^{\mu_{\rho(1)}} \, \bar{\ga}^{\mu_{\rho(2)}} \, ... \, \bar{\ga}^{\mu_{\rho(2\ell)}} \, \ga^{\mu_{\rho(2\ell+1)}} \, C\bigr)_{\al} \, ^{\dbe} \, \prod_{k=1}^{2\ell+1} \tspin \left( \tfrac{1}{2} \smallint^{z_A} _{z_{\rho(k)}} \ve{\om} \, + \, \tfrac{1}{2} \smallint^{z_B} _{z_{\rho(k)}} \ve{\om} \right) \notag \\
& \ \ \ \ \times \  \prod_{j=1}^{n-\ell-1} \frac{\eta^{\mu_{\rho(2\ell+2j)} \mu_{\rho(2\ell+2j+1)}}}{E_{\rho(2\ell+2j),\rho(2\ell+2j+1)} } \; E_{\rho(2\ell+2j),A} \, E_{\rho(2\ell+2j+1),B}  \, \tspin \left( \smallint^{z_{\rho(2\ell+2j)}}_{z_{\rho(2\ell+2j+1)}} \ve{\om} \, + \, \tfrac{1}{2} \smallint^{z_A} _{z_B} \ve{\om} \right) \notag \\
& \ \ \equiv \ \ \frac{  \left[ \tspin \left( \tfrac{1}{2} \smallint^{z_A} _{z_B} \ve{\om} \right) \right]^2 }{\bigl[ \, \tspin ( \ve{0} ) \, \bigr]^2 \, E_{AB}^{1/2}} \; \Om_{(n,D=4)}^{\mu_{1} ... \mu_{2n-1}}\,_{\al } \, \! ^{\dbe}(z_{i})\,,
\label{Omega8} \\
& \ \ \om_{(n,D=8)}^{\mu_{1} ... \mu_{2n-2}}\,_{\al \be}(z_{i})  \ \ := \ \
\langle \psi^{\mu_{1}}(z_{1}) \, \psi^{\mu_{2}}(z_{2}) \, ... \, \psi^{\mu_{2n-2}}(z_{2n-2}) \, S_{\al}(z_{A}) \, S_{\be}(z_{B}) \rangle \spin \ \Bigl. \Bigr|_{D=8} \notag \\
& \ \  = \ \ \frac{ \left[ \tspin \left( \tfrac{1}{2} \smallint^{z_A} _{z_B} \ve{\om} \right) \right]^{5-n}}{ \bigl[ \, \tspin ( \ve{0} ) \, \bigr]^4 \, E_{AB}  \, \prod_{i=1}^{2n-2} (E_{iA} \, E_{iB})^{1/2} } \, \sum_{\ell = 0}^{n-1} \, \biggl( \frac{E_{AB}}{2 \, \tspin \left( \tfrac{1}{2} \smallint^{z_A} _{z_B} \ve{\om} \right)} \biggr)^{\ell} \notag \\
& \ \ \ \ \times \sum_{\rho \in S_{2n-2}/{\cal Q}_{n,\ell}} \! \! \!  \te{sgn}(\rho) \, \bigl(\ga^{\mu_{\rho(1)}} \, \bar{\ga}^{\mu_{\rho(2)}} \, ... \, \bar{\ga}^{\mu_{\rho(2\ell)} } \, C \bigr)_{\al \be} \, \prod_{k=1}^{2\ell} \tspin \left( \tfrac{1}{2} \smallint^{z_A} _{z_{\rho(k)}} \ve{\om} \, + \, \tfrac{1}{2} \smallint^{z_B} _{z_{\rho(k)}} \ve{\om} \right) \notag \\
& \ \ \ \ \times \  \prod_{j=1}^{n-\ell-1} \frac{\eta^{\mu_{\rho(2\ell+2j-1)} \mu_{\rho(2\ell+2j)}}}{E_{\rho(2\ell+2j-1),\rho(2\ell+2j)} } \; E_{\rho(2\ell+2j-1),A} \, E_{\rho(2\ell+2j),B}  \, \tspin \left( \smallint^{z_{\rho(2\ell+2j-1)}}_{z_{\rho(2\ell+2j)}} \ve{\om} \, + \, \tfrac{1}{2} \smallint^{z_A} _{z_B} \ve{\om} \right) \notag \\
& \ \ \equiv \ \ \frac{  \left[ \tspin \left( \tfrac{1}{2} \smallint^{z_A} _{z_B} \ve{\om} \right) \right]^2 }{\bigl[ \, \tspin ( \ve{0} ) \, \bigr]^2 \, E_{AB}^{1/2}} \; \om_{(n,D=4)}^{\mu_{1} ... \mu_{2n-2}}\,_{\al \be}(z_{i})\,.
\label{omega8}
\end{align}

\medskip In $D=6$ and $D=10$ dimensions, an odd number of NS fermions requires alike spin fields $S_\al S_\be$ for
nonzero correlations (and correspondingly, spin fields of opposite chirality $S_\al S^{\dbe}$ are needed for an even
number of fermions $\psi^{2n}$):
\begin{align}
& \ \ \Om_{(n,D=6)}^{\mu_{1} ... \mu_{2n-1}}\,_{\al \be}(z_{i})  \ \ := \ \
\langle \psi^{\mu_{1}}(z_{1}) \, \psi^{\mu_{2}}(z_{2}) \, ... \, \psi^{\mu_{2n-1}}(z_{2n-1}) \, S_{\al}(z_{A}) \, S_{\be}(z_{B}) \rangle \spin  \ \Bigl. \Bigr|_{D=6} \notag \\
& \ \  = \ \ \frac{\left[ \tspin \left( \tfrac{1}{2} \smallint^{z_A} _{z_B} \ve{\om} \right) \right]^{3-n}}{\sqrt{2} \, \bigl[ \, \tspin ( \ve{0} ) \, \bigr]^3 \, E_{AB}^{1/4} \, \prod_{i=1}^{2n-1} (E_{iA} \, E_{iB})^{1/2} } \, \sum_{\ell = 0}^{n-1} \, \biggl( \frac{E_{AB}}{2 \, \tspin \left( \tfrac{1}{2} \smallint^{z_A} _{z_B} \ve{\om} \right)} \biggr)^{\ell} \notag \\
& \ \ \ \ \times \sum_{\rho \in S_{2n-1}/{\cal P}_{n,\ell}} \! \! \!  \te{sgn}(\rho) \, \bigl(\ga^{\mu_{\rho(1)}} \, \bar{\ga}^{\mu_{\rho(2)}} \, ... \, \bar{\ga}^{\mu_{\rho(2\ell)}} \, \ga^{\mu_{\rho(2\ell+1)}} \, C\bigr)_{\al \be} \, \prod_{k=1}^{2\ell+1} \tspin \left( \tfrac{1}{2} \smallint^{z_A} _{z_{\rho(k)}} \ve{\om} \, + \, \tfrac{1}{2} \smallint^{z_B} _{z_{\rho(k)}} \ve{\om} \right) \notag \\
& \ \ \ \ \times \  \prod_{j=1}^{n-\ell-1} \frac{\eta^{\mu_{\rho(2\ell+2j)} \mu_{\rho(2\ell+2j+1)}}}{E_{\rho(2\ell+2j),\rho(2\ell+2j+1)} } \; E_{\rho(2\ell+2j),A} \, E_{\rho(2\ell+2j+1),B}  \, \tspin \left( \smallint^{z_{\rho(2\ell+2j)}}_{z_{\rho(2\ell+2j+1)}} \ve{\om} \, + \, \tfrac{1}{2} \smallint^{z_A} _{z_B} \ve{\om} \right)\,,
\label{Omega6} \\
& \ \ \om_{(n,D=6)}^{\mu_{1} ... \mu_{2n-2}}\,_{\al} \, \! ^{ \dbe}(z_{i})  \ \ := \ \
\langle \psi^{\mu_{1}}(z_{1}) \, \psi^{\mu_{2}}(z_{2}) \, ... \, \psi^{\mu_{2n-2}}(z_{2n-2}) \, S_{\al}(z_{A}) \, S^{\dbe}(z_{B}) \rangle \spin \ \Bigl. \Bigr|_{D=6} \notag \\
& \ \  = \ \ \frac{ \left[ \tspin \left( \tfrac{1}{2} \smallint^{z_A} _{z_B} \ve{\om} \right) \right]^{4-n}}{  \bigl[ \, \tspin ( \ve{0} ) \, \bigr]^3 \, E_{AB}^{3/4}  \, \prod_{i=1}^{2n-2} (E_{iA} \, E_{iB})^{1/2} } \, \sum_{\ell = 0}^{n-1} \, \biggl( \frac{E_{AB}}{2 \, \tspin \left( \tfrac{1}{2} \smallint^{z_A} _{z_B} \ve{\om} \right)} \biggr)^{\ell} \notag \\
& \ \ \ \ \times \sum_{\rho \in S_{2n-2}/{\cal Q}_{n,\ell}} \! \! \!  \te{sgn}(\rho) \, \bigl(\ga^{\mu_{\rho(1)}} \, \bar{\ga}^{\mu_{\rho(2)}} \, ... \, \bar{\ga}^{\mu_{\rho(2\ell)} } \, C \bigr)_{\al} \, \! ^{\dbe} \, \prod_{k=1}^{2\ell} \tspin \left( \tfrac{1}{2} \smallint^{z_A} _{z_{\rho(k)}} \ve{\om} \, + \, \tfrac{1}{2} \smallint^{z_B} _{z_{\rho(k)}} \ve{\om} \right) \notag \\
& \ \ \ \ \times \  \prod_{j=1}^{n-\ell-1} \frac{\eta^{\mu_{\rho(2\ell+2j-1)} \mu_{\rho(2\ell+2j)}}}{E_{\rho(2\ell+2j-1),\rho(2\ell+2j)} } \; E_{\rho(2\ell+2j-1),A} \, E_{\rho(2\ell+2j),B}  \, \tspin \left( \smallint^{z_{\rho(2\ell+2j-1)}}_{z_{\rho(2\ell+2j)}} \ve{\om} \, + \, \tfrac{1}{2} \smallint^{z_A} _{z_B} \ve{\om} \right)\,. 
\label{omega6}
\end{align}
In progressing towards $D=10$ dimensions, two additional factors of (\ref{mini}) for the extra spin systems appear:
\begin{align}
& \ \ \Om_{(n,D=10)}^{\mu_{1} ... \mu_{2n-1}}\,_{\al \be}(z_{i})  \ \ := \ \
\langle \psi^{\mu_{1}}(z_{1}) \, \psi^{\mu_{2}}(z_{2}) \, ... \, \psi^{\mu_{2n-1}}(z_{2n-1}) \, S_{\al}(z_{A}) \, S_{\be}(z_{B}) \rangle \spin \ \Bigl. \Bigr|_{D=10} \notag \\
& \ \  = \ \ \frac{\left[ \tspin \left( \tfrac{1}{2} \smallint^{z_A} _{z_B} \ve{\om} \right) \right]^{5-n}}{\sqrt{2} \, \bigl[ \, \tspin ( \ve{0} ) \, \bigr]^5 \, E_{AB}^{3/4} \, \prod_{i=1}^{2n-1} (E_{iA} \, E_{iB})^{1/2} } \, \sum_{\ell = 0}^{n-1} \, \biggl( \frac{E_{AB}}{2 \, \tspin \left( \tfrac{1}{2} \smallint^{z_A} _{z_B} \ve{\om} \right)} \biggr)^{\ell} \notag \\
& \ \ \ \ \times \sum_{\rho \in S_{2n-1}/{\cal P}_{n,\ell}} \! \! \!  \te{sgn}(\rho) \, \bigl(\ga^{\mu_{\rho(1)}} \, \bar{\ga}^{\mu_{\rho(2)}} \, ... \, \bar{\ga}^{\mu_{\rho(2\ell)}} \, \ga^{\mu_{\rho(2\ell+1)}} \, C\bigr)_{\al \be} \, \prod_{k=1}^{2\ell+1} \tspin \left( \tfrac{1}{2} \smallint^{z_A} _{z_{\rho(k)}} \ve{\om} \, + \, \tfrac{1}{2} \smallint^{z_B} _{z_{\rho(k)}} \ve{\om} \right) \notag \\
& \ \ \ \ \times \  \prod_{j=1}^{n-\ell-1} \frac{\eta^{\mu_{\rho(2\ell+2j)} \mu_{\rho(2\ell+2j+1)}}}{E_{\rho(2\ell+2j),\rho(2\ell+2j+1)} } \; E_{\rho(2\ell+2j),A} \, E_{\rho(2\ell+2j+1),B}  \, \tspin \left( \smallint^{z_{\rho(2\ell+2j)}}_{z_{\rho(2\ell+2j+1)}} \ve{\om} \, + \, \tfrac{1}{2} \smallint^{z_A} _{z_B} \ve{\om} \right)\,,
\label{Omega10} \\
& \ \ \om_{(n,D=10)}^{\mu_{1} ... \mu_{2n-2}}\,_{\al} \, \!^{ \dbe}(z_{i})  \ \ := \ \
\langle \psi^{\mu_{1}}(z_{1}) \, \psi^{\mu_{2}}(z_{2}) \, ... \, \psi^{\mu_{2n-2}}(z_{2n-2}) \, S_{\al}(z_{A}) \, S^{\dbe}(z_{B}) \rangle \spin \ \Bigl. \Bigr|_{D=10} \notag \\
& \ \  = \ \ \frac{\left[ \tspin \left( \tfrac{1}{2} \smallint^{z_A} _{z_B} \ve{\om} \right) \right]^{6-n}}{  \bigl[ \, \tspin ( \ve{0} ) \, \bigr]^5 \, E_{AB}^{ 5/4 }  \, \prod_{i=1}^{2n-2} (E_{iA} \, E_{iB})^{1/2} } \, \sum_{\ell = 0}^{n-1} \, \biggl( \frac{E_{AB}}{2 \, \tspin \left( \tfrac{1}{2} \smallint^{z_A} _{z_B} \ve{\om} \right)} \biggr)^{\ell} \notag \\
& \ \ \ \ \times \sum_{\rho \in S_{2n-2}/{\cal Q}_{n,\ell}} \! \! \!  \te{sgn}(\rho) \, \bigl(\ga^{\mu_{\rho(1)}} \, \bar{\ga}^{\mu_{\rho(2)}} \, ... \, \bar{\ga}^{\mu_{\rho(2\ell)} } \, C \bigr)_{\al} \, \! ^{\dbe} \, \prod_{k=1}^{2\ell} \tspin \left( \tfrac{1}{2} \smallint^{z_A} _{z_{\rho(k)}} \ve{\om} \, + \, \tfrac{1}{2} \smallint^{z_B} _{z_{\rho(k)}} \ve{\om} \right) \notag \\
& \ \ \ \ \times \  \prod_{j=1}^{n-\ell-1} \frac{\eta^{\mu_{\rho(2\ell+2j-1)} \mu_{\rho(2\ell+2j)}}}{E_{\rho(2\ell+2j-1),\rho(2\ell+2j)} } \; E_{\rho(2\ell+2j-1),A} \, E_{\rho(2\ell+2j),B}  \, \tspin \left( \smallint^{z_{\rho(2\ell+2j-1)}}_{z_{\rho(2\ell+2j)}} \ve{\om} \, + \, \tfrac{1}{2} \smallint^{z_A} _{z_B} \ve{\om} \right)\,. 
\label{omega10}
\end{align}
To describe both chirality structures in a unified manner, we have to use Dirac spinor notation with indices $_K = \{
_\ka, ^{\dka} \}$, see Appendix \ref{appA}. Then, our previous results for $D=4,6,8,10$ for $\langle \psi^n S S \rangle$
generalize as follows to even dimensions $D$: \bigskip

\framebox{\begin{minipage}{6.3in}
\begin{align}
& \ \ \Om_{(n,D)}^{\mu_{1} ... \mu_{2n-1}}\,_{KL}(z_{i})  \ \ := \ \
\langle \psi^{\mu_{1}}(z_{1}) \, \psi^{\mu_{2}}(z_{2}) \, ... \, \psi^{\mu_{2n-1}}(z_{2n-1}) \, S_{K}(z_{A}) \, S_{L}(z_{B}) \rangle \spin \notag \\
& \ \  = \ \ \frac{\left[ \tspin \left( \tfrac{1}{2} \smallint^{z_A} _{z_B} \ve{\om} \right) \right]^{D/2-n}}{\sqrt{2} \, \bigl[ \, \tspin ( \ve{0} ) \, \bigr]^{D/2} \, E_{AB}^{D/8 - 1/2} \, \prod_{i=1}^{2n-1} (E_{iA} \, E_{iB})^{1/2} } \, \sum_{\ell = 0}^{n-1} \, \biggl( \frac{E_{AB}}{2 \, \tspin \left( \tfrac{1}{2} \smallint^{z_A} _{z_B} \ve{\om} \right)} \biggr)^{\ell} \notag \\
& \ \ \ \ \times \sum_{\rho \in S_{2n-1}/{\cal P}_{n,\ell}} \! \! \!  \te{sgn}(\rho) \, \bigl(\Ga^{\mu_{\rho(1)}} \, \Ga^{\mu_{\rho(2)}} \, ... \, \Ga^{\mu_{\rho(2\ell)}} \, \Ga^{\mu_{\rho(2\ell+1)}} \, {\cal C} \bigr)_{KL} \, \prod_{k=1}^{2\ell+1} \tspin \left( \tfrac{1}{2} \smallint^{z_A} _{z_{\rho(k)}} \ve{\om} \, + \, \tfrac{1}{2} \smallint^{z_B} _{z_{\rho(k)}} \ve{\om} \right) \notag \\
& \ \ \ \ \times \  \prod_{j=1}^{n-\ell-1} \frac{\eta^{\mu_{\rho(2\ell+2j)} \mu_{\rho(2\ell+2j+1)}}}{E_{\rho(2\ell+2j),\rho(2\ell+2j+1)} } \; E_{\rho(2\ell+2j),A} \, E_{\rho(2\ell+2j+1),B}  \, \tspin \left( \smallint^{z_{\rho(2\ell+2j)}}_{z_{\rho(2\ell+2j+1)}} \ve{\om} \, + \, \tfrac{1}{2} \smallint^{z_A} _{z_B} \ve{\om} \right)\,, 
\label{OmegaD} \\
& \ \ \om_{(n,D)}^{\mu_{1} ... \mu_{2n-2}}\,_{KL}(z_{i})  \ \ := \ \
\langle \psi^{\mu_{1}}(z_{1}) \, \psi^{\mu_{2}}(z_{2}) \, ... \, \psi^{\mu_{2n-2}}(z_{2n-2}) \, S_{K}(z_{A}) \, S_L(z_{B}) \rangle \spin \notag \\
& \ \  = \ \ \frac{\left[ \tspin \left( \tfrac{1}{2} \smallint^{z_A} _{z_B} \ve{\om} \right) \right]^{D/2+1-n}}{  \bigl[ \, \tspin ( \ve{0} ) \, \bigr]^{D/2} \, E_{AB}^{ D/8 }  \, \prod_{i=1}^{2n-2} (E_{iA} \, E_{iB})^{1/2} } \, \sum_{\ell = 0}^{n-1} \, \biggl( \frac{E_{AB}}{2 \, \tspin \left( \tfrac{1}{2} \smallint^{z_A} _{z_B} \ve{\om} \right)} \biggr)^{\ell} \notag \\
& \ \ \ \ \times \sum_{\rho \in S_{2n-2}/{\cal Q}_{n,\ell}} \! \! \!  \te{sgn}(\rho) \, \bigl(\Ga^{\mu_{\rho(1)}} \, \Ga^{\mu_{\rho(2)}} \, ... \, \Ga^{\mu_{\rho(2\ell)} } \, {\cal C} \bigr)_{KL} \, \prod_{k=1}^{2\ell} \tspin \left( \tfrac{1}{2} \smallint^{z_A} _{z_{\rho(k)}} \ve{\om} \, + \, \tfrac{1}{2} \smallint^{z_B} _{z_{\rho(k)}} \ve{\om} \right) \notag \\
& \ \ \ \ \times \  \prod_{j=1}^{n-\ell-1} \frac{\eta^{\mu_{\rho(2\ell+2j-1)} \mu_{\rho(2\ell+2j)}}}{E_{\rho(2\ell+2j-1),\rho(2\ell+2j)} } \; E_{\rho(2\ell+2j-1),A} \, E_{\rho(2\ell+2j),B}  \, \tspin \left( \smallint^{z_{\rho(2\ell+2j-1)}}_{z_{\rho(2\ell+2j)}} \ve{\om} \, + \, \tfrac{1}{2} \smallint^{z_A} _{z_B} \ve{\om} \right)\,.
\label{omegaD}
\end{align}
\end{minipage}}

\bigskip
\noindent
The proof in $D$ dimensions can be carried over almost literally from the four-dimensional case in \cite{loop4D}. The
only explicit $D$ dependence lies in the pre-factors $\frac{ \left[ \tspin \left( \tfrac{1}{2} \smallint^{z_A} _{z_B}
      \ve{\om} \right) \right]^{D/2 - n} }{\bigl[ \, \tspin ( \ve{0} ) \, \bigr]^{D/2} \, E_{AB}^{D/8 - 1/2}}$ and
$\frac{ \left[ \tspin \left( \tfrac{1}{2} \smallint^{z_A} _{z_B} \ve{\om} \right) \right]^{D/2 + 1 - n} }{\bigl[ \,
  \tspin ( \ve{0} ) \, \bigr]^{D/2} \, E_{AB}^{D/8}}$ which are designed to match the leading $z_A \rightarrow z_B$
behavior in the OPE of $S_K(z_A) S_L(z_B)$, see (\ref{ope20}).

\section{Four and more spin fields in $\bm{D=6}$}
\label{sec_d6}

In this section we first give the results for all $D=6$ correlators involving four or more spin fields up to six-point
level. Later, we generalize to correlators of the type $\vevs{(S_\al\,S^{\dbe})^N}$ and provide an explicit formula for arbitrary
$N$. These correlation functions are important because a variety of correlators involving fermions can be derived from
them by means of the factorization prescription $\psi^\mu \sim (C^{-1} \gab^\mu)^{\be \al} S_\al S_\be$ introduced in
subsection \ref{review1}.

\subsection{Lower point results}

The simplest correlators in six dimensions involving four spin fields only are
\begin{align}
  \vevs{S_\al(z_1)\,S_\be(z_2)\,S_\ga(z_3)\,S_\de(z_4)}\, =& \, \frac{\teta{1&2\\3&4}\,\teta{1&3\\2&4}\,\teta{1&4\\2&3}}{2 \, \ttt}\,
  \frac{(\ga^\mu\,C)_{\al\be}(\ga_\mu\,C)_{\ga\de}}{(E_{12}\,E_{13}\,E_{14}\,E_{23}\,E_{24}\,E_{34})^{1/4}}\,,\\
  \vevs{S_\al(z_1)\,S_\be(z_2)\,S^{\dga}(z_3)\,S^{\dde}(z_4)}\, =& \, \frac{\teta{1&2\\3&4}}{\ttt}\,
  \bigg(\frac{E_{13}\,E_{14}\,E_{23}\,E_{24}}{E_{12}\,E_{34}}\bigg)^{1/4}\,\notag\\
  &\hspace{1cm}\times\bigg[
  \frac{\cc[\al]{\dga}\,\cc[\be]{\dde}}{E_{13}\,E_{24}}\,\teta{1&4\\2&3}^2-\frac{\cc[\al]{\dde}\,\cc[\be]{\dga}}{E_{14}\,E_{23}}\,\teta{1&3\\2&4}^2
  \bigg]\,.
  \label{4s}
\end{align}
The only non-vanishing five-point correlator with four spin fields and one fermion involves three alike chiralities:
\begin{align}
  \vevs{\psi^\mu(z_1)&\,S_\al(z_2)\,S_\be(z_3)\,S_\ga(z_4)\,S^{\dde}(z_5)}\,=\,
  \frac{1}{\sqrt{2}\,\ttt}\,\frac{(E_{25}\,E_{35}\,E_{45})^{1/4}}{(E_{12}\,E_{13}\,E_{14}\,E_{15})^{1/2}\,(E_{23}\,E_{24}\,E_{34})^{1/4}}\notag\\
  \times \bigg[
  &(\ga^\mu\,C)_{\al\be}\,\frac{\cc[\ga]{\dde}}{E_{45}}\,E_{14}\,\teta{1&1&4\\2&3&5}\,\teta{2&4\\3&5}\,\teta{2&5\\3&4}
  -(\ga^\mu\,C)_{\al\ga}\,\frac{\cc[\be]{\dde}}{E_{35}}\,E_{13}\,\teta{1&1&3\\2&4&5}\,\teta{2&3\\4&5}\,\teta{2&5\\3&4}\notag\\
  &\phantom{(\ga^\mu\,C)_{\al\be}\,\frac{\cc[\ga]{\dde}}{E_{45}}\,E_{14}\,\teta{1&1&4\\2&3&5}\,\teta{2&4\\3&5}\,\teta{2&5\\3&4}}
  +(\ga^\mu\,C)_{\be\ga}\,\frac{\cc[\al]{\dde}}{E_{25}}\,E_{12}\,\teta{1&1&2\\3&4&5}\,\teta{2&3\\4&5}\,\teta{2&4\\3&5}
  \bigg]\,.
\end{align}
There are two six-point correlators involving only spin fields. The first one consists of five left- and one
right-handed spin-field:
\begin{align}
  \vevs{S_\al(z_1)\,S_\be(z_2)&\,S_\ga(z_3)\,S_\de(z_4)\,S_\ep(z_5)\,S^{\di}(z_6)}\,=\,
  \frac{1}{2\,\ttt}\,\bigg(\frac{E_{16}\,E_{26}\,E_{36}\,E_{46}\,E_{56}}{E_{12}\,E_{13}\,E_{14}\,E_{15}\,E_{23}\,E_{24}\,E_{25}\,E_{34}\,E_{35}\,E_{45}}\bigg)^{1/4}\notag\\
  \times\bigg[
  &(\ga^\mu\,C)_{\al\be}(\ga_\mu\,C)_{\ga\ep}\,\frac{\cc[\de]{\di}}{E_{46}}\,\frac{E_{45}}{E_{56}}
  \teta{1&2&6\\3&4&5}\,\teta{1&3&6\\2&4&5}\,\teta{1&4&5\\2&3&6}\notag\\
  +\,&(\ga^\mu\,C)_{\al\be}(\ga_\mu\,C)_{\ep\de}\,\frac{\cc[\ga]{\di}}{E_{36}}\,\frac{E_{35}}{E_{56}}
  \teta{1&2&6\\3&4&5}\,\teta{1&3&5\\2&4&6}\,\teta{1&4&6\\2&3&5}\notag\\
  +\,&(\ga^\mu\,C)_{\al\ep}(\ga_\mu\,C)_{\ga\de}\,\frac{\cc[\be]{\di}}{E_{26}}\,\frac{E_{25}}{E_{56}}
  \teta{1&2&5\\3&4&6}\,\teta{1&3&6\\2&4&5}\,\teta{1&4&6\\2&3&5}\notag\\
  +\,&(\ga^\mu\,C)_{\ep\be}(\ga_\mu\,C)_{\ga\de}\,\frac{\cc[\al]{\di}}{E_{16}}\,\frac{E_{15}}{E_{56}}
  \teta{1&2&5\\3&4&6}\,\teta{1&3&5\\2&4&6}\,\teta{1&4&5\\2&3&6}
  \bigg]\,.
\end{align}
In addition we have the correlator with three left- and right-handed spin fields each
\begin{align}
  \vevs{S_\al(z&_1)\,S_\be(z_2)\,S_\ga(z_3)\,S^{\dde}(z_4)\,S^{\dep}(z_5)\,S^{\di}(z_6)}\,=\,
  \frac{1}{\ttt}\,\bigg(\frac{E_{14}\,E_{15}\,E_{16}\,E_{24}\,E_{25}\,E_{26}\,E_{34}\,E_{35}\,E_{36}}{E_{12}\,E_{13}\,E_{23}\,E_{45}\,E_{46}\,E_{56}}\bigg)^{1/4}\notag\\
  \times\bigg[
   &\frac{\cc[\al]{\dde}\,\cc[\be]{\dep}\,\cc[\ga]{\di}}{E_{14}\,E_{25}\,E_{36}}\,
  \teta{1&2&6\\3&4&5}\,\teta{1&3&5\\2&4&6}\,\teta{1&5&6\\2&3&4}
  -\frac{\cc[\al]{\dde}\,\cc[\be]{\di}\,\cc[\ga]{\dep}}{E_{14}\,E_{26}\,E_{35}}\,
  \teta{1&2&5\\3&4&6}\,\teta{1&3&6\\2&4&5}\,\teta{1&5&6\\2&3&4}\notag\\
  +\,&\frac{\cc[\al]{\dep}\,\cc[\be]{\di}\,\cc[\ga]{\dde}}{E_{15}\,E_{26}\,E_{34}}\,
  \teta{1&2&4\\3&5&6}\,\teta{1&3&6\\2&4&5}\,\teta{1&4&6\\2&3&5}
  -\frac{\cc[\al]{\dep}\,\cc[\be]{\dde}\,\cc[\ga]{\di}}{E_{15}\,E_{24}\,E_{36}}\,
  \teta{1&2&6\\3&4&5}\,\teta{1&3&4\\2&5&6}\,\teta{1&4&6\\2&3&5}\notag\\
  +\,&\frac{\cc[\al]{\di}\,\cc[\be]{\dde}\,\cc[\ga]{\dep}}{E_{16}\,E_{24}\,E_{35}}\,
  \teta{1&2&5\\3&4&6}\,\teta{1&3&4\\2&5&6}\,\teta{1&4&5\\2&3&6}
  -\frac{\cc[\al]{\di}\,\cc[\be]{\dep}\,\cc[\ga]{\dde}}{E_{16}\,E_{25}\,E_{34}}\,
  \teta{1&2&4\\3&5&6}\,\teta{1&3&5\\2&4&6}\,\teta{1&4&5\\2&3&6}
  \bigg]\,.
\end{align}
Furthermore, two NS fermions can be accompanied by four spin fields, either with uniform chirality 
\begin{align}
  \vevs{\psi^\mu(z_1)\,\psi^\nu&(z_2)\,S_\al(z_3)\,S_\be(z_4)\,S_\ga(z_5)\,S_\de(z_6)}\,=\,
  -\frac{1}{2\,\ttt}\,\frac{(E_{13}\,E_{14}\,E_{15}\,E_{16}\,E_{23}\,E_{24}\,E_{25}\,E_{26})^{-1/2}}{E_{12} \, (E_{34}\,E_{35}\,E_{36}\,E_{45}\,E_{46}\,E_{56}\,)^{1/4}}\notag\\
  \bigg[
  &(\ga^\mu\,C)_{\al\be}(\ga^\nu\,C)_{\ga\de}\,E_{15}\,E_{16}\,E_{23}\,E_{24}\,
  \teta{1&1&5&6\\2&2&3&4}\,\teta{3&5\\4&6}\,\teta{3&6\\4&5}\notag\\
  +\,&(\ga^\mu\,C)_{\ga\de}(\ga^\nu\,C)_{\al\be}\,E_{13}\,E_{14}\,E_{25}\,E_{26}\,
  \teta{1&1&3&4\\2&2&5&6}\,\teta{3&5\\4&6}\,\teta{3&6\\4&5}\notag\\
  -\,&(\ga^\mu\,C)_{\al\ga}(\ga^\nu\,C)_{\be\de}\,E_{14}\,E_{16}\,E_{23}\,E_{25}\,
  \teta{1&1&4&6\\2&2&3&5}\,\teta{3&4\\5&6}\,\teta{3&6\\4&5}\notag\\
  -\,&(\ga^\mu\,C)_{\be\de}(\ga^\nu\,C)_{\al\ga}\,E_{13}\,E_{15}\,E_{24}\,E_{26}\,
  \teta{1&1&3&5\\2&2&4&6}\,\teta{3&4\\5&6}\,\teta{3&6\\4&5}\notag\\
  +\,&(\ga^\mu\,C)_{\al\de}(\ga^\nu\,C)_{\be\ga}\,E_{14}\,E_{15}\,E_{23}\,E_{26}\,
  \teta{1&1&4&5\\2&2&3&6}\,\teta{3&4\\5&6}\,\teta{3&5\\4&6}\notag\\
  +\,&(\ga^\mu\,C)_{\be\ga}(\ga^\nu\,C)_{\al\de}\,E_{13}\,E_{16}\,E_{24}\,E_{25}\,
  \teta{1&1&3&6\\2&2&4&5}\,\teta{3&4\\5&6}\,\teta{3&5\\4&6}
  \bigg]\ ,
  \label{6s}
\end{align}
or with mixed chiralities:
\begin{align}
  \vevs{\psi^\mu(z_1)\,\psi^\nu&(z_2)\,S_\al(z_3)\,S_\be(z_4)\,S^{\dga}(z_5)\,S^{\dde}(z_6)}\,=\,
  \frac{1}{\ttt}\,\frac{(E_{35}\,E_{36}\,E_{45}\,E_{46})^{1/4}\,(E_{34}\,E_{56})^{-1/4}}{(E_{13}\,E_{14}\,E_{15}\,E_{16}\,E_{23}\,E_{24}\,E_{25}\,E_{26})^{1/2}}\notag\\
  \bigg[
  &\frac{\eta^{\mu\nu}\,\cc[\al]{\dga}\,\cc[\be]{\dde}}{E_{12}\,E_{35}\,E_{46}}\,E_{13}\,E_{14}\,E_{25}\,E_{26}\,
  \teta{1&1&3&4\\2&2&5&6}\,\teta{3&6\\4&5}^2\notag\\
  -\,&\frac{\eta^{\mu\nu}\,\cc[\al]{\dde}\,\cc[\be]{\dga}}{E_{12}\,E_{36}\,E_{45}}\,E_{13}\,E_{14}\,E_{25}\,E_{26}\,
  \teta{1&1&3&4\\2&2&5&6}\,\teta{3&5\\4&6}^2\notag\\
  +\,&\frac{1}{2}\,(\ga^\mu\,C)_{\al\be}(\gab^\nu\,C)^{\dga\dde}\,E_{12}\,
  \teta{1&1&2&2\\3&4&5&6}\,\teta{3&5\\4&6}\,\teta{3&6\\4&5}\notag\\
  +\,&\frac{1}{2}\,(\ga^\mu\,\gab^\nu\,C)_\al{}^{\dga}\,\frac{\cc[\be]{\dde}}{E_{46}}\,E_{14}\,E_{26}\,
  \teta{1&1&4\\3&5&6}\,\teta{2&2&6\\3&4&5}\,\teta{3&6\\4&5}\notag\\
  -\,&\frac{1}{2}\,(\ga^\mu\,\gab^\nu\,C)_\al{}^{\dde}\,\frac{\cc[\be]{\dga}}{E_{45}}\,E_{14}\,E_{25}\,
  \teta{1&1&4\\3&5&6}\,\teta{2&2&5\\3&4&6}\,\teta{3&5\\4&6}\notag\\
  -\,&\frac{1}{2}\,(\ga^\mu\,\gab^\nu\,C)_\be{}^{\dga}\,\frac{\cc[\al]{\dde}}{E_{36}}\,E_{13}\,E_{26}\,
  \teta{1&1&3\\4&5&6}\,\teta{2&2&6\\3&4&5}\,\teta{3&5\\4&6}\notag\\
  +\,&\frac{1}{2}\,(\ga^\mu\,\gab^\nu\,C)_\be{}^{\dde}\,\frac{\cc[\al]{\dga}}{E_{35}}\,E_{13}\,E_{25}\,
  \teta{1&1&3\\4&5&6}\,\teta{2&2&5\\3&4&6}\,\teta{3&6\\4&5}
  \bigg]\,.
\end{align}
The seven-point correlator consisting of one NS fermion, four left- and two right-handed spin fields is found to be
\begin{align}
  \vevs{\psi^\mu(z_1)\,S&_\al(z_2)\,S_\be(z_3)\,S_\ga(z_4)\,S_\de(z_5)\,S^{\dep}(z_6)\,S^{\di}(z_7)}\,=\,\notag\\
  &\hspace{-.52cm}\frac{1}{\sqrt{2}\,\ttt}\,
  \frac{(E_{26}\,E_{27}\,E_{36}\,E_{37}\,E_{46}\,E_{47}\,E_{56}\,E_{57})^{1/4}}{(E_{12}\,E_{13}\,E_{14}\,E_{15}\,E_{16}\,E_{17})^{1/2}(E_{23}\,E_{24}\,E_{25}\,E_{34}\,E_{35}\,E_{45}\,E_{67})^{1/4}}\notag\\
  \times\Bigg[
  &(\ga^\mu\,C)_{\al\be}\,E_{14}\,E_{15}\,\teta{1&1&4&5\\2&3&6&7}\,\bigg(
   \frac{\cc[\ga]{\dep}\,\cc[\de]{\di}}{E_{46}\,E_{57}}\,\teta{2&4&7\\3&5&6}\,\teta{2&5&6\\3&4&7}
  -\frac{\cc[\de]{\dep}\,\cc[\ga]{\di}}{E_{47}\,E_{56}}\,\teta{2&4&6\\3&5&7}\,\teta{2&5&7\\3&4&6}\bigg)\notag\\
  +\,&(\ga^\mu\,C)_{\al\ga}\,E_{13}\,E_{15}\,\teta{1&1&3&5\\2&4&6&7}\,\bigg(
   \frac{\cc[\de]{\dep}\,\cc[\be]{\di}}{E_{37}\,E_{56}}\,\teta{2&3&6\\4&5&7}\,\teta{2&5&7\\3&4&6}
  -\frac{\cc[\be]{\dep}\,\cc[\de]{\di}}{E_{36}\,E_{57}}\,\teta{2&3&7\\4&5&6}\,\teta{2&5&6\\3&4&7}\bigg)\notag\\
  +\,&(\ga^\mu\,C)_{\al\de}\,E_{13}\,E_{14}\,\teta{1&1&3&4\\2&5&6&7}\,\bigg(
   \frac{\cc[\be]{\dep}\,\cc[\ga]{\di}}{E_{36}\,E_{47}}\,\teta{2&3&7\\4&5&6}\,\teta{2&4&6\\3&5&7}
  -\frac{\cc[\ga]{\dep}\,\cc[\be]{\di}}{E_{37}\,E_{46}}\,\teta{2&3&6\\4&5&7}\,\teta{2&4&7\\3&5&6}\bigg)\notag\\
  +\,&(\ga^\mu\,C)_{\be\ga}\,E_{12}\,E_{15}\,\teta{1&1&2&5\\3&4&6&7}\,\bigg(
   \frac{\cc[\al]{\dep}\,\cc[\de]{\di}}{E_{26}\,E_{57}}\,\teta{2&3&7\\4&5&6}\,\teta{2&4&7\\3&5&6}
  -\frac{\cc[\de]{\dep}\,\cc[\al]{\di}}{E_{27}\,E_{56}}\,\teta{2&3&6\\4&5&7}\,\teta{2&4&6\\3&5&7}\bigg)\notag\\
  +\,&(\ga^\mu\,C)_{\be\de}\,E_{12}\,E_{14}\,\teta{1&1&2&4\\3&5&6&7}\,\bigg(
   \frac{\cc[\ga]{\dep}\,\cc[\al]{\di}}{E_{27}\,E_{46}}\,\teta{2&3&6\\4&5&7}\,\teta{2&5&6\\3&4&7}
  -\frac{\cc[\al]{\dep}\,\cc[\ga]{\di}}{E_{26}\,E_{47}}\,\teta{2&3&7\\4&5&6}\,\teta{2&5&7\\3&4&6}\bigg)\notag\\
  +\,&(\ga^\mu\,C)_{\ga\de}\,E_{12}\,E_{13}\,\teta{1&1&2&3\\4&5&6&7}\,\bigg(
   \frac{\cc[\al]{\dep}\,\cc[\be]{\di}}{E_{26}\,E_{37}}\,\teta{2&4&7\\3&5&6}\,\teta{2&5&7\\3&4&6}
  -\frac{\cc[\be]{\dep}\,\cc[\al]{\di}}{E_{27}\,E_{36}}\,\teta{2&4&6\\3&5&7}\,\teta{2&5&6\\3&4&7}\bigg)
  \Bigg]\,.
\end{align}

\subsection{Generalizations to higher point}

In six space-time dimensions, the structure of Fierz identities is still sufficiently simple that the $2N$-point
correlation function with $N$ left- and right-handed spin fields each can be expressed in terms of
$C_{\al_i}{}^{\dbe_j}$ matrices only: \bigskip

\framebox{\begin{minipage}{6.3in}
\begin{align}
  \bigg\langle{\prod_{i=1}^N S_{\al_i}(z_{2i-1})\,S^{\dbe_i}(z_{2i})\bigg\rangle\spin}&\eq 
  \frac{\big[\Theta\spin(\sum_{i=1}^N\frac{1}{2}\int_{z_{2i-1}}^{z_{2i}}\vec\omega)\big]^{N-3}}{\ttt}\,\bigg(\frac{\prod_{i,j=1}^N
    E_{2i-1,2j}}{\prod_{i<j}^N E_{2i-1,2j-1}\,E_{2i,2j}}\bigg)^{1/4}\notag\\
  & \! \! \! \! \! \! \! \! \! \! \! \times \sum_{\rho \in S_N}\te{sgn}(\rho)\,\prod_{m=1}^N\frac{\cc[\al_{2m-1}]{\dbe_{\rho(2m)}}}{E_{2m-1,\rho(2m)}}\,
  \Theta\spin\bigg(\sum_{i=1}^N\frac{1}{2}\int_{z_{2i-1}}^{z_{2i}}\vec\omega-\int_{z_{\rho(2m)}}^{2m-1}\vec\omega\bigg)\ .
  \label{nn}
\end{align}
\end{minipage}}

\bigskip
\noindent
The results \eqref{4s} and \eqref{6s} coincide with the above formula for the cases $N=2,3$. For higher $N$ Fay's
trisecant identities \cite{HS,fay} might be needed to make specific $\al_i,\dbe_j$ choices compatible with the result
above. Due to the chirality structure of the charge conjugation matrix in $D=6$ the correlator above is the direct
relative of $\vevs{S_{\al_1}(z_1)\dots S_{\al_n}(z_n)}$ in $D=4$. Therefore the proof of \eqref{nn} proceeds in the same
way as in \cite{tree,loop4D}.

\medskip Having the explicit formula \eqref{nn} for this class of correlators is a great benefit in view of the
factorization prescription for NS fermions. One can combine two spin fields of alike chirality to a NS fermion via \beq
\psi^\mu(w) \eq - \, \frac{1}{2 \sqrt{2}} \; \lim_{z \rightarrow w} (z-w)^{1/4} \, (C^{-1} \, \gab^\mu)^{\be \al} \,
S_\al (z) \, S_\be(w)
\label{factorD=6}
\eeq
and thereby derive the following further classes of correlation functions:
\begin{equation}
  \vevs{\psi^{\mu_1}(z_1)\dots\psi^{\mu_{k+l}}(z_{k+l})\,S_{\al_1}(x_1)\dots S_{\al_{n-2k}}(x_{n-2k})\,S^{\dbe_1}(y_1)\dots
    S^{\dbe_{n-2l}}(y_{n-2l})}\,.
  \end{equation}
For example the last three correlators calculated in the previous subsection can be derived from the $N=4$ version of \eqref{nn}.

\section{Four and more spin fields in $\bm{D=8}$}
\label{sec_d8}

\begin{figure}
  \centering
  \includegraphics[width=.35\linewidth]{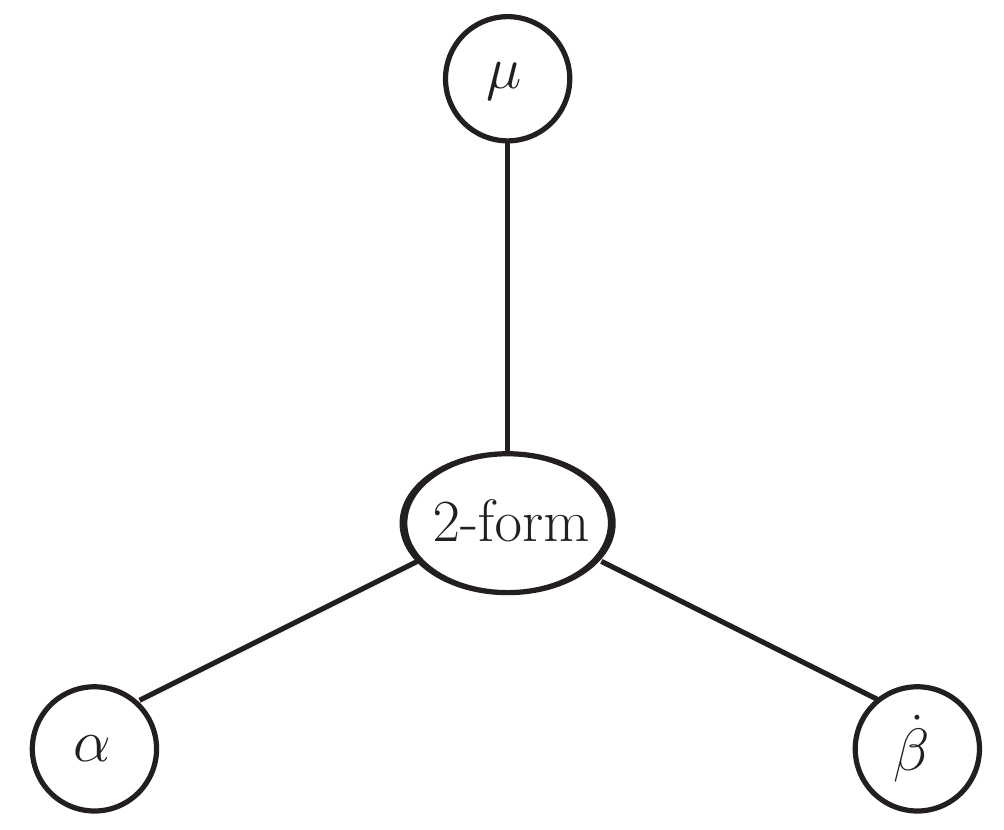}
  \caption{\small Dynkin diagram for $SO(8)$.}
  \label{fig_dynkin}
\end{figure}

The $S_3$ permutation symmetry of the Mercedes star--shaped $SO(8)$ Dynkin diagram in Figure \ref{fig_dynkin} -- also referred to as triality --
plays an important role for the RNS CFT in $D=8$.
In eight dimensions NS fermions and spin fields have equal conformal dimension $h=\frac{D}{16}=\frac{1}{2}$. Therefore,
the OPEs (\ref{ope1}), (\ref{ope2}), (\ref{ope43}) and (\ref{ope44}) become particularly symmetric and we will make use
of $SO(8)$ triality to rewrite them in unified fashion.

\medskip The short distance behavior of conformal fields is sufficient input to determine their correlations on the
sphere. This is why triality covariance of OPEs is inherited by tree-level correlators. At higher genus, however, the
different global properties of the $\psi^\mu$ and $S_{\al}, S^{\dbe}$ fields under transport around the world-sheet's
homology cycles will break this covariance, at least as far as the spin structure dependent factors
$\Theta^{\vec{a}}_{\vec{b}}$ are concerned. Hence, triality does not hold for correlators at loop-level.

\subsection[$SO(8)$ triality for tree-level correlations]{$\bm{SO(8)}$ triality for tree-level correlations}
\label{sec:triality}

In this subsection, we will present eight-dimensional tree-level correlators in manifestly triality covariant form. Let
us introduce some notation for this purpose. Firstly, it is convenient to work with generalized fields $P^i,Q^j,R^k$ of
conformal dimension $h=\half$, that can either be $\psi^\mu,S_\al$ or $S^{\dbe}$:
\begin{equation}
(P^i,Q^j,R^k) \, = \, \bigl(\rho(\psi^\mu),\rho(S_\al),\rho(S^{\dbe}) \bigr) \quad \te{for} \ \rho \in S_3\,.
\label{so8,0}
\end{equation}
On the level of Clebsch--Gordan coefficients, we introduce a universal metric $g$ and three-point couplings $G$ with
generalized indices $i,j,k \in \{ \mu,\al ,\dbe \}$:
\begin{align}
  g^{ij} \ \ &:= \ \ \left\{ \begin{array}{cl} \eta^{\mu \nu} &: \ (i,j) = (\mu,\nu) \\ C_{\al \be} &: \ (i,j) =
      (\al,\be) \\ C^{\dal \dbe} &: \ (i,j) = (\dal,\dbe) \\ 0 &: \ \te{otherwise} \end{array} \right.  \co g_{ij} \ \
  := \ \ \left\{ \begin{array}{cl} \eta_{\mu \nu} &: \ (i,j) = (\mu,\nu) \\ C^{\al \be} &: \ (i,j) = (\al,\be) \\
      C_{\dal \dbe} &: \ (i,j) = (\dal,\dbe) \\ 0 &: \ \te{otherwise}\end{array} \right.\,,
  \label{so8,1} \\
  G^{ijk} \ \ &:= \ \ \left\{ \begin{array}{cl} \frac{1}{\sqrt{2}} \; (\ga^{\mu} \, C)_\al \, \! ^{\dbe} & : \ (i,j,k) = \bigl( \rho(\mu),\rho(\al),\rho(\dbe) \bigr) \ \te{for some} \ \rho \in S_3 \\
      0 &: \ \te{otherwise}\end{array} \right.\,. \label{so8,2}
\end{align}
In this notation, both the Dirac algebra and the Fierz identity $C_{\al \be} C^{\dga \dde} = \frac{1}{2} (\ga^\mu \,
C)_\al \, \! ^{\dga} (\ga_\mu \, C)_\be \, \! ^{\dde} \,+ \, \frac{1}{2} (\ga^\mu \, C)_\be \, \! ^{\dga} (\ga_\mu \,
C)_\al \, \! ^{\dde}$ are special cases of the triality covariant tensor equation
\begin{equation}
g^{i_1 i_2} \, g^{j_1 j_2} \eq G^{i_1 j_1 k_1} \, G^{ i_2 j_2 k_2} \, g_{k_1 k_2} \ + \ G^{i_1 j_2 k_1} \, G^{ i_2 j_1 k_2} \, g_{k_1 k_2} \,.
\label{so8,2a}
\end{equation}
Note the minus sign $\sqrt{2} G^{ijk} \equiv (\ga^\mu C)_\al {}^{\dbe} = -(\gab^\mu C)^{\dbe}{}_\al$ due to the
antisymmetry of $(\Ga ^\mu {\cal C})$.

\medskip 
The definitions (\ref{so8,0}), (\ref{so8,1}) and (\ref{so8,2}) allow to rewrite the $D=8$ OPEs (\ref{ope1}), (\ref{ope2}), (\ref{ope43}) and (\ref{ope44}) in the unified fashion:
\begin{align}
P^{i}(z) \, P^{j}(w) \ \ &= \ \ \frac{g^{ij}}{z-w} \ + \ ... \ , \label{so8,3} \\
P^i(z) \, Q^j(w) \ \ &= \ \ \frac{G^{ijk}}{(z-w)^{1/2}} \; g_{kl} \, R^l(w) \ + \ ... \,. \label{so8,4}
\end{align}
As we have mentioned above, this is all the input necessary to derive the tree-level correlation function $\langle
P^{i_1}(x_1) ... P^{i_p}(x_p) \, Q^{j_1}(y_1) ... Q^{j_q}(y_q) R^{k_1}(z_1) ... R^{k_r}(z_r) \rangle$ in triality
covariant form -- every assignment $\rho \in S_3$ for $(P^i,Q^j,R^k) = \bigl( \rho(\psi^\mu) , \rho(S_\al) ,
\rho(S^{\dbe}) \bigr)$ is possible.

\medskip This circumstance can be used as a strong tool to derive new correlators: Suppose we know $\langle \psi^{\mu_1}
... \psi^{\mu_\ell} S_{\al_1} ... S_{\al_m} S^{\dbe_1} ... S^{\dbe_n} \rangle$ for some $\ell,m,n \in \NN_0$, then one
can rewrite this result in a covariant way as $\langle P^{i_1} ... P^{i_\ell} Q^{j_1} ... Q^{j_m} R^{k_1} ... R^{k_n}
\rangle$ via $(\psi^\mu,S_\al,S^{\dbe}) \equiv (P^i,Q^j, R^k)$ as well as $\eta^{\mu \nu},C_{\al \be},C^{\dal \dbe}
\mapsto g^{ij}$ and $(\ga^\mu C)_\al{}^{\dbe} \mapsto \sqrt{2} G^{ijk}$. One is then free to pick a different
assignment, e.g.\ $(P^i,Q^j, R^k) \equiv (S_\al, S^{\dbe}, \psi^\mu)$ which yields the correlator $\langle \psi^{\mu_1}
... \psi^{\mu_n} S_{\al_1} ... S_{\al_\ell} S^{\dbe_1} ... S^{\dbe_m} \rangle$ with $(\ell,m,n)$ traded for
$(n,\ell,m)$.

\medskip Uniform correlators like $\langle (\psi^\mu)^{2n} \rangle $ or $\langle (S_\al)^{2n} \rangle$ with one type of
field only are trivially determined by Wick's theorem. Each field $\psi^\mu, S_\al, S^{\dbe}$ by itself is a free
world-sheet fermion, hence $\langle S_{\al_1}(z_1) ... S_{\al_{2n}} (z_{2n}) \rangle$ can be reduced to two-point
contractions in the same manner as $\langle \psi^{\mu_1}(z_1) ... \psi^{\mu_{2n}} (z_{2n}) \rangle$.

\medskip As a first nontrivial example, let us apply the triality-based methods to the five-point functions $\langle P P P Q R
\rangle$. From (\ref{Omega8}) for $n=2$ we know the result for three vectors $P^i \equiv \psi^\mu$:
\begin{align}
\langle &\psi^\mu (z_1) \, \psi^\nu(z_2) \, \psi^\la(z_3)\, S_\al(z_4) \, S^{\dbe}(z_5) \rangle \eq \frac{1}{\sqrt{2} \, (z_{14} \, z_{15} \, z_{24} \, z_{25} \, z_{34} \, z_{35} \, z_{45})^{1/2}} \notag \\
& \times \ \left[ \frac{\eta^{\mu \nu }}{z_{12}} \; (\ga^\la \, C)_\al \, \!^{\dbe} \, z_{14} \, z_{25} \ - \ \frac{\eta^{\mu \la }}{z_{13}} \; (\ga^\nu \, C)_\al \, \!^{\dbe} \, z_{14} \, z_{35} \ + \ \frac{\eta^{\nu \la }}{z_{23}} \; (\ga^\mu \, C)_\al \, \!^{\dbe} \, z_{24} \, z_{35} \ + \ \frac{z_{45}}{2} \; (\ga^\mu \, \gab^\nu \, \ga^\la \, C)_\al \, \!^{\dbe} \right]\,.
\label{5pt1}
\end{align}
This can be translated into triality covariant language,
\begin{align}
\langle &P^{i_1} (z_1) \, P^{i_2}(z_2) \, P^{i_3}(z_3)\, Q^{j_4}(z_4) \, R^{k_5}(z_5) \rangle \eq \frac{1}{ (z_{14} \, z_{15} \, z_{24} \, z_{25} \, z_{34} \, z_{35} \, z_{45})^{1/2}} \; \times \ \biggl[ \frac{g^{i_1 i_2}}{z_{12}} \; G^{i_3 j_4 k_5} \, z_{14} \, z_{25} \biggr. \notag \\
\biggl.& \ \ \ \ \ \ \ \ - \ \frac{g^{i_1 i_3}}{z_{13}} \; G^{i_2 j_4 k_5} \, z_{14} \, z_{35} \ + \ \frac{g^{i_2 i_3}}{z_{23}} \; G^{i_1 j_4 k_5} \, z_{24} \, z_{35}  - \ z_{45} \; G^{i_1 j_4 k} \,  g_{kk'} G^{i_2 j k'} \, g_{jj'} \, G^{i_3 j' k_5} \biggr] \,,
\label{5pt2}
\end{align}
where $(\ga^\mu \gab^\nu \ga^\la C)_\al \, \!^{\dbe} \equiv - 2\sqrt{2} \, G^{i_1 j_4 k} g_{kk'} G^{i_2 j k'} g_{jj'}
G^{i_3 j' k_5}$. The minus sign arises from the conversion $\gab^\nu \rightarrow \ga^\nu$. By specializing to a
configuration with three left-handed spin fields $P^i \equiv S_\al$ (and $Q^j \equiv \psi^\mu$ and $R^k \equiv S^{\dbe}$
or vice versa), one arrives at the so far unknown result
\begin{align}
\langle &\psi^\mu (z_1) \, S_{\al}(z_2) \, S_{\be}(z_3)\, S_{\ga}(z_4) \, S^{\dde}(z_5) \rangle \eq  \frac{ - \, 1}{ (z_{12} \, z_{13} \, z_{14} \, z_{15} \, z_{25} \, z_{35} \, z_{45})^{1/2}} \ \times \ \biggl[  \, \frac{z_{15}}{2} \; (\ga^\mu \, \gab^\la \, C)_{\ga \be} \, (\ga_\la \, C)_{\al } \, \! ^{\dde} \biggr. \notag \\
& \biggl. \ \ \ \ \ \ \ \ \ \  \ \ \ \ \ \ \ \ \ \ + \ \frac{C_{\al \be}}{z_{23}} \; (\ga^\mu \, C)_\ga \, \! ^{\dde} \, z_{13} \, z_{25} \ - \ \frac{C_{\al \ga}}{z_{24}} \; (\ga^\mu \, C)_\be \, \! ^{\dde} \, z_{14} \, z_{25} \ + \ \frac{C_{ \be \ga }}{z_{34}} \; (\ga^\mu \, C)_\al \, \! ^{\dde} \, z_{14} \, z_{35}  \biggr] \ .
\label{5pt3}
\end{align}

\medskip Identities for the Clebsch--Gordan coefficients in (\ref{5pt3}) can be derived from the Dirac algebra.
Generalizing $2\eta^{\mu \nu} (\ga^\la C)_\al \, \! ^{\dbe} = -(\ga^\mu \gab^\nu \ga^\la C)_\al \, \! ^{\dbe} - (\ga^\nu
\gab^\mu \ga^\la C)_\al \, \! ^{\dbe}$ to
\begin{equation}
G^{i_4 j_3 k} \, g_{kk'} \, G^{i j_1 k'} \, g_{ii'} \, G^{i' j_2 k_5} \ + \ G^{i_4 j_3 k} \, g_{kk'} \, G^{i j_2 k'} \, g_{ii'} \, G^{i' j_1 k_5} \eq g^{j_1 j_2} \, G^{i_4 j_3 k_5}\,,
\label{5pt4}
\end{equation}
and then specializing $(j_1,j_2,j_3) = (\al,\be,\ga)$ and $i_4 = \mu$, $k_5 = \dde$ yields
\begin{equation}
- \,C_{\al \be} \, (\ga^\mu \, C)_{\ga }\, \! ^{\dde} \eq \frac{1}{2} \; (\ga^{\la} \, \gab^\mu \, C)_{\be \ga} \, (\ga_\la \, C)_{\al} \, \!^{\dde} \ + \ \frac{1}{2} \; (\ga^{\la} \, \gab^\mu \, C)_{\al \ga} \, (\ga_\la \, C)_{\be} \, \!^{\dde} \ .
\label{5pt5}
\end{equation}
This result also follows from (\ref{F8.3}) upon multiplication with $\ga^\mu_{\ga \dga}$.

\medskip The most interesting application of this procedure is to relate the $2n$- and $(2n+1)$-point correlation
functions from subsection \ref{sec:highpt} to so far unknown correlators with a large number of spin field insertions:
\begin{equation*}
  \vev{\psi^{2\ell}\,S^2}\ \ \leftrightarrow \ \ \vev{\psi^2\,S^{2\ell}}\,,\qquad
  \vev{\psi^{2\ell}\,S^2}\ \ \leftrightarrow \ \ \vev{S^{2\ell}\,\dot S^2}\,,\qquad
  \vev{\psi^{2\ell-1}\,S\,\dot S} \ \ \leftrightarrow \ \  \vev{\psi\,S^{2\ell-1}\,\dot S}\,.
\end{equation*}
For this purpose, we need the $SO(8)$ triality covariant version of the tree-level correlators (\ref{Omega8}) and
(\ref{omega8})\footnote{The $\left( z_{AB} /2 \right)^\ell$ factors from (\ref{Omega8}) and (\ref{omega8}) are converted
  to $(-z_{AB})^\ell$ by means of $(\ga^\mu C)_\al \, \!^{\dbe} \equiv \sqrt{2}G^{ijk}$ as well as $(\gab^\mu C)^{\dbe}
  \, \!_{\al} \equiv -\sqrt{2}G^{ijk}$.}:
\begin{align}
& \ \ \Om_{(n,D=8)}^{i_{1} ... i_{2n-1} jk}(z_i) \ \ := \ \
\langle P^{i_{1}}(z_{1}) \, P^{i_{2}}(z_{2}) \, ... \, P^{i_{2n-1}}(z_{2n-1}) \, Q^{j}(z_{A}) \, R^{k}(z_{B}) \rangle  \notag \\
& \ \  = \ \ \frac{1}{z_{AB}^{1/2} \, \prod_{i=1}^{2n-1} (z_{iA} \, z_{iB})^{1/2} } \, \sum_{\ell = 0}^{n-1} \, \bigl(  - \, z_{AB}  \bigr)^{\ell} \! \! \! \! \! \sum_{\rho \in S_{2n-1}/{\cal P}_{n,\ell}} \! \! \! \! \!
 \, G^{i_{\rho(1)} j r_1} \, G^{i_{\rho(2)} q_1} {} _{r_1} \, G^{i_{\rho(3)}} {}_{q_1} {}^{r_2} \, G^{i_{\rho(4)} q_2}{}_{r_2} \, ... \, G^{i_{\rho(2\ell+1)}} {}_{q_{\ell}} {}^{k} \notag \\
& \ \ \ \  \ \ \ \ \times  \te{sgn}(\rho)  \, \prod_{j=1}^{n-\ell-1} \frac{g^{i_{\rho(2\ell+2j)} i_{\rho(2\ell+2j+1)}}}{z_{\rho(2\ell+2j),\rho(2\ell+2j+1)} } \; z_{\rho(2\ell+2j),A} \, z_{\rho(2\ell+2j+1),B} \,,
\label{Omega8c} \\
& \ \ \om_{(n,D=8)}^{i_{1} ... i_{2n-2} j_1 j_2}(z_{i})  \ \ := \ \
\langle P^{i_1}(z_{1}) \, P^{i_{2}}(z_{2}) \, ... \, P^{i_{2n-2}}(z_{2n-2}) \, Q^{j_1}(z_{A}) \, Q^{j_2}(z_{B}) \rangle  \notag \\
& \ \  = \ \ \frac{1}{ z_{AB}  \, \prod_{i=1}^{2n-2} (z_{iA} \, z_{iB})^{1/2} } \, \sum_{\ell = 0}^{n-1} \, \bigl( - \, z_{AB}  \bigr)^{\ell}  \! \! \! \sum_{\rho \in S_{2n-2}/{\cal Q}_{n,\ell}} \! \! \!   G^{i_{\rho(1)} j_1 r_1} \, G^{i_{\rho(2)} q_1}{}_{r_1} \, G^{i_{\rho(3)}}{}_{q_1} {}^{r_2} \, ... \, G^{i_{\rho(2\ell)} j_2}{}_{r_{\ell}} \notag \\
& \ \ \ \  \ \ \ \ \times \ \te{sgn}(\rho) \, \prod_{j=1}^{n-\ell-1} \frac{g^{i_{\rho(2\ell+2j-1)} i_{\rho(2\ell+2j)}}}{z_{\rho(2\ell+2j-1),\rho(2\ell+2j)} } \; z_{\rho(2\ell+2j-1),A} \, z_{\rho(2\ell+2j),B}  \,.
\label{omega8c}
\end{align}
The former gives rise to a new result under $(P^{i_\ell},Q^j,R^k) \equiv (S_{\al_\ell}, \psi^\mu, S^{\dbe})$:
\begin{align}
\langle &\psi^\mu(z_A) \, S_{\al_1}(z_1) \, ... \, S_{\al_{2n-1}}(z_{2n-1}) \, S^{\dbe}(z_B) \rangle \eq
\frac{(-1)^{n-1}}{\sqrt{2} \, z_{AB}^{1/2} \, \prod_{i=1}^{2n-1} (z_{Ai} \, z_{iB})^{1/2} } \, \sum_{\ell = 0}^{n-1} \,
\left( \frac{z_{AB}}{2}  \right)^{\ell} \notag \\
& \! \! \! \! \! \sum_{\rho \in S_{2n-1}/{\cal P}_{n,\ell}} \! \! \! \! \!
 \, \te{sgn}(\rho)  \, (\ga^\mu)_{\al_{\rho(1)} \dde_1} \, (\ga^{\la_1} \, C)_{ \al_{\rho(2)}}{}^{\dde_1} \, (\ga_{\la_1})_{\al_{\rho(3)} \dde_2} \, (\ga^{\la_2} \, C)_{\al_{\rho(4)} }{}^{\dde_2} \, ... \, (\ga^{\la_\ell} \, C)_{\al_{\rho(2\ell)} }{}^{\dde_{\ell}} \, (\ga_{\la_\ell} \, C)_{\al_{\rho(2\ell+1)}} {}^{\dbe}
 \notag \\
& \ \ \ \  \ \ \ \ \times \  \prod_{j=1}^{n-\ell-1} \frac{C_{\al_{\rho(2\ell+2j)} \al_{\rho(2\ell+2j+1)}}}{z_{\rho(2\ell+2j),\rho(2\ell+2j+1)} } \; z_{A,\rho(2\ell+2j)} \, z_{\rho(2\ell+2j+1),B} \,.
\label{spec1}
\end{align}
The latter yields two classes of correlation functions. Firstly we assign $(P^{i_\ell},Q^{j_1},Q^{j_2}) \equiv
(S_{\al_\ell}, \psi^\mu, \psi^\nu)$
\begin{align}
&\langle \psi^\mu(z_A) \, \psi^\nu(z_B) \, S_{\al_1}(z_1) \, ... \, S_{\al_{2n-2}}(z_{2n-2}) \rangle \eq \frac{1}{ z_{AB}  \, \prod_{i=1}^{2n-2} (z_{Ai} \, z_{Bi})^{1/2} } \, \sum_{\ell = 0}^{n-1} \, \left( - \;  \frac{ z_{AB} }{2}  \right)^{\ell}  \notag \\
& \! \! \! \sum_{\rho \in S_{2n-2}/{\cal Q}_{n,\ell}} \! \! \! \te{sgn}(\rho) \,  (\ga^\mu)_{\al_{\rho(1)} \dde_1} \, (\ga^{\la_1} \, C)_{\al_{\rho(2)}} {}^{\dde_1} \, (\ga_{\la_1})_{\al_{\rho(3)} \dde_2} \, (\ga^{\la_2} \, C)_{\al_{\rho(4)}} {}^{\dde_2} \, ... \, (\ga_{\la_{\ell-1}})_{\al_{\rho(2\ell-1)} \dde_{\ell}} \, (\ga^\nu \, C)_{\al_{\rho(2\ell)}}{}^{\dde_{\ell}} 
 \notag \\
& \ \ \ \  \ \ \ \ \times \  \prod_{j=1}^{n-\ell-1} \frac{C_{\al_{\rho(2\ell+2j-1)} \al_{\rho(2\ell+2j)}}}{z_{\rho(2\ell+2j-1),\rho(2\ell+2j)} } \; z_{A,\rho(2\ell+2j-1)} \, z_{B,\rho(2\ell+2j)}  \,,
\label{spec2}
\end{align}
and secondly we can specify $(P^{i_\ell},Q^{j_1},Q^{j_2}) \equiv (S_{\al_\ell}, S^{\dbe}, S^{\dga})$:
\begin{align}
&\langle S_{\al_1}(z_1) \, ... \, S_{\al_{2n-2}}(z_{2n-2}) \, S^{\dbe}(z_A) \, S^{\dga}(z_B) \rangle \eq \frac{1}{ z_{AB}  \, \prod_{i=1}^{2n-2} (z_{Ai} \, z_{Bi})^{1/2} } \, \sum_{\ell = 0}^{n-1} \, \left( - \;  \frac{ z_{AB} }{2}  \right)^{\ell}  \notag \\
& \! \! \! \sum_{\rho \in S_{2n-2}/{\cal Q}_{n,\ell}} \! \! \! \te{sgn}(\rho) \, (\ga^{\la_1} \, C)_{\al_{\rho(1)}} {}^{\dbe} \, (\ga_{\la_1})_{\al_{\rho(2)} \dde_1} \, (\ga^{\la_2} \, C)_{\al_{\rho(3)}}{}^{\dde_1} \, (\ga_{\la_2})_{\al_{\rho(4)} \dde_2} \, ... \, (\ga^{\la_\ell} \, C)_{\al_{\rho(2\ell-1)}}{}^{\dde_{\ell-1}} \, (\ga_{\la_\ell} \, C)_{\al_{\rho(2\ell)}}{}^{\dga}
 \notag \\
& \ \ \ \  \ \ \ \ \times \  \prod_{j=1}^{n-\ell-1} \frac{C_{\al_{\rho(2\ell+2j-1)} \al_{\rho(2\ell+2j)}}}{z_{\rho(2\ell+2j-1),\rho(2\ell+2j)} } \; z_{\rho(2\ell+2j-1),A} \, z_{\rho(2\ell+2j),B}  \,.
\label{spec3}
\end{align}
Not all correlation functions can be derived from (\ref{Omega8}) and (\ref{omega8}) via triality. Even the six-point function $\langle \psi^\mu \psi^\nu S_\al
S_\be S^{\dga} S^{\dde} \rangle$ must be derived from first principles, since this field configuration is a triality
fixed point. Let us give the tree-level result in a triality covariant way
\begin{align}
\langle &P^{i_1}(z_1) \, P^{i_2}(z_2) \, Q^{j_1}(z_3) \, Q^{j_2}(z_4) \, R^{k_1}(z_5) \, R^{k_2}(z_6) \rangle  \eq \frac{1}{ (z_{13} \, z_{14} \, z_{15} \, z_{16} \, z_{23} \, z_{24} \, z_{25} \, z_{26} \, z_{35} \, z_{36} \, z_{45} \, z_{46})^{1/2}} \notag \\
&\times \ \left[ \, \frac{ g^{i_1 i_2} \, g^{j_1 j_2} \, g^{k_1 k_2}}{4 \, z_{12} \, z_{34} \, z_{56}} \;  \Bigl( \, z_{36} \, z_{45} \, \bigl( \, z_{13} \, z_{16} \, z_{24} \, z_{25} \, \ + \ z_{14} \, z_{15} \, z_{23} \, z_{26} \, \bigr) \Bigr. \right. \notag \\
& \ \ \ \ \ \ \ \ \ \ \ \ \ \ \ \ \ \ \ \ \ \ \ \ \ \ \ \ \ \ \ \ \left. \Bigl. \ + \ z_{35} \, z_{46} \,  \bigl( \, z_{13} \, z_{15} \, z_{24} \, z_{26}   \ + \ z_{14} \, z_{16} \, z_{23} \, z_{25} \, \bigr) \, \Bigr) \right. \notag \\
& \ \ \ \, \left. + \ \frac{1}{2}  \; G^{i_1 j_1 k_1} \, G^{i_2 j_2 k_2} \, \Bigl( \, z_{14} \, z_{25} \, z_{36}  \ - \ z_{16} \, z_{23} \, z_{45} \,   \Bigr)  \  + \ \frac{1}{2}  \; G^{i_1 j_1 k_2} \, G^{i_2 j_2 k_1} \,  \Bigl( \, z_{15} \, z_{23} \, z_{46}  \ - \ z_{14} \, z_{26} \, z_{35}   \, \Bigr) \right. \notag \\
& \ \ \ \, \left. + \ \frac{1}{2}  \; G^{i_1 j_2 k_1} \, G^{i_2 j_1 k_2} \, \Bigl( \, z_{16} \, z_{24} \, z_{35} \ - \ z_{13} \, z_{25} \, z_{46}  \, \Bigr) \  + \ \frac{1}{2}  \; G^{i_1 j_2 k_2} \, G^{i_2 j_1 k_1} \, \Bigl( \, z_{13} \, z_{26} \, z_{45} \ - \ z_{15} \, z_{24} \, z_{36} \, \Bigr) \right. \notag \\
& \ \ \ \, \left. - \ \frac{g^{k_1 k_2}}{2 \, z_{56}} \;  G^{i_1 [ j_1}{}_r \, G^{|i_2|j_2 ] r} \,
\Bigl( \, z_{15} \, z_{25} \, z_{36} \, z_{46}   \ - \ z_{16} \, z_{26} \, z_{35} \, z_{45} \, \Bigr) \right. \notag \\
& \ \ \ \, \left. - \   \frac{g^{j_1 j_2}}{2 \, z_{34}} \; G^{i_1 q [ k_1} \, G^{|i_2 |}{}_q {}^{ k_2] } \, \Bigl( \, z_{13} \, z_{23} \, z_{45} \, z_{46}  \ - \ z_{14} \, z_{24} \, z_{35} \, z_{36} \, \Bigr) \right. \notag \\
& \ \ \ \, \left. - \  \frac{g^{i_1 i_2}}{2 \,z_{12}} \; G^{p j_1 [ k_1} \, G_p{}^{|j_2| k_2]} \, \Bigl( \, z_{13} \, z_{14} \, z_{25} \, z_{26} \ + \ z_{15} \, z_{16} \, z_{23} \, z_{24} \, \Bigr) \right. \notag \\
& \ \ \ \ \ \! \biggl. + \  G^{[i_1 | [ j_1 | r} \, G^{p | j_2]}{}_r \, G^{| i_2] q [k_1} \, G_{pq}{}^{k_2]} \, \Bigl( \, z_{14} \, z_{25} \, z_{36}  \ + \ z_{16} \, z_{23} \, z_{45} \,   \Bigr) \, \biggr]\,. 
\end{align}

\subsection{The loop completion}

As explained above, $SO(8)$ triality covariance of the RNS CFT breaks down at nonzero genus because spin fields have
different global periodicity properties than the NS fermions. Therefore, the loop generalizations of the tree-level
correlators in the previous subsection cannot be derived by means of triality. In particular, we cannot give the higher
point functions $\langle \psi^\mu S_{\al_1} ... S_{\al_{2n-1}} S^{\dbe} \rangle \spin$, $\langle \psi^\mu \psi^\nu S_{\al_1}
... S_{\al_{2n-2}} \rangle \spin$ and $\langle S_{\al_1} ... S_{\al_{2n-2}} S^{\dbe} S^{\dga} \rangle \spin$ with arbitrary $n$ at
loop-level even though they are available at tree-level with (\ref{spec1}), (\ref{spec2}) and (\ref{spec3}). Hence, we can only study individual cases with a fixed number of fields at higher genus.

\medskip This section contains any four-, five- and six-point function with at least four spin fields on higher
genus. They are obtained by the usual method of testing various index configurations. The tree-level results derived
from the triality analysis are a good starting point for this procedure since all the $z_{ij}$ can be replaced by
$E_{ij}$ and only the $\Theta^{\vec{a}}_{\vec{b}}$ arguments are left to determine.

\medskip
Let us start with four spin field correlations, firstly
\begin{align}
\langle S_\al(z_1) \, &S_{\be}(z_2) \, S^{\dga}(z_3) \, S^{\dde}(z_4) \rangle \spin \eq \frac{ \tspinb 1 &3 \\ 2 &4 \tspine \, \tspinb 1 &4 \\2 &3 \tspine}{2 \, \tttt \, E_{12} \, E_{34} \, (E_{13} \, E_{14} \, E_{23} \, E_{24})^{1/2}} \notag \\
&\times \ \Bigl[ \, (\ga^\mu \, C)_\al {}^{\dga} \, (\ga_\mu \, C)_\be {}^{\dde} \, E_{14} \, E_{23} \, \tspinb 1 &4 \\ 2 &3 \tspine^2 \ + \ (\ga^\mu \, C)_\al {}^{\dde} \, (\ga_\mu \, C)_\be {}^{\dga} \, E_{13} \, E_{24} \, \tspinb 1 &3 \\ 2 &4 \tspine^2 \, \Bigr]\,,
\end{align} 
and secondly
\begin{align}
&\langle S_\al(z_1) \, S_{\be}(z_2) \, S_{\ga}(z_3) \, S_{\de}(z_4) \rangle \spin \eq \frac{1}{\tttt} \notag \\
& \times \ \bigg[ \frac{ C_{\al \be} \, C_{\ga \de}}{E_{12} \, E_{34}}\,\tspinb 1 &3 \\ 2 &4 \tspine^2 \,  \tspinb 1 &4 \\ 2 &3 \tspine^2 \ - \ \frac{ C_{\al \ga} \, C_{\be \de}}{E_{13} \, E_{24}}\,\tspinb 1 &2 \\ 3 &4 \tspine^2 \,  \tspinb 1 &4 \\ 2 &3 \tspine^2 \ + \ \frac{ C_{\al \de} \, C_{\be \ga}}{E_{14} \, E_{23}}\tspinb 1 &2 \\ 3 &4 \tspine^2 \,  \tspinb 1 &3 \\ 2 &4 \tspine^2\bigg] \ . 
\end{align}
In presence of an additional NS fermion, we find
\begin{align}
\langle \psi^\mu(z_1) \, &S_{\al}(z_2) \, S_{\be}(z_3) \, S_{\ga}(z_4) \, S^{\dde}(z_5) \rangle \spin \eq \frac{- \, 1}{\sqrt{2} \, \tttt \, (E_{12} \, E_{13} \, E_{14} \, E_{15} \, E_{25} \, E_{35} \, E_{45})^{1/2}} \notag \\
&\times \ \left[ \, \frac{C_{\al \be}}{E_{23}} \; (\ga^\mu \, C)_\ga {}^{\dde} \,E_{13} \, E_{25} \, \tspinb 1 &1 &3 \\ 2 &4 &5 \tspine \, \tspinb 2 &4 \\ 3 &5 \tspine \, \tspinb 2 &5 \\ 3 &4 \tspine^2  \right. \notag \\
& \ \ \ \, \left. + \ \frac{C_{\ga \be}}{E_{34}} \; (\ga^\mu \, C)_\al {}^{\dde} \,E_{13} \, E_{45} \, \tspinb 1 &1 &3 \\ 2 &4 &5 \tspine \, \tspinb 2 &4 \\ 3 &5 \tspine \, \tspinb 2 &3 \\ 4 &5 \tspine^2 \right. \notag \\
& \ \ \ \, \left. + \ \frac{1}{2}\,(\ga^\la \, \gab^\mu \, C)_{\al \be} \, (\ga_\la \, C)_\ga {}^{\dde} \; \frac{E_{14} \, E_{25}}{E_{24}} \; \tspinb 1 &1 &4 \\ 2 &3 &5 \tspine \, \tspinb 2 &3 \\ 4 &5 \tspine \, \tspinb 2 &5 \\ 3 &4 \tspine^2 \right. \notag \\
& \ \ \ \, \left. + \ \frac{1}{2}\,(\ga^\la \, \gab^\mu \, C)_{\ga \be} \, (\ga_\la \, C)_\al {}^{\dde} \; \frac{E_{12} \, E_{45}}{E_{24}} \; \tspinb 1 &1 &2 \\ 3 &4 &5 \tspine \, \tspinb 2 &5 \\ 3 &4 \tspine \, \tspinb 2 &3 \\ 4 &5 \tspine^2 \, \right]\,.
\end{align}
Here we have chosen a different basis of tensors than in (\ref{5pt3}) to make antisymmetry in $S_\al(z_2) \leftrightarrow S_\ga(z_4)$ manifest.

\medskip The triality symmetric tree-level correlator $\langle P^{i_1} P^{i_2} Q^{j_1} Q^{j_2} R^{k_1} R^{k_2} \rangle$
generalizes as follows to higher genus:
\begin{align}
\langle &\psi^{\mu}(z_1) \, \psi^\nu(z_2) \, S_\al(z_3) \, S_\be(z_4) \, S^{\dga}(z_5) \, S^{\dde}(z_6) \rangle \spin \eq \frac{( E_{35} \, E_{36} \, E_{45} \, E_{46} )^{-1/2}}{4 \, \tttt \, (E_{13} \, E_{14} \, E_{15} \, E_{16} \, E_{23} \, E_{24} \, E_{25} \, E_{26} )^{1/2}} \notag \\
&\times \ \left[ \, \frac{ \eta^{\mu \nu} \, C_{\al \be} \, C^{\dga \dde}}{E_{12} \, E_{34} \, E_{56}} \; \tspinb 3 &5 \\ 4 &6 \tspine \, \tspinb 3 &6 \\ 4 &5 \tspine \right. \notag \\
& \ \ \ \ \ \ \ \ \ \ \ \ \ \ \Bigl( \, E_{36} \, E_{45} \, \tspinb 3 &6 \\ 4& 5 \tspine \, \bigl( \, E_{13} \, E_{16} \, E_{24} \, E_{25} \, \tspinb 1 &1 &3 &6 \\ 2 &2 &4 &5 \tspine \ + \ E_{14} \, E_{15} \, E_{23} \, E_{26} \, \tspinb 1 &1 &4 &5 \\ 2 &2 &3 &6 \tspine \, \bigr) \Bigr. \notag \\
& \ \ \ \ \ \ \ \ \  \ \ \ \ \Bigl. + \ E_{35} \, E_{46} \, \tspinb 3 &5 \\ 4& 6 \tspine \, \bigl( \, E_{13} \, E_{15} \, E_{24} \, E_{26} \, \tspinb 1 &1 &3 &5 \\ 2 &2 &4 &6 \tspine \ + \ E_{14} \, E_{16} \, E_{23} \, E_{25} \, \tspinb 1 &1 &4 &6 \\ 2 &2 &3 &5 \tspine \, \bigr) \, \Bigr) \notag \\
& \ \ \ \, \left. + \ (\ga^\mu \, C)_\al {}^{\dga} \, (\ga^\nu \, C)_\be {}^{\dde} \, \tspinb 3 &4 \\ 5 &6 \tspine \, \tspinb 3 &6 \\ 4 &5 \tspine \right. \notag \\
& \ \ \ \ \ \ \ \ \ \ \ \ \ \ \Bigl( \, E_{14} \, E_{25} \, E_{36} \, \tspinb 1 &1 &4 \\ 3 &5 &6 \tspine \, \tspinb 2 &2 &5 \\ 3 &4 &6 \tspine \ - \ E_{16} \, E_{23} \, E_{45} \, \tspinb 1 &1 &6 \\ 3 &4 &5 \tspine \, \tspinb 2 &2 &3 \\ 4 &5 &6 \tspine \, \Bigr) \notag \\
& \ \ \ \, \left. + \ (\ga^\mu \, C)_\al {}^{\dde} \, (\ga^\nu \, C)_\be {}^{\dga} \, \tspinb 3 &4 \\ 5 &6 \tspine \, \tspinb 3 &5 \\ 4 &6 \tspine \right. \notag \\
& \ \ \ \ \ \ \ \ \ \ \ \ \ \ \Bigl( \, E_{15} \, E_{23} \, E_{46} \, \tspinb 1 &1 &5 \\ 3 &4 &6 \tspine \, \tspinb 2 &2 &3 \\ 4 &5 &6 \tspine \ - \ E_{14} \, E_{26} \, E_{35} \, \tspinb 1 &1 &4 \\ 3 &5 &6 \tspine \, \tspinb 2 &2 &6 \\ 3 &4 &5 \tspine \, \Bigr) \notag \\
& \ \ \ \, \left. + \ (\ga^\mu \, C)_\be {}^{\dga} \, (\ga^\nu \, C)_\al {}^{\dde} \, \tspinb 3 &4 \\ 5 &6 \tspine \, \tspinb 3 &5 \\ 4 &6 \tspine \right. \notag \\
& \ \ \ \ \ \ \ \ \ \ \ \ \ \ \Bigl( \, E_{16} \, E_{24} \, E_{35} \, \tspinb 1 &1 &6 \\ 3 &4 &5 \tspine \, \tspinb 2 &2 &4 \\ 3 &5 &6 \tspine \ - \ E_{13} \, E_{25} \, E_{46} \, \tspinb 1 &1 &3 \\ 4 &5 &6 \tspine \, \tspinb 2 &2 &5 \\ 3 &4 &6 \tspine \, \Bigr) \notag \\
& \ \ \ \, \left. + \ (\ga^\mu \, C)_\be {}^{\dde} \, (\ga^\nu \, C)_\al {}^{\dga} \, \tspinb 3 &4 \\ 5 &6 \tspine \, \tspinb 3 &6 \\ 4 &5 \tspine \right. \notag \\
& \ \ \ \ \ \ \ \ \ \ \ \ \ \ \Bigl( \, E_{13} \, E_{26} \, E_{45} \, \tspinb 1 &1 &3 \\ 4 &5 &6 \tspine \, \tspinb 2 &2 &6 \\ 3 &4 &5 \tspine \ - \ E_{15} \, E_{24} \, E_{36} \, \tspinb 1 &1 &5 \\ 3 &4 &6 \tspine \, \tspinb 2 &2 &4 \\ 3 &5 &6 \tspine \, \Bigr) \notag \\
& \ \ \ \, \left. + \ \frac{C^{\dga \dde}}{E_{56}} \; (\ga^{\mu \nu} \, C)_{\al \be} \, \tspinb 3 &5 \\ 4 &6 \tspine \, \tspinb 3 &6 \\ 4 &5 \tspine\right. \notag \\
& \ \ \ \ \ \ \ \ \ \ \ \ \ \ \Bigl( \, E_{15} \, E_{25} \, E_{36} \, E_{46} \, \tspinb 1 &1 &5 \\ 3 &4 &6 \tspine \, \tspinb 2 &2 &5 \\ 3 &4 &6 \tspine \ - \ E_{16} \, E_{26} \, E_{35} \, E_{45} \, \tspinb 1 &1 &6 \\ 3 &4 &5 \tspine \, \tspinb 2 &2 &6 \\ 3 &4 &5 \tspine \, \Bigr) \notag \\
& \ \ \ \, \left. + \ \frac{C_{\al \be}}{E_{34}} \; (\gab^{\mu \nu} \, C)^{\dga \dde} \, \tspinb 3 &5 \\ 4 &6 \tspine \, \tspinb 3 &6 \\ 4 &5 \tspine\right. \notag \\
& \ \ \ \ \ \ \ \ \ \ \ \ \ \ \Bigl( \, E_{13} \, E_{23} \, E_{45} \, E_{46} \, \tspinb 1 &1 &3 \\ 4 &5 &6 \tspine \, \tspinb 2 &2 &3 \\ 4 &5 &6 \tspine \ - \ E_{14} \, E_{24} \, E_{35} \, E_{36} \, \tspinb 1 &1 &4 \\ 3 &5 &6 \tspine \, \tspinb 2 &2 &4 \\ 3 &5 &6 \tspine \, \Bigr) \notag \\
& \ \ \ \, \left. - \ \frac{\eta^{\mu \nu}}{E_{12}} \; (\ga_\la \, C)_{[\al}{}^{\dga} \, (\ga^\la \, C)_{\be]} {}^{\dde} \, \tspinb 3 &4 \\ 5 &6 \tspine \, \tspinb 3 &5 \\ 4 &6 \tspine \, \tspinb 3 &6 \\ 4 &5 \tspine \right. \notag \\
& \ \ \ \ \ \ \ \ \ \ \ \ \ \ \Bigl( \, E_{13} \, E_{14} \, E_{25} \, E_{26} \, \tspinb 1 &1 &3 &4 \\ 2 &2 &5 &6 \tspine  \ + \ E_{15} \, E_{16} \, E_{23} \, E_{24} \, \tspinb 1 &1 &5 &6 \\ 2 &2 &3 &4 \tspine \, \Bigr) \notag \\
& \ \ \ \, \left. + \  (\ga^{[\mu} {}_\la \, C)_{\al \be} \, (\gab^{\nu]\la } \, C)^{\dga \dde} \, \tspinb 3 &4 \\ 5 &6 \tspine  \, \tspinb 3 &6 \\ 4 &5 \tspine \right. \notag \\
& \ \ \ \ \ \ \ \ \ \ \ \ \ \!  \ \biggl. \, \Bigl( \,  E_{14} \, E_{25} \, E_{36} \, \tspinb 1 &1 &4 \\ 3 &5 &6 \tspine \, \tspinb 2 &2 &5 \\ 3 &4 &6 \tspine  \ + \ E_{16} \, E_{23} \, E_{45} \, \tspinb 1 &1 &6 \\ 3 &4 &5 \tspine \, \tspinb 2 &2 &3 \\ 4 &5 &6 \tspine \, \Bigr)  \, \biggr]\,. 
\label{6pt,tri}
\end{align}
Using Fay's trisecant identity, one can alternatively represent the last two lines as
\begin{align}
\frac{1}{E_{12}} \; (\ga^{[\mu} {}_\la \, C)_{\al \be} \, &(\gab^{\nu]\la } \, C)^{\dga \dde} \, \tspinb 3 &4 \\ 5 &6 \tspine \, \tspinb 3 &5 \\ 4 &6 \tspine \, \tspinb 3 &6 \\ 4 &5 \tspine  \notag \\
& \times \  \Bigl( \, E_{13} \, E_{14} \, E_{25} \, E_{26} \, \tspinb 1 &1 &3 &4 \\ 2 &2 &5 &6 \tspine  \ - \ E_{15} \, E_{16} \, E_{23} \, E_{24} \, \tspinb 1 &1 &5 &6 \\ 2 &2 &3 &4 \tspine \, \Bigr) \ .
\end{align}
The second non-vanishing six-point function of this type reads:
\begin{align}
\langle &\psi^{\mu}(z_1) \, \psi^\nu(z_2) \, S_\al(z_3) \, S_\be(z_4) \, S_{\ga}(z_5) \, S_{\de}(z_6) \rangle \spin \eq \frac{1}{4 \, \tttt \, (E_{13} \, E_{14} \, E_{15} \, E_{16}  \, E_{23} \, E_{24} \, E_{25} \, E_{26})^{1/2}} \notag \\
&\times \ \left[ \, \frac{ \eta^{\mu \nu} \, C_{\al \be} \, C_{\ga \de}}{E_{12} \, E_{34} \, E_{56}} \; \tspinb 3 &5 \\ 4 &6 \tspine \, \tspinb 3 &6 \\ 4 &5 \tspine  \right. \notag \\
& \ \ \ \ \ \ \ \ \ \ \ \ \ \ \Bigl( \, E_{13} \, E_{15} \, E_{24} \, E_{26} \, \tspinb 3 &6 \\ 4 &5 \tspine \, \tspinb 1 &1 &3 &5 \\ 2 &2 &4 &6 \tspine \ + \ E_{14} \, E_{16} \, E_{23} \, E_{25} \, \tspinb 3 &6 \\ 4 &5 \tspine \, \tspinb 1 &1 &4 &6 \\ 2 &2 &3 &5 \tspine \bigr. \notag \\
& \ \ \ \ \ \ \ \ \ \ \ \ \ \Bigl. + \ E_{13} \, E_{16} \, E_{24} \, E_{25} \, \tspinb 3 &5 \\ 4 &6 \tspine \, \tspinb 1 &1 &3 &6 \\ 2 &2 &4 &5 \tspine \ + \ E_{14} \, E_{15} \, E_{23} \, E_{26} \, \tspinb 3 &5 \\ 4 &6 \tspine \, \tspinb 1 &1 &4 &5 \\ 2 &2 &3 &6 \tspine \,
 \Bigr) \notag \\
& \ \ \ \, \left. - \ \frac{2\,\eta^{\mu \nu} \, C_{\al \ga} \, C_{\be \de}}{E_{12} \, E_{35} \, E_{46}} \; \tspinb 3 &6 \\ 4 &5 \tspine \, \tspinb 3 &4 \\ 5 &6 \tspine^2 \, \Bigl( \, E_{13} \, E_{24} \, E_{25} \, E_{16} \, \tspinb 1 &1 &3 &6 \\ 2 &2 &4 &5 \tspine \ + \ E_{14} \, E_{15} \, E_{23} \, E_{26} \, \tspinb 1 &1 &4 &5 \\ 2 &2 &3 &6 \tspine \, \Bigr) \right. \notag \\
& \ \ \ \, \left. + \ \frac{2\,\eta^{\mu \nu} \, C_{\al \de} \, C_{\be \ga}}{E_{12} \, E_{36} \, E_{45}} \; \tspinb 3 &5 \\ 4 &6 \tspine \, \tspinb 3 &4 \\ 5 &6 \tspine^2 \, \Bigl( \, E_{13} \, E_{24} \, E_{15} \, E_{26} \, \tspinb 1 &1 &3 &5 \\ 2 &2 &4 &6 \tspine \ + \ E_{14} \, E_{16} \, E_{23} \, E_{25} \, \tspinb 1 &1 &4 &6 \\ 2 &2 &3 &5 \tspine \, \Bigr) \right. \notag \\
& \ \ \ \, \left. + \ \frac{C_{\al \be} \, (\ga^{\mu \nu} \, C)_{\ga \de}}{E_{34}} \; \tspinb 3 &5 \\ 4 &6 \tspine \, \tspinb 3 &6 \\ 4 &5 \tspine \, \Bigl( \, E_{13} \, E_{24} \, \tspinb 1 &1 &3 \\ 4 &5 &6 \tspine \, \tspinb 2 &2 &4 \\ 3 &5 &6 \tspine \ + \ E_{14} \, E_{23} \, \tspinb 1 &1 &4 \\ 3 &5 &6 \tspine \, \tspinb 2 &2 &3 \\ 4 &5 &6 \tspine \, \Bigr) \right. \notag \\
& \ \ \ \, \left. + \ \frac{C_{\ga \de} \, (\ga^{\mu \nu} \, C)_{\al \be}}{E_{56}} \; \tspinb 3 &5 \\ 4 &6 \tspine \, \tspinb 3 &6 \\ 4 &5 \tspine \, \Bigl( \, E_{15} \, E_{26} \, \tspinb 1 &1 &5 \\ 3 &4 &6 \tspine \, \tspinb 2 &2 &6 \\ 3 &4 &5 \tspine \ + \ E_{16} \, E_{25} \, \tspinb 1 &1 &6 \\ 3 &4 &5 \tspine \, \tspinb 2 &2 &5 \\ 3 &4 &6 \tspine \, \Bigr) \right.  \notag \\
& \ \ \ \, \left. - \ \frac{C_{\al \ga} \, (\ga^{\mu \nu} \, C)_{\be \de}}{E_{35}} \; \tspinb 3 &4 \\ 5 &6 \tspine \, \tspinb 3 &6 \\4&5 \tspine \; \Bigl( \, E_{13} \, E_{25} \, \tspinb 1 &1 &3 \\ 4 &5 &6 \tspine \, \tspinb 2 &2 &5 \\ 3 &4 &6 \tspine \ + \, E_{15} \, E_{23} \,\tspinb 1 &1 &5 \\ 3 &4 &6 \tspine \,\, \tspinb 2 &2 &3 \\ 4 &5 &6 \tspine \, \Bigr) \right.  \notag \\
& \ \ \ \, \left. - \ \frac{C_{\be \de} \, (\ga^{\mu \nu} \, C)_{\al \ga}}{E_{46}} \; \tspinb 3 &4 \\ 5 &6 \tspine \, \tspinb 3 &6 \\4&5 \tspine \; \Bigl( \, E_{14} \, E_{26} \, \tspinb 1 &1 &4 \\ 3 &5 &6 \tspine \, \tspinb 2 &2 &6 \\ 3 &4 &5 \tspine \ + \ E_{16} \, E_{24} \, \tspinb 1 &1 &6 \\ 3 &4 &5 \tspine \, \tspinb 2 &2 &4 \\ 3 &5 &6 \tspine \, \Bigr) \right. \notag \\
& \ \ \ \, \left. + \ \frac{C_{\al \de} \, (\ga^{\mu \nu} \, C)_{\be \ga}}{E_{36}} \; \tspinb 3 &4 \\ 5 &6 \tspine \, \tspinb 3 &5 \\4&6 \tspine \; \Bigl( \, E_{13} \, E_{26} \, \tspinb 1 &1 &3 \\ 4 &5 &6 \tspine \, \tspinb 2 &2 &6 \\ 3 &4 &5 \tspine \ + \ E_{16} \, E_{23} \, \tspinb 1 &1 &6 \\ 3 &4 &5 \tspine \, \tspinb 2 &2 &3 \\ 4 &5 &6 \tspine \, \Bigr) \right. \notag \\
& \ \ \ \, \left. + \ \frac{C_{\be \ga} \, (\ga^{\mu \nu} \, C)_{\al \de}}{E_{45}} \; \tspinb 3 &4 \\ 5 &6 \tspine \, \tspinb 3 &5 \\4&6 \tspine \; \Bigl( \, E_{14} \, E_{25} \, \tspinb 1 &1 &4 \\ 3 &5 &6 \tspine \, \tspinb 2 &2 &5 \\ 3 &4 &6 \tspine \ + \ E_{15} \, E_{24} \,\tspinb 1 &1 &5 \\ 3 &4 &6 \tspine \, \tspinb 2 &2 &4 \\ 3 &5 &6 \tspine \,\Bigr) \right. \notag \\
& \ \ \ \ \, \, \! \biggl. + \ (\ga^{\la(\mu} \, C)_{\al \be} \, (\ga^{\nu)}{}_{\la} \, C)_{\ga \de}\,E_{12}\, \tspinb 1 &1 &2 &2 \\ 3 &4 &5 &6 \tspine \, \tspinb 3 &4 \\ 5 &6 \tspine \, \tspinb 3 &5 \\ 4 &6 \tspine \, \tspinb 3 &6 \\ 4 &5 \tspine \, \biggr]\,. 
\end{align}
On the level of six spin fields, there is firstly the rather trivial correlator whose tree level limit is determined by Wick's theorem
\begin{align}
  \langle S_\al(z_1) \, S_\be(z_2) \, S_\ga(z_3)& \, S_\de(z_4)\, S_{\ep}(z_5) \, S_{\io}(z_6) \rangle \spin \eq
  \frac{1}{\tttt}\notag\\
  \times \, \bigg[&\frac{C_{\al \be} \, C_{\ga \de} \, C_{\ep \io}}{E_{12} \, E_{34} \, E_{56}}\;\tspinb 1 &3 &5 \\2 &4 &6 \tspine \, \tspinb 1 &3 &6 \\2 &4 &5 \tspine \, \tspinb 1 &4 &6 \\ 2 &3 &5 \tspine \, \tspinb 1 &4 &5 \\2 &3 &6 \tspine \notag \\
  -\ &\frac{C_{\al \be} \, C_{\ga \ep} \, C_{\de \io}}{E_{12} \, E_{35} \, E_{46}}\;\tspinb 1 &3 &4 \\2 &5 &6 \tspine \, \tspinb 1 &3 &6 \\2 &4 &5 \tspine \, \tspinb 1 &5 &6 \\ 2 &3 &4 \tspine \, \tspinb 1 &4 &5 \\2 &3 &6 \tspine \notag \\
  +\ &\frac{C_{\al \be} \, C_{\ga \io} \, C_{\de \ep}}{E_{12} \, E_{36} \, E_{45}}\; \tspinb 1 &3 &4 \\2 &5 &6 \tspine \, \tspinb 1 &3 &5 \\2 &4 &6 \tspine \, \tspinb 1 &5 &6 \\ 2 &3 &4 \tspine \, \tspinb 1 &4 &6 \\2 &3 &5 \tspine \notag \\
  + \ & \te{cyclic completion in} \ (2,\be), (3,\ga), (4,\de), (5,\ep), (6,\io)\bigg] \,.
\end{align}
Its relative with mixed chiralities has a more involved structure due to the $S_\al \leftrightarrow S^{\dbe}$ interaction:
\begin{align}
&\langle S_\al(z_1) \, S_\be(z_2) \, S_\ga(z_3) \, S_\de(z_4) \, S^{\dep}(z_5) \, S^{\di}(z_6) \rangle \spin \eq \frac{1}{4 \, \tttt \, (E_{15} \, E_{16} \, E_{25} \, E_{26} \, E_{35} \, E_{36}\, E_{45} \, E_{46})^{1/2}} \notag \\
&\times \ \left[ \, \frac{ C_{\al \de} \, C_{\be \ga} \, C^{\dep \di}}{E_{14} \, E_{23} \, E_{56}}  \right.
\Bigl( \, E_{15} \, E_{25} \, E_{36} \, E_{46} \, \tspinb 1 &2 &5 \\ 3 &4 &6 \tspine^2 \, \tspinb 1 &3 &5 \\ 2 &4 &6\tspine \, \tspinb 1 &3 &6 \\ 2 &4 &5 \tspine\notag\\
&\hspace{2.83cm}+ E_{15} \, E_{26} \, E_{35} \, E_{46} \, \tspinb 1 &3 &5 \\ 2 &4 &6 \tspine^2 \, \tspinb 1 &2 &5 \\ 3 &4 &6 \tspine \, \tspinb 1 &2 &6 \\ 3 &4 &5 \tspine \, \Bigr. \notag \\
&\hspace{2.83cm}+ E_{16} \, E_{25} \, E_{36} \, E_{45} \, \tspinb 1 &3 &6 \\ 2 &4 &5 \tspine^2 \, \tspinb 1 &2 &5 \\ 3 &4 &6
\tspine \, \tspinb 1 &2 &6 \\ 3 &4 &5 \tspine\notag\\
&\hspace{2.83cm}+ E_{16} \, E_{26} \, E_{35} \, E_{45} \, \tspinb 1 &2 &6 \\ 3 &4 &5 \tspine^2 \, \tspinb 1 &3 &5 \\ 2 &4 &6 \tspine \, \tspinb 1 &3 &6 \\ 2 &4 &5 \tspine \, \Bigr) \notag \\
& \ \ \ \, \left. + 2\, \left( \, \frac{C_{\al \be} \, C_{\ga \de}\, C^{\dep \di} }{E_{12} \, E_{34} \, E_{56}} \; \tspinb 1 &3 &5 \\2 &4 &6 \tspine \, \tspinb 1&3 &6 \\ 2 &4 &5 \tspine \ - \ \frac{C_{\al \ga} \, C_{\be \de}\, C^{\dep \di} }{E_{13} \, E_{24} \, E_{56}} \; \tspinb 1 &2 &5 \\3 &4 &6 \tspine \, \tspinb 1&2 &6 \\ 3 &4 &5 \tspine \, \right) \right. \notag \\
& \hspace{2.83cm} \Bigl( \, E_{15} \, E_{26} \,E_{36} \, E_{45} \, \tspinb 1 &4 &5 \\ 2 &3 &6 \tspine^2 \ + \ E_{16} \, E_{25} \,E_{35} \, E_{46} \, \tspinb 1 &4 &6 \\ 2 &3 &5 \tspine^2 \, \Bigr) \notag \\
& \ \ \ \, \left. - \ \frac{C_{\al \be}}{E_{12}} \; (\ga_\la \, C)_{[\ga} {}^{\dep} \, (\ga^\la \, C)_{\de]} {}^{\di}  \, \tspinb 1 &3 &4 \\ 2 &5 &6 \tspine\, \tspinb 1 &5 &6 \\ 2 &3 &4 \tspine \right. \notag \\
& \hspace{2.83cm}\Bigl( \, E_{15} \, E_{26} \, \tspinb 1 &3 &5 \\ 2 &4 &6 \tspine \, \tspinb 1 &4 &5 \\ 2 &3 &6 \tspine \ + \   E_{16} \, E_{25} \, \tspinb 1 &3 &6 \\ 2 &4 &5 \tspine \, \tspinb 1 &4 &6 \\ 2 &3 &5 \tspine \, \Bigr)  \notag \\
& \ \ \ \, \left. + \ \frac{C_{\al \ga}}{E_{13}} \; (\ga_\la \, C)_{[\be} {}^{\dep} \, (\ga^\la \, C)_{\de]} {}^{\di} \, \tspinb 1 &2 &4 \\ 3 &5 &6 \tspine\, \tspinb 1 &5 &6 \\ 2 &3 &4 \tspine \right. \notag \\
& \hspace{2.83cm} \Bigl( \, E_{15} \, E_{36} \, \tspinb 1 &2 &5 \\ 3 &4 &6 \tspine \, \tspinb 1 &4 &5 \\ 2 &3 &6 \tspine \ + \   E_{16} \, E_{35} \, \tspinb 1 &2 &6 \\ 3 &4 &5 \tspine \, \tspinb 1 &4 &6 \\ 2 &3 &5 \tspine \, \Bigr)  \notag \\
& \ \ \ \, \left. - \ \frac{C_{\al \de}}{E_{14}} \; (\ga_\la \, C)_{[\be} {}^{\dep} \, (\ga^\la \, C)_{\ga]} {}^{\di} \, \tspinb 1 &2 &3 \\ 4 &5 &6 \tspine \, \tspinb 1 &5 &6 \\ 2 &3 &4 \tspine  \right. \notag \\
& \hspace{2.83cm} \Bigl( \, E_{15} \, E_{46} \, \tspinb 1 &2 &5 \\ 3 &4 &6 \tspine \, \tspinb 1 &3 &5 \\ 2 &4 &6 \tspine \ + \   E_{16} \, E_{45} \, \tspinb 1 &2 &6 \\ 3 &4 &5 \tspine \, \tspinb 1 &3 &6 \\ 2 &4 &5 \tspine \, \Bigr)  \notag \\
& \ \ \ \, \left. - \ \frac{C_{\be \ga}}{E_{23}} \; (\ga_\la \, C)_{[\al} {}^{\dep} \, (\ga^\la \, C)_{\de]} {}^{\di} \, \tspinb 1 &2 &4 \\ 3 &5 &6 \tspine\, \tspinb 1 &3 &4 \\ 2 &5 &6 \tspine \right. \notag \\
& \hspace{2.83cm} \Bigl( \, E_{25} \, E_{36} \, \tspinb 1 &2 &5 \\ 3 &4 &6 \tspine \, \tspinb 1 &3 &6 \\ 2 &4 &5 \tspine \ + \   E_{26} \, E_{35} \, \tspinb 1 &2 &6 \\ 3 &4 &5 \tspine \, \tspinb 1 &3 &5 \\ 2 &4 &6 \tspine \, \Bigr)  \notag \\
& \ \ \ \, \left. + \ \frac{C_{\be \de}}{E_{24}} \; (\ga_\la \, C)_{[\al} {}^{\dep} \, (\ga^\la \, C)_{\ga]} {}^{\di}  \, \tspinb 1 &2 &3 \\ 4 &5 &6 \tspine\, \tspinb 1 &3 &4 \\ 2 &5 &6 \tspine \right. \notag \\
& \hspace{2.83cm}  \Bigl( \, E_{25} \, E_{46} \, \tspinb 1 &2 &5 \\ 3 &4 &6 \tspine \, \tspinb 1 &4 &6 \\ 2 &3 &5 \tspine \ + \   E_{26} \, E_{45} \, \tspinb 1 &2 &6 \\ 3 &4 &5 \tspine \, \tspinb 1 &4 &5 \\ 2 &3 &6 \tspine \, \Bigr)  \notag \\
& \ \ \ \, \left. - \ \frac{C_{\ga \de}}{E_{34}} \; (\ga_\la \, C)_{[\al} {}^{\dep} \, (\ga^\la \, C)_{\be]} {}^{\di}  \, \tspinb 1 &2 &3 \\ 4 &5 &6 \tspine\, \tspinb 1 &2 &4 \\ 3 &5 &6 \tspine \right. \notag \\
& \hspace{2.83cm}   \Bigl( \, E_{35} \, E_{46} \, \tspinb 1 &3 &5 \\ 2 &4 &6 \tspine \, \tspinb 1 &4 &6 \\ 2 &3 &5 \tspine \ + \   E_{36} \, E_{45} \, \tspinb 1 &3 &6 \\ 2 &4 &5 \tspine \, \tspinb 1 &4 &5 \\ 2 &3 &6 \tspine \, \Bigr)  \notag \\
& \ \ \ \ \, \, \! \biggl. - \ E_{56} \, (\ga_\la \, C)_{\al} {}^{(\dep} \, (\ga^{| \la \rho |} \, C)_{\be \ga} \, (\ga_\rho \, C)_\de {}^{\di)} \,  \tspinb 1 &2 &3 \\ 4 &5 &6 \tspine \,\tspinb 1 &2 &4 \\ 3 &5 &6 \tspine \, \tspinb 1 &3 &4 \\ 2 &5 &6 \tspine \,\tspinb 1 &5 &6 \\ 2 &3 &4 \tspine \,  \biggr]\,.
\end{align}
Seven-point correlations require a basis of at least 24 independent Lorentz tensors, so we refrain from computing higher point examples without systematics.

\section{Four and more spin fields in $\bm{D=10}$}
\label{sec_d10}

Ten-dimensional correlators with four spin fields only are given by
\begin{align}
  \vevs{S_{\al}(z_1)\,S_{\be}(z_2)\,&S_{\ga}(z_3)\,S_{\de}(z_4)}\,=\,
  \frac{1}{2\,(E_{12}\,E_{13}\,E_{14}\,E_{23}\,E_{24}\,E_{34})^{3/4}}\,\frac{\teta{1&2\\3&4}\,\teta{1&3\\2&4}\,\teta{1&4\\2&3}}{\ttttt}\notag\\
  &\times \left[(\ga^\mu\,C)_{\al\be}\,(\ga_\mu\,C)_{\ga\de}\,E_{14}\,E_{23}\,\teta{1&4\\2&3}^2
              -(\ga^\mu\,C)_{\al\de}\,(\ga_\mu\,C)_{\ga\be}\,E_{12}\,E_{34}\,\teta{1&2\\3&4}^2\right]\,, \notag \\
  \vevs{S_{\al}(z_1)\,S_{\be}(z_2)\,&S^{\dga}(z_3)\,S^{\dde}(z_4)}\,=\,
  \bigg(\frac{E_{12}\,E_{34}}{E_{13}\,E_{14}\,E_{23}\,E_{24}}\bigg)^{1/4}\,\frac{\teta{1&2\\3&4}}{\ttttt}\notag\\
  &\times\Bigg[
  \frac{\cc[\al]{\dde}\,\cc[\be]{\dga}}{E_{14}\,E_{23}}\,\teta{1&2\\3&4}^2\teta{1&3\\2&4}^2
  -\frac{\cc[\al]{\dga}\,\cc[\be]{\dde}}{E_{13}\,E_{24}}\,\teta{1&2\\3&4}^2\teta{1&4\\2&3}^2\notag\\
  &\hspace{5.67cm}+\frac{1}{2}\,\frac{(\ga^\mu\,C)_{\al\be}\,(\gab_\mu\,C)^{\dga\dde}}{E_{12}\,E_{34}}\,\teta{1&3\\2&4}^2\teta{1&4\\2&3}^2 \Bigg]\,.
\end{align}
In presence of one NS fermion, we find
\begin{align}
\langle \psi^\mu(z_1) \,&S_\al(z_2) \, S_\be(z_3) \, S_\ga(z_4) \, S^{\dde}(z_5) \rangle \spin \eq \frac{(E_{23} \, E_{24} \, E_{34})^{-3/4}}{\sqrt{2} \, \ttttt \, (E_{12} \, E_{13} \, E_{14} \, E_{15})^{1/2}  \, (E_{25} \, E_{35} \, E_{45})^{1/4}} \notag \\
&\times \, \biggl[ \, \frac{C_\ga {}^{\dde} }{E_{45}} \; (\ga^\mu \, C)_{\al \be} \, E_{15} \, E_{24} \, E_{34} \, \tspinb 2 &5 \\ 3 &4 \tspine^2 \, \tspinb 2 &4 \\ 3&5 \tspine^2 \,  \tspinb 1 &1 &5 \\ 2 &3 &4 \tspine   \biggr. \notag \\
& \ \ \, \ \biggl. + \ \frac{C_\al {}^{\dde} }{E_{25}} \; (\ga^\mu \, C)_{\be \ga} \, E_{15} \, E_{23} \,  E_{24} \,\tspinb 2 & 3 \\ 4 &5 \tspine^2 \, \tspinb 2 &4 \\ 3 &5 \tspine^2 \,   \tspinb 1 &1 &5 \\ 2 &3 &4 \tspine  \biggr. \notag \\
& \ \ \, \ \biggl. - \ \frac{C_\be {}^{\dde} }{E_{35}} \; (\ga^\mu \, C)_{\al \ga} \, E_{15} \, E_{23} \, E_{34}  \, \tspinb 1 &1 &5 \\ 2 &3 &4 \tspine \, \tspinb 2 &3 \\ 4 &5 \tspine^2 \, \tspinb 2 &5 \\ 3 &4 \tspine^2 \biggr. \notag \\
& \ \ \, \ \biggl. - \ \frac{1}{2} \; (\ga^{\nu } \, \gab^{\mu} \, C)_\ga {}^{\dde} \, (\ga_\nu \, C)_{\al \be} \, E_{12} \, E_{34} \, \tspinb 1 &1 &2 \\ 3 &4 &5 \tspine \, \tspinb 2 &3 \\ 4 &5 \tspine \, \tspinb 2 &4 \\ 3 &5 \tspine \, \tspinb 2 &5 \\ 3 &4  \tspine^2 \biggr. \notag \\
& \ \ \, \ \biggl. + \ \frac{1}{2} \; (\ga^{\nu } \, \gab^{\mu} \, C)_\al {}^{\dde} \, (\ga_\nu \, C)_{\be \ga} \, E_{14} \, E_{23} \, \tspinb 1 &1 &4 \\ 2 &3 &5 \tspine \, \tspinb 2 &4 \\ 3 &5 \tspine \, \tspinb 2 &5 \\ 3 & 4 \tspine \, \tspinb 2 &3 \\ 4 &5 \tspine^2 \, \biggr]\,.
\end{align}
The identity $(\ga_\nu \, \gab^\mu \, C)_{(\al} {}^{\dde} (\ga^\nu \, C)_{\be \ga)}= 0$ admits to recast the last two lines in the form
\begin{align}
\frac{1}{2} \; (\ga^{\nu } \, \gab^{\mu} \, C)_\be {}^{\dde}& \, (\ga_\nu \, C)_{\al \ga} \, E_{12} \, E_{34} \, \tspinb 1 &1 &2 \\ 3 &4 &5 \tspine \, \tspinb 2 &3 \\ 4 &5 \tspine \, \tspinb 2 &4 \\ 3 &5 \tspine \, \tspinb 2 &5 \\ 3 &4  \tspine^2 \notag \\
&+ \ \frac{1}{2} \; (\ga^{\nu } \, \gab^{\mu} \, C)_\al {}^{\dde} \, (\ga_\nu \, C)_{\be \ga} \, E_{13} \, E_{24} \, \tspinb 1 &1 &3 \\ 2 &4 &5 \tspine \, \tspinb 2 &4 \\ 3 &5 \tspine^2 \, \tspinb 2 &5 \\ 3 & 4 \tspine \, \tspinb 2 &3 \\ 4 &5 \tspine\,.
\end{align}
The following correlator has been partially computed in \cite{AS3} for the purpose of four fermion scattering at
$g=1$-loop. Let us give the complete $g$-loop result here:
\begin{align}
\langle \psi^\mu(z_1) \, \psi&^\nu(z_2) \,  S_{\al}(z_3) \, S_{\be}(z_4) \, S_{\ga}(z_5) \, S_{\de}(z_6) \rangle \spin \eq \frac{(E_{34} \, E_{35} \, E_{36} \, E_{45} \, E_{46} \, E_{56})^{-3/4}}{2 \, \ttttt \, (E_{13} \, E_{14} \, E_{15} \, E_{16} \, E_{23} \, E_{24} \, E_{25} \, E_{26})^{1/2} } \notag \\
\biggl[ \, &\frac{ \eta^{\mu \nu}}{E_{12} } \; (\ga^\la \, C)_{\al \be} \, (\ga_\la \, C)_{\ga \de} \, E_{36} \, E_{45} \, \tspinb 3 &4 \\ 5 &6 \tspine \,\tspinb 3 &5 \\ 4 &6 \tspine \, \tspinb 3 &6 \\ 4 &5 \tspine^2 \biggr. \notag \\
& \ \ \ \ \ \ \ \ \ \Bigl( \, E_{13} \, E_{16} \, E_{24} \, E_{25} \, \tspinb 1 &1 &3 &6 \\ 2 &2 &4 &5 \tspine \ + \ E_{14} \, E_{15} \, E_{23} \, E_{26} \, \tspinb 1 &1 &4 &5 \\ 2 &2 &3 &6 \tspine \, \Bigr) \notag \\
- \ &\frac{ \eta^{\mu \nu}}{E_{12} } \; (\ga^\la \, C)_{\al \de} \, (\ga_\la \, C)_{\ga \be} \, E_{34} \, E_{56} \, \tspinb 3 &6 \\ 5 &4 \tspine \,\tspinb 3 &5 \\ 6 &4 \tspine \, \tspinb 3 &4 \\ 6 &5 \tspine^2 \biggr. \notag \\
& \ \ \ \ \ \ \ \ \ \Bigl( \, E_{13} \, E_{14} \,  E_{25} \, E_{26} \, \tspinb 1 &1 &3 &4 \\ 2 &2 &6 &5 \tspine \ + \ E_{15} \, E_{16} \, E_{23} \, E_{24} \, \tspinb 1 &1 &6 &5 \\ 2 &2 &3 &4 \tspine \, \Bigr) \notag \\
+ \ &\frac{1}{2} \; (\ga^\mu \, C)_{\ga \be} \, (\ga^\nu \, C)_{\al \de} \, E_{34} \, E_{56} \, \tspinb 3 &4 \\ 5 &6 \tspine^2 \, \tspinb 3 &5 \\ 4 &6 \tspine \notag \\
& \ \ \ \ \ \ \ \ \ 
  \Bigl( \, E_{13} \, E_{25} \, E_{46} \, \tspinb 1 &1 &3 \\ 4 &5 &6 \tspine \, \tspinb 2 &2 &5 \\ 3 &4 &6 \tspine \ - \ E_{16} \, E_{24} \, E_{35} \, \tspinb 1 &1 &6 \\ 3 &4 &5 \tspine \, \tspinb 2 &2 &4 \\ 3 &5 &6 \tspine \, \Bigr) \notag \\
+ \ &\frac{1}{2} \; (\ga^\mu \, C)_{\al \de} \, (\ga^\nu \, C)_{\ga \be} \, E_{34} \, E_{56} \, \tspinb 3 &4 \\ 5 &6 \tspine^2 \, \tspinb 3 &5 \\ 4 &6 \tspine \notag \\
& \ \ \ \ \ \ \ \ \  \Bigl( \, E_{14} \, E_{26} \, E_{35} \, \tspinb 1 &1 &4 \\ 3 &5 &6 \tspine \, \tspinb 2 &2 &6 \\ 3 &4 &5 \tspine \ - \ E_{15} \, E_{23} \, E_{46} \, \tspinb 1 &1 &5 \\ 3 &4 &6 \tspine \, \tspinb 2 &2 &3 \\ 4 &5 &6 \tspine \, \Bigr) \notag \\
- \ &\frac{1}{2} \; (\ga^\mu \, C)_{\al \be} \, (\ga^\nu \, C)_{\ga \de} \, E_{36} \, E_{45} \, \tspinb 3 &6 \\ 4 &5 \tspine^2 \, \tspinb 3 &5 \\ 4 &6 \tspine \notag \\
& \ \ \ \ \ \ \ \ \  \Bigl( \, E_{16} \, E_{24} \, E_{35} \, \tspinb 1 &1 &6 \\ 3 &4 &5 \tspine \, \tspinb 2 &2 &4 \\ 3 &5 &6 \tspine \ + \ E_{15} \, E_{23} \, E_{46} \, \tspinb 1 &1 &5 \\ 3 &4 &6 \tspine \, \tspinb 2 &2 &3 \\ 4 &5 &6 \tspine \, \Bigr) \notag \\
+ \ &\frac{1}{2} \; (\ga^\mu \, C)_{\ga \de} \, (\ga^\nu \, C)_{\al \be} \, E_{36} \, E_{45} \, \tspinb 3 &6 \\ 4 &5 \tspine^2 \, \tspinb 3 &5 \\ 4 &6 \tspine \notag \\
& \ \ \ \ \ \ \ \ \  \Bigl( \, E_{13} \, E_{25} \, E_{46} \, \tspinb 1 &1 &3 \\ 4 &5 &6 \tspine \, \tspinb 2 &2 &5 \\ 3 &4 &6 \tspine \ + \ E_{14} \, E_{26} \, E_{35} \, \tspinb 1 &1 &4 \\ 3 &5 &6 \tspine \, \tspinb 2 &2 &6 \\ 3 &4 &5 \tspine \, \Bigr) \notag \\
+ \ &\frac{1}{4} \; (\ga^{\mu \nu \la} \, C)_{\al \be} \, (\ga_\la \, C)_{\ga \de} \; E_{34} \, E_{36} \, E_{45} \, \tspinb 3 &4 \\ 5 &6 \tspine \, \tspinb 3 &6 \\ 4 &5 \tspine^2 \notag \\
& \ \ \ \ \ \ \ \ \  \Bigl( \, E_{15} \, E_{26} \, \tspinb 1 &1 &5 \\ 3 &4 &6 \tspine \, \tspinb 2 &2 &6 \\ 3 &4 &5 \tspine  \ + \ E_{16} \, E_{25} \, \tspinb 1 &1 &6 \\ 3 &4 &5 \tspine \, \tspinb 2 &2 &5 \\ 3 &4 &6 \tspine\, \Bigr) \notag \\
+ \ &\frac{1}{4} \; (\ga^{\mu \nu \la} \, C)_{\ga \de} \, (\ga_\la \, C)_{\al \be} \; E_{36} \, E_{45} \, E_{56} \, \tspinb 3 &4 \\ 5 &6 \tspine \, \tspinb 3 &6 \\ 4 &5 \tspine^2 \notag \\
& \ \ \ \ \ \ \ \ \ \Bigl( \, E_{13} \, E_{24} \, \tspinb 1 &1 &3 \\ 4 &5 &6 \tspine \, \tspinb 2 &2 &4 \\ 3  &5 &6 \tspine \ + \ E_{14} \, E_{23} \, \tspinb 1 &1 &4 \\ 3 &5 &6 \tspine \, \tspinb 2 &2 &3 \\ 4 &5 &6 \tspine \, \Bigr) \notag \\
- \ &\frac{1}{4} \; (\ga^{\mu \nu \la} \, C)_{\al \de} \, (\ga_\la \, C)_{\ga \be} \; E_{34} \, E_{36} \, E_{56} \, \tspinb 3 &4 \\ 5 &6 \tspine^2 \, \tspinb 3 &6 \\ 4 &5 \tspine \notag \\
& \ \ \ \ \ \ \ \ \  \Bigl( \, E_{15} \, E_{24} \, \tspinb 1 &1 &5 \\ 3 &4 &6 \tspine \, \tspinb 2 &2 &4 \\ 3 &5 &6 \tspine \  + \ E_{14} \, E_{25} \, \tspinb 1 &1 &4 \\ 3 &5 &6 \tspine \, \tspinb 2 &2 &5 \\ 3 &4 &6 \tspine \, \Bigr) \notag \\
+ \ &\frac{1}{4} \; (\ga^{\mu \nu \la} \, C)_{\ga \be} \, (\ga_\la \, C)_{\al \de} \; E_{34} \, E_{45} \, E_{56} \, \tspinb 3 &4 \\ 5 &6 \tspine^2 \, \tspinb 3 &6 \\ 4 &5 \tspine \notag \\
& \ \ \ \ \ \ \ \ \ 
 \Bigl( \, E_{13} \, E_{26} \, \tspinb 1 &1 &3 \\ 4 &5 &6 \tspine \, \tspinb 2 &2 &6 \\ 3 &4 &5 \tspine \ + \ E_{16} \, E_{23} \, \tspinb 1 &1 &6 \\ 3 &4 &5 \tspine \, \tspinb 2 &2 &3 \\ 4 &5 &6 \tspine \, \Bigr) \notag \\
\biggl. - \ &(\ga^{(\mu} \, C)_{\al \ga} \, (\ga^{\nu)} \, C)_{\be \de} \, E_{12} \, E_{34} \, E_{36} \, E_{45}\, E_{56} \, \tspinb 1 &1 &2 &2 \\ 3 &4 &5 &6 \tspine \, \tspinb 3 &4 \\ 5 &6 \tspine^2 \, \tspinb 3 &6 \\ 4 &5 \tspine^2 \, \biggr]\,.
\label{sen}
\end{align}
This representation in terms of antisymmetric products $\ga^{\mu \nu \la}$ rather than $\ga^\mu \gab^\nu \ga^\la$ was
chosen in order to make antisymmetry under the exchange of $\psi^\mu(z_1) \leftrightarrow \psi^\nu(z_2)$ and $S_{\al_i}(z_i) \leftrightarrow S_{\al_j}(z_j)$ manifest (up to a pre-factor $E_{ij}^{1/4}$ in the latter case).

\medskip
If the four spin fields have mixed chirality, one has
\begin{align}
  \vevs{\psi^\mu(z_1)&\,\psi^\nu(z_2)\,S_{\al}(z_3)\,S_{\be}(z_4)\,S^{\dga}(z_5)\,S^{\dde}(z_6)}\,=\,
  \frac{1}{\ttttt}\,
  \frac{(E_{34}\,E_{56})^{-3/4}\,(E_{35}\,E_{36}\,E_{45}\,E_{46})^{-1/4}}{(E_{13}\,E_{14}\,E_{15}\,E_{16}\,E_{23}\,E_{24}\,E_{25}\,E_{26})^{1/2}}\notag\\
  \times\bigg[\,
  &\frac{1}{2}\,\frac{\eta^{\mu\nu}}{E_{12}}\,(\ga^\la\,C)_{\al\be}(\gab_\la\,C)^{\dga\dde}\,E_{13}\,E_{15}\,E_{24}\,E_{26}\,
  \teta{1&1&3&5\\2&2&4&6}\,\teta{3&4\\5&6}\,\teta{3&5\\4&6}\,\teta{3&6\\4&5}^2\notag\\
  +\,&\frac{\eta^{\mu\nu}\,\cc[\al]{\dde}\,\cc[\be]{\dga}}{E_{12}\,E_{36}\,E_{45}}\,E_{13}\,E_{14}\,E_{25}\,E_{26}\,E_{34}\,E_{56}\,
  \teta{1&1&3&4\\2&2&5&6}\,\teta{3&4\\5&6}^2\,\teta{3&5\\4&6}^2\notag\\
  -\,&\frac{\eta^{\mu\nu}\,\cc[\al]{\dga}\,\cc[\be]{\dde}}{E_{12}\,E_{35}\,E_{46}}\,E_{13}\,E_{14}\,E_{24}\,E_{26}\,E_{34}\,E_{56}\,
  \teta{1&1&3&4\\2&2&5&6}\,\teta{3&4\\5&6}^2\,\teta{3&6\\4&5}^2\notag\\
  -\,&\frac{1}{2}\,(\ga^\mu\,\gab^\nu\,C)_\al{}^{\dga}\,\frac{\cc[\be]{\dde}}{E_{46}}\,E_{16}\,E_{24}\,E_{34}\,E_{56}\,
  \teta{1&1&6\\3&4&5}\,\teta{2&2&4\\3&5&6}\,\teta{3&6\\4&5}\,\teta{3&4\\5&6}^2\notag\\
  +\,&\frac{1}{2}\,(\ga^\mu\,\gab^\nu\,C)_\al{}^{\dde}\,\frac{\cc[\be]{\dga}}{E_{45}}\,E_{15}\,E_{24}\,E_{34}\,E_{56}\,
  \teta{1&1&5\\3&4&6}\,\teta{2&2&4\\3&5&6}\,\teta{3&5\\4&6}\,\teta{3&4\\5&6}^2\notag\\
  +\,&\frac{1}{2}\,(\ga^\mu\,\gab^\nu\,C)_\be{}^{\dga}\,\frac{\cc[\al]{\dde}}{E_{36}}\,E_{13}\,E_{26}\,E_{34}\,E_{56}\,
  \teta{1&1&3\\4&5&6}\,\teta{2&2&6\\3&4&5}\,\teta{3&5\\4&6}\,\teta{3&4\\5&6}^2\notag\\
  -\,&\frac{1}{2}\,(\ga^\mu\,\gab^\nu\,C)_\be{}^{\dde}\,\frac{\cc[\al]{\dga}}{E_{35}}\,E_{13}\,E_{25}\,E_{34}\,E_{56}\,
  \teta{1&1&3\\4&5&6}\,\teta{2&2&5\\3&4&6}\,\teta{3&6\\4&6}\,\teta{3&4\\4&5}^2\notag\\
  -\,&\frac{1}{2}\,(\ga^\mu\,C)_{\al\be}(\gab^\nu\,C)^{\dga\dde}\,E_{15}\,E_{24}\,E_{36}\,
  \teta{1&1&5\\3&4&6}\,\teta{2&2&4\\3&5&6}\,\teta{3&5\\4&6}\,\teta{3&6\\4&5}^2\notag\\
  +\,&\frac{1}{2}\,(\ga^\nu\,C)_{\al\be}(\gab^\mu\,C)^{\dga\dde}\,E_{13}\,E_{25}\,E_{46}\,
  \teta{1&1&3\\4&5&6}\,\teta{2&2&5\\3&4&6}\,\teta{3&6\\4&5}\,\teta{3&5\\4&6}^2\notag\\
  -\,&\frac{1}{4}\,(\ga^\mu\,\gab^\la\,C)_\al{}^{\dga}(\ga^\nu\,\gab_\la\,C)_\be{}^{\dde}\,E_{16}\,E_{25}\,E_{34}\,
  \teta{1&1&6\\3&4&5}\,\teta{2&2&5\\3&4&6}\,\teta{3&4\\5&6}\,\teta{3&5\\4&6}\,\teta{3&6\\4&5}\notag\\
  -\,&\frac{1}{4}\,(\ga^\mu\,\gab^\la\,C)_\al{}^{\dde}(\ga^\nu\,\gab_\la\,C)_\be{}^{\dga}\,E_{15}\,E_{26}\,E_{34}\,
  \teta{1&1&5\\3&4&6}\,\teta{2&2&6\\3&4&5}\,\teta{3&4\\5&6}\,\teta{3&5\\4&6}\,\teta{3&6\\4&5}\notag\\
  +\,&\frac{1}{4}\,(\gab^\mu\,\ga^\nu\,\gab^\la\,C)^{\dga\dde}(\ga_\la\,C)_{\al\be}\,E_{13}\,E_{24}\,E_{56}\,
  \teta{1&1&3\\4&5&6}\,\teta{2&2&4\\3&5&6}\,\teta{3&4\\5&6}\,\teta{3&5\\4&6}\,\teta{3&6\\4&5}
  \bigg]\,.
\end{align}
The correlator with five left-handed spin fields and one right-handed spin field has appeared in the literature before,
namely in \cite{LE2} at tree-level for the purpose of six fermion scattering. Let us give its loop generalization here:
\begin{align}
  \vevs{S_{\al}\,(z_1)&\,S_{\be}(z_2)\,S_{\ga}(z_3)\,S_{\de}(z_4)\,S_{\ep}(z_5)\,S^{\di}(z_6)}\,=\,
  \frac{1}{2\,\ttttt}\,
  \frac{(E_{12}\,E_{13}\,E_{14}\,E_{15}\,E_{23}\,E_{24}\,E_{25}\,E_{34}\,E_{35}\,E_{45})^{-3/4}}{(E_{16}\,E_{26}\,E_{36}\,E_{46}\,E_{56})^{1/4}}\notag\\
  \times\bigg[\,
  &(\ga^\mu\,C)_{\al\be}(\ga_\mu\,C)_{\ga\de}\,\frac{\cc[\ep]{\di}}{E_{56}}\,E_{14}\,E_{15}\,E_{23}\,E_{25}\,E_{35}\,E_{46}\,
  \teta{1&2&5\\3&4&6}\,\teta{1&3&5\\2&4&6}\,\teta{1&4&5\\2&3&6}\,\teta{1&4&6\\2&3&5}^2\notag\\
  -\,&(\ga^\mu\,C)_{\al\de}(\ga_\mu\,C)_{\be\ga}\,\frac{\cc[\ep]{\di}}{E_{56}}\,E_{12}\,E_{15}\,E_{25}\,E_{34}\,E_{35}\,E_{46}\,
  \teta{1&2&6\\3&4&5}\,\teta{1&3&5\\2&4&6}\,\teta{1&4&6\\2&3&5}\,\teta{1&2&5\\3&4&6}^2\notag\\
  +\,&(\ga^\mu\,C)_{\al\be}(\ga_\mu\,C)_{\ga\ep}\,\frac{\cc[\de]{\di}}{E_{46}}\,E_{14}\,E_{15}\,E_{23}\,E_{24}\,E_{34}\,E_{56}\,
  \teta{1&2&4\\3&5&6}\,\teta{1&3&4\\2&5&6}\,\teta{1&4&5\\2&3&6}\,\teta{1&5&6\\2&3&4}^2\notag\\
  -\,&(\ga^\mu\,C)_{\al\ep}(\ga_\mu\,C)_{\be\ga}\,\frac{\cc[\de]{\di}}{E_{46}}\,E_{12}\,E_{14}\,E_{24}\,E_{34}\,E_{35}\,E_{56}\,
  \teta{1&2&6\\3&4&5}\,\teta{1&3&4\\2&5&6}\,\teta{1&5&6\\2&3&4}\,\teta{1&2&4\\3&5&6}^2\notag\\
  +\,&(\ga^\mu\,C)_{\al\be}(\ga_\mu\,C)_{\de\ep}\,\frac{\cc[\ga]{\di}}{E_{36}}\,E_{13}\,E_{15}\,E_{24}\,E_{26}\,E_{34}\,E_{35}\,
  \teta{1&2&6\\3&4&5}\,\teta{1&3&4\\2&5&6}\,\teta{1&5&6\\2&3&4}\,\teta{1&3&5\\2&4&6}^2\notag\\
  -\,&(\ga^\mu\,C)_{\al\ep}(\ga_\mu\,C)_{\be\de}\,\frac{\cc[\ga]{\di}}{E_{36}}\,E_{12}\,E_{13}\,E_{26}\,E_{34}\,E_{35}\,E_{45}\,
  \teta{1&2&3\\4&5&6}\,\teta{1&3&4\\2&5&6}\,\teta{1&3&5\\2&4&6}\,\teta{1&2&6\\3&4&5}^2\notag\\
  +\,&(\ga^\mu\,C)_{\al\ga}(\ga_\mu\,C)_{\de\ep}\,\frac{\cc[\be]{\di}}{E_{26}}\,E_{12}\,E_{15}\,E_{24}\,E_{25}\,E_{34}\,E_{36}\,
  \teta{1&2&4\\3&5&6}\,\teta{1&3&6\\2&4&5}\,\teta{1&5&6\\2&3&4}\,\teta{1&2&5\\3&4&6}^2\notag\\
  -\,&(\ga^\mu\,C)_{\al\ep}(\ga_\mu\,C)_{\ga\de}\,\frac{\cc[\be]{\di}}{E_{26}}\,E_{12}\,E_{13}\,E_{24}\,E_{25}\,E_{36}\,E_{45}\,
  \teta{1&2&3\\4&5&6}\,\teta{1&2&4\\3&5&6}\,\teta{1&2&5\\3&4&6}\,\teta{1&3&6\\2&4&5}^2\notag\\
  +\,&(\ga^\mu\,C)_{\be\ep}(\ga_\mu\,C)_{\ga\de}\,\frac{\cc[\al]{\di}}{E_{16}}\,E_{12}\,E_{14}\,E_{15}\,E_{23}\,E_{36}\,E_{45}\,
  \teta{1&2&3\\4&5&6}\,\teta{1&2&4\\3&5&6}\,\teta{1&2&5\\3&4&6}\,\teta{1&4&5\\2&3&6}^2\notag\\
  -\,&(\ga^\mu\,C)_{\be\ga}(\ga_\mu\,C)_{\de\ep}\,\frac{\cc[\al]{\di}}{E_{16}}\,E_{12}\,E_{14}\,E_{15}\,E_{25}\,E_{34}\,E_{36}\,
  \teta{1&2&4\\3&5&6}\,\teta{1&3&4\\2&5&6}\,\teta{1&4&5\\2&3&6}\,\teta{1&2&5\\3&4&6}^2\notag\\
  -\,&\frac{1}{2}\,(\ga^\mu\,\gab^\nu\,C)_\be{}^{\di}(\ga_\mu\,C)_{\al\ep}(\ga_\nu\,C)_{\ga\de}\,E_{12}\,E_{14}\,E_{24}\,E_{35}\,E_{35}\notag\\
  &\hspace{5cm}\times\teta{1&2&6\\3&4&5}\,\teta{1&3&5\\2&4&6}\,\teta{1&4&6\\2&3&5}\,\teta{1&2&4\\3&5&6}^2\notag\\
  +\,&\frac{1}{2}\,(\ga^\mu\,\gab^\nu\,C)_\al{}^{\di}(\ga_\mu\,C)_{\be\de}(\ga_\nu\,C)_{\ga\ep}\,E_{12}\,E_{15}\,E_{25}\,E_{34}\,E_{34}\notag\\
  &\hspace{5cm}\times\teta{1&2&6\\3&4&5}\,\teta{1&3&4\\2&5&6}\,\teta{1&5&6\\2&3&4}\,\teta{1&2&5\\3&4&6}^2\notag\\
  +\,&\frac{1}{2}\,(\ga^\mu\,\gab^\nu\,C)_\ep{}^{\di}(\ga_\mu\,C)_{\al\be}(\ga_\nu\,C)_{\ga\de}\,E_{14}\,E_{15}\,E_{23}\,E_{24}\,E_{35}\notag\\
  &\hspace{5cm}\times\teta{1&2&4\\3&5&6}\,\teta{1&3&5\\2&4&6}\,\teta{1&4&5\\2&3&6}\,\teta{1&4&6\\2&3&5}\,\teta{1&5&6\\2&3&4}\notag\\
  +\,&\frac{1}{2}\,(\ga^\mu\,\gab^\nu\,C)_\de{}^{\di}(\ga_\mu\,C)_{\al\be}(\ga_\nu\,C)_{\ga\ep}\,E_{13}\,E_{15}\,E_{24}\,E_{25}\,E_{34}\notag\\
  &\hspace{5cm}\times\teta{1&2&5\\3&4&6}\,\teta{1&3&4\\2&5&6}\,\teta{1&3&5\\2&4&6}\,\teta{1&3&6\\2&4&5}\,\teta{1&5&6\\2&3&4}\notag\\
  -\,&\frac{1}{2}\,(\ga^\mu\,\gab^\nu\,C)_\ga{}^{\di}(\ga_\mu\,C)_{\al\de}(\ga_\nu\,C)_{\be\ep}\,E_{12}\,E_{15}\,E_{24}\,E_{34}\,E_{35}\notag\\
  &\hspace{5cm}\times\teta{1&2&4\\3&5&6}\,\teta{1&2&5\\3&4&6}\,\teta{1&2&6\\3&4&5}\,\teta{1&3&5\\2&4&6}\,\teta{1&5&6\\2&3&4}\notag\\
  -\,&\frac{1}{2}\,(\ga^\mu\,\gab^\nu\,C)_\ga{}^{\di}(\ga_\mu\,C)_{\al\ep}(\ga_\nu\,C)_{\be\de}\,E_{12}\,E_{14}\,E_{25}\,E_{34}\,E_{35}\notag\\
  &\hspace{5cm}\times\teta{1&2&4\\3&5&6}\,\teta{1&2&5\\3&4&6}\,\teta{1&2&6\\3&4&5}\,\teta{1&3&4\\2&5&6}\,\teta{1&4&6\\2&3&5}
  \bigg]\,.
\end{align}
There is also a non-vanishing correlation with three left- and right-handed spin fields each:
\begin{align}
  \vevs{S_{\al}&(z_1)\,S_{\be}(z_2)\,S_{\ga}(z_3)\,S^{\dde}(z_4)\,S^{\dep}(z_5)\,S^{\di}(z_6)}\,=\,-
  \frac{1}{\ttttt}\,
  \bigg(\frac{E_{12}\,E_{13}\,E_{23}\,E_{45}\,E_{46}\,E_{56}}{E_{14}\,E_{15}\,E_{16}\,E_{24}\,E_{25}\,E_{26}\,E_{34}\,E_{35}\,E_{36}}\bigg)^{1/4}\notag\\
  \times\Bigg[\,
   &\frac{\cc[\al]{\dde}\,\cc[\be]{\dep}\,\cc[\ga]{\di}}{E_{14}\,E_{25}\,E_{36}}\,\teta{1&2&3\\4&5&6}^2\,\teta{1&5&6\\2&3&4}\,\teta{1&3&5\\2&4&6}\,\teta{1&2&6\\3&4&5}\notag\\
  -\,&\frac{\cc[\al]{\dde}\,\cc[\be]{\di}\,\cc[\ga]{\dep}}{E_{14}\,E_{26}\,E_{35}}\,\teta{1&2&3\\4&5&6}^2\,\teta{1&5&6\\2&3&4}\,\teta{1&3&6\\2&4&5}\,\teta{1&2&5\\3&4&6}\notag\\
  +\,&\frac{\cc[\al]{\dep}\,\cc[\be]{\di}\,\cc[\ga]{\dde}}{E_{15}\,E_{26}\,E_{34}}\,\teta{1&2&3\\4&5&6}^2\,\teta{1&4&6\\2&3&5}\,\teta{1&3&6\\2&4&5}\,\teta{1&2&4\\3&5&6}\notag\\
  -\,&\frac{\cc[\al]{\dep}\,\cc[\be]{\dde}\,\cc[\ga]{\di}}{E_{15}\,E_{24}\,E_{36}}\,\teta{1&2&3\\4&5&6}^2\,\teta{1&4&6\\2&3&5}\,\teta{1&3&4\\2&5&6}\,\teta{1&2&6\\3&4&5}\notag\\
  +\,&\frac{\cc[\al]{\di}\,\cc[\be]{\dde}\,\cc[\ga]{\dep}}{E_{16}\,E_{24}\,E_{35}}\,\teta{1&2&3\\4&5&6}^2\,\teta{1&4&5\\2&3&6}\,\teta{1&3&4\\2&5&6}\,\teta{1&2&5\\3&4&6}\notag\\
  -\,&\frac{\cc[\al]{\di}\,\cc[\be]{\dep}\,\cc[\ga]{\dde}}{E_{16}\,E_{25}\,E_{34}}\,\teta{1&2&3\\4&5&6}^2\,\teta{1&4&5\\2&3&6}\,\teta{1&3&5\\2&4&6}\,\teta{1&2&4\\3&5&6}\notag\\
  -\,&\frac{1}{2}\,\frac{(\ga^\mu\,C)_{\al\be}(\gab_\mu\,C)^{\dde\dep}\,\cc[\ga]{\di}}{E_{12}\,E_{36}\,E_{45}}\,
  \teta{1&2&3\\4&5&6}\,\teta{1&3&4\\2&5&6}\,\teta{1&3&5\\2&4&6}\,\teta{1&4&6\\2&3&5}\,\teta{1&5&6\\2&3&4}\notag\\
  +\,&\frac{1}{2}\,\frac{(\ga^\mu\,C)_{\al\be}(\gab_\mu\,C)^{\dde\di}\,\cc[\ga]{\dep}}{E_{12}\,E_{35}\,E_{46}}\,
  \teta{1&2&3\\4&5&6}\,\teta{1&3&4\\2&5&6}\,\teta{1&3&6\\2&4&5}\,\teta{1&4&5\\2&3&6}\,\teta{1&5&6\\2&3&4}\notag\\
  -\,&\frac{1}{2}\,\frac{(\ga^\mu\,C)_{\al\be}(\gab_\mu\,C)^{\dep\di}\,\cc[\ga]{\dde}}{E_{12}\,E_{34}\,E_{56}}\,
  \teta{1&2&3\\4&5&6}\,\teta{1&3&5\\2&4&6}\,\teta{1&3&6\\2&4&5}\,\teta{1&4&5\\2&3&6}\,\teta{1&4&6\\2&3&5}\notag\\
  +\,&\frac{1}{2}\,\frac{(\ga^\mu\,C)_{\al\ga}(\gab_\mu\,C)^{\dde\dep}\,\cc[\be]{\di}}{E_{13}\,E_{26}\,E_{45}}\,
  \teta{1&2&3\\4&5&6}\,\teta{1&2&4\\3&5&6}\,\teta{1&2&5\\3&4&6}\,\teta{1&4&6\\2&3&5}\,\teta{1&5&6\\2&3&4}\notag\\
  -\,&\frac{1}{2}\,\frac{(\ga^\mu\,C)_{\al\ga}(\gab_\mu\,C)^{\dde\di}\,\cc[\be]{\dep}}{E_{13}\,E_{25}\,E_{46}}\,
  \teta{1&2&3\\4&5&6}\,\teta{1&2&4\\3&5&6}\,\teta{1&2&6\\3&4&5}\,\teta{1&4&5\\2&3&6}\,\teta{1&5&6\\2&3&4}\notag\\
  +\,&\frac{1}{2}\,\frac{(\ga^\mu\,C)_{\al\ga}(\gab_\mu\,C)^{\dep\di}\,\cc[\be]{\dde}}{E_{13}\,E_{24}\,E_{56}}\,
  \teta{1&2&3\\4&5&6}\,\teta{1&2&5\\3&4&6}\,\teta{1&2&6\\3&4&5}\,\teta{1&4&5\\2&3&6}\,\teta{1&4&6\\2&3&5}\notag\\
  -\,&\frac{1}{2}\,\frac{(\ga^\mu\,C)_{\be\ga}(\gab_\mu\,C)^{\dde\dep}\,\cc[\al]{\di}}{E_{16}\,E_{23}\,E_{45}}\,
  \teta{1&2&3\\4&5&6}\,\teta{1&2&4\\3&5&6}\,\teta{1&2&5\\3&4&6}\,\teta{1&3&4\\2&5&6}\,\teta{1&3&5\\2&4&6}\notag\\
  +\,&\frac{1}{2}\,\frac{(\ga^\mu\,C)_{\be\ga}(\gab_\mu\,C)^{\dde\di}\,\cc[\al]{\dep}}{E_{15}\,E_{23}\,E_{46}}\,
  \teta{1&2&3\\4&5&6}\,\teta{1&2&4\\3&5&6}\,\teta{1&2&6\\3&4&5}\,\teta{1&3&4\\2&5&6}\,\teta{1&3&6\\2&4&5}\notag\\
  -\,&\frac{1}{2}\,\frac{(\ga^\mu\,C)_{\be\ga}(\gab_\mu\,C)^{\dep\di}\,\cc[\al]{\dde}}{E_{14}\,E_{23}\,E_{56}}\,
  \teta{1&2&3\\4&5&6}\,\teta{1&2&5\\3&4&6}\,\teta{1&2&6\\3&4&5}\,\teta{1&3&5\\2&4&6}\,\teta{1&3&6\\2&4&5}\notag\\
  +\,&\frac{1}{4}\,(\ga^\mu\gab^\nu\,C)_\al{}^{\dep}(\ga_\mu\,C)_{\be\ga}(\gab_\nu\,C)^{\dde\di}\,\frac{E_{36}}{E_{13}\,E_{23}\,E_{46}\,E_{56}}\,
  \teta{1&2&4\\3&5&6}\,\teta{1&2&5\\3&4&6}\,\teta{1&2&6\\3&4&5}\,\teta{1&3&6\\2&4&5}\,\teta{1&4&5\\2&3&6}\notag\\
  -\,&\frac{1}{4}\,(\ga^\mu\gab^\nu\,C)_\ga{}^{\dep}(\ga_\mu\,C)_{\al\be}(\gab_\nu\,C)^{\dde\di}\,\frac{E_{16}}{E_{12}\,E_{13}\,E_{46}\,E_{56}}\,
  \teta{1&2&6\\3&4&5}\,\teta{1&3&6\\2&4&5}\,\teta{1&4&5\\2&3&6}\,\teta{1&4&6\\2&3&5}\,\teta{1&5&6\\2&3&4}\notag\\
  +\,&\frac{1}{4}\,(\ga^\mu\gab^\nu\,C)_\al{}^{\di}(\ga_\mu\,C)_{\be \ga}(\gab_\nu\,C)^{\dde\dep}\,\frac{E_{35}}{E_{13}\,E_{23}\,E_{45}\,E_{56}}\,
  \teta{1&2&4\\3&5&6}\,\teta{1&2&5\\3&4&6}\,\teta{1&2&6\\3&4&5}\,\teta{1&3&5\\2&4&6}\,\teta{1&4&6\\2&3&5}\notag\\
  -\,&\frac{1}{4}\,(\ga^\mu\gab^\nu\,C)_\ga{}^{\di}(\ga_\mu\,C)_{\al\be}(\gab_\nu\,C)^{\dde\dep}\,\frac{E_{15}}{E_{12}\,E_{13}\,E_{45}\,E_{56}}\,
  \teta{1&2&5\\3&4&6}\,\teta{1&4&5\\2&3&6}\,\teta{1&3&5\\2&4&6}\,\teta{1&5&6\\2&3&4}\,\teta{1&4&6\\2&3&5}
  \Bigg]\,.
\end{align}
Note that also this result exhibits manifest antisymmetry under $S_\al(z_1) \leftrightarrow S_\ga(z_3)$ and
$S^{\dep}(z_5) \leftrightarrow S^{\di}(z_6)$ up to the powers $(E_{13} E_{56})^{1/4}$.

\section{Conclusions}

In this work, we have provided a toolkit which plays an essential role in computing tree-level or multi-loop amplitudes
in $D=6$-- and $D=8$--dimensional superstring compactifications but also in the non-compactified $D=10$ situation. The
applicability of these SCFT results ranges from superstring theories to the heterotic string theories allowing for a CFT
description.

\medskip One of the highlights in this work are equations (\ref{OmegaD}) and (\ref{omegaD}), which pave the way to
compute two fermion, $N$ boson amplitudes in every even space-time dimension on arbitrary genus. Another very general
result (\ref{nn}) is given in six space-time dimensions, describing a $2N$-point function of spin fields at
every loop level. This can be used for scattering amplitudes involving many fermions.

\medskip The superstring amplitudes computed in this work play an important role in testing various aspects of duality
symmetries relating different string vacua.  Furthermore, these amplitudes are phenomenologically relevant to
investigate low energy dynamics of various compactification geometries. In particular, the $D=4$--dimensional situation
gives rise to stringy predictions for hadron colliders \cite{LHC1,LHC2,LHC3}. Another, rather theoretical motivation
lies in the study of loop generalizations of supersymmetric Ward identities which were found to extend to all $\alpha'$
orders at tree-level \cite{Stieberger:2007jv}. It would be interesting to see whether one can still relate amplitudes of
different fermion numbers on higher genus. Our results are essential to obtain further data points in this
investigation.

\vskip1cm
\goodbreak
\centerline{\noindent{\bf Acknowledgments} }\vskip 2mm

We wish to thank Stephan Stieberger for triggering this project, continuous support as well as careful reading and
commenting on later versions of the draft. Furthermore we are indebted to Ioannis Florakis and Alois Kabelschacht for
many fruitful discussion.

\section*{Appendix}
\appendix

\section{Gamma matrices in $\bm{D}$ dimensions}
\label{appA}

Gamma matrices $\Ga^\mu$ play a key role in the interplay of vector- and spinor representation of the Lorentz group
$SO(1,D-1)$ in $D$ dimensions. First of all, they can be be viewed as operators acting on spinor spaces whose
anti-commutation relations are given by the Clifford algebra. On the other hand, antisymmetric products $\Ga^{\mu_1
  ... \mu_p}$ of $p$ gamma matrices, multiplied by the charge conjugation matrix ${\cal C}$, are Clebsch--Gordan
coefficients which take bi-spinors to $p$ forms. The tensor structure of correlation functions involving spinorial
fields is expressed in terms of these products $(\Ga^{\mu_1 ... \mu_p} {\cal C})$, the charge conjugation matrix ${\cal C}$ guaranteeing Ramond charge conservation. Hence, we need to keep their
manipulation under control in deriving and applying spin field correlators. The reader might refer to
\cite{gamma1,gamma2,gamma3} for further information on the spinor algebra in higher dimensions.

\subsection{Notation and conventions}
\label{appA}

Let us first of all fix our notation and conventions. We use the sign convention of Wess \& Bagger for the Clifford algebra
\begin{equation}
\bigl\{ \Ga^{\mu} \, , \, \Ga^{\nu} \bigr\} \eq - \, 2 \, \eta^{\mu \nu}\,,
\label{cliff}
\end{equation}
as has been done in \cite{LHC1, LHC2, LHC3, tree, loop4D} for the four-dimensional calculations. The signature of the Minkowski
metric only matters if an explicit representation for gamma matrices needs to be given. For the purposes of this paper, however,
knowledge of the vanishing gamma matrix entries is sufficient.

\medskip Spinors in even space-time dimensions $D \in 2 \NN$ form a $2^{D/2}$-dimensional vector space which decomposes
into two irreducible Weyl representations of $SO(1,D-1)$ of dimension $2^{D/2-1}$ each. We will refer to them as the
left- and right-handed representations. Their elements are called Weyl spinors of positive or negative chirality. Generic Dirac spinors $\Xi$ live in the direct sum of both irreducible subspaces and will be written in component notation:
\begin{equation}
\Xi_A \ = \ \vecb \psi_\al \\ \chi^{\dal} \vece \co \psi_\al \ \equiv \ \te{left-handed} \co  \chi^{\dal} \ \equiv \ \te{right-handed}\,.
\label{cpt}
\end{equation}

\medskip Gamma matrices transform left-handed spinors into right-handed ones and vice versa. Hence, one can write them
in block off-diagonal form whose non-vanishing blocks $\ga^\mu$ and $\gab^\mu$ are also known as generalized
$\si$-matrices. Their action on a Dirac spinor $\Xi_B$ in chiral index notation reads
\begin{equation}
(\Ga^\mu)_{A}\,^{B} \, \Xi_B \eq \ccb 0 &\ga^\mu_{\al \dbe} \\ \gab^{\mu \dal \be} & 0 \cce \, \vecb \psi_\be \\ \chi^{\dbe} \vece \eq \vecb \ga^\mu_{\al \dbe} \, \chi^{\dbe} \\ \gab^{\mu \dal \be} \, \psi_{\be} \vece  \ ,
\label{block}
\end{equation}
where the Clifford algebra of full Dirac matrices $\Ga^\mu$ translates into
\begin{equation}
\ga^\mu_{\al \dbe} \, \gab^{\nu \dbe \ka} \ + \ \ga^\nu_{\al \dbe} \, \gab^{\mu \dbe \ka} \eq - \, 2 \, \de^\ka_\al \, \eta^{\mu \nu} \co \gab^{\mu \dal \be} \, \ga^{\nu}_{\be \dka} \ + \ \gab^{\nu \dal \be} \, \ga^{\mu}_{\be \dka} \eq -\,  2 \, \de^{\dal}_{\dka} \, \eta^{\mu \nu} \ .
\label{cliff2}
\end{equation}
More generally, odd (even) products of $\Ga$ matrices carry alternating products of $\ga$ and $\gab$ matrices in their off-diagonal (diagonal) block:
\begin{equation}
(\Ga^{\mu_1} \, \Ga^{\mu_2} \, ... \, \Ga^{ \mu_{p}})_A \, ^B \eq \left\{ \begin{array}{cl} \ccb 0 &(\ga^{\mu_1} \, \gab^{\mu_2} \, ... \, \ga^{\mu_p} )_{\al \dbe} \\   (\gab^{\mu_1} \, \ga^{\mu_2} \, ... \, \gab^{\mu_p} )^{\dal \be} &0 \cce  &: \, p \ \te{odd}\,, \\ 
\ccb (\ga^{\mu_1} \, \gab^{\mu_2} \, ... \, \gab^{\mu_p} )_{\al} \, ^{\be} &0 \\ 0 &   (\gab^{\mu_1} \, \ga^{\mu_2} \, ... \, \ga^{\mu_p} )^{\dal}\,_{\dbe} \cce  &: \, p \ \te{even}\,.
\end{array} \right.
 \label{gaprod}
\end{equation}
The $\Ga^\mu$ matrices alone obviously have the wrong index structure to serve as Clebsch--Gordan coefficient for bi-spinors. Some kind of metric on spinor space is needed -- the charge conjugation matrix ${\cal C}_{AB}$:
\begin{equation}
(\Ga^\mu)_A \,^B \, {\cal C}_{BD} \ \equiv  \ (\Ga^\mu \, {\cal C})_{AD}\,.
\label{cc}
\end{equation}
The chirality structure of ${\cal C}$ now depends on the dimension $D$ due to the representation theory of the associated $SO(1,D-1)$ group. In dimensions $D=0 \ \te{mod} \ 4$, only spinor representation of alike chirality contain a scalar in their tensor product whereas $D= 2 \ \te{mod} \ 4$ dimensions require opposite chiralities to form a singlet:
\begin{equation}
\begin{array}{rll}
D=0 \ \te{mod} \ 4 : \ \  &{\cal C}_{AB} \eq \ccb C_{\al \be} &0 \\ 0 &C^{\dal \dbe} \cce \,, \ \ \  &(\Ga^{\mu} \, {\cal C})_{AB} \eq \ccb 0 &(\ga^\mu \, C)_{\al} \, ^{\dbe} \\ (\gab^\mu C)^{\dal} \, _{\be} &0 \cce\,, \\
D=2 \ \te{mod} \ 4 : \ \  &{\cal C}_{AB} \eq \ccb 0 &C_\al \, ^{\dbe} \\ C^{\dal} \, _ {\be}&0 \cce \,, \ \ \  &(\Ga^{\mu} \, {\cal C})_{AB} \eq \ccb (\ga^\mu \, C)_{\al \be} &0 \\ 0 & (\gab^\mu C)^{\dal \dbe} \cce\,.
\end{array}
\label{gacc}
\end{equation}
The inverse ${\cal C}^{-1}$ of the charge conjugation matrix
again has a chiral structure which varies from $D=0 \ \te{mod} \ 4$ to $D= 2 \ \te{mod}
\ 4$ dimensions:
\begin{align}
D = 0 \ \te{mod} \ 4 : \ \ \ \ ({\cal C}^{-1})^{AB} \ \ &= \ \ \ccb (C^{-1})^{\al \be} &0 \\ 0 &(C^{-1})_{\dal \dbe} \cce\,, \notag \\
 ( {\cal C}^{-1} \, \Ga^{\mu} )^{AB} \ \ &= \ \ \ccb 0 &(C^{-1} \, \ga^\mu )^{\al} \, _{\dbe} \\ (C^{-1} \, \gab^\mu )_{\dal} \, ^{\be} &0 \cce \notag\,, \\
D = 2 \ \te{mod} \ 4 : \ \  \ \ ({\cal C}^{-1})^{AB} \ \ &= \ \ \ccb 0 &(C^{-1})^\al \, _{\dbe} \\ (C^{-1})_{\dal} \, ^ {\be}&0 \cce \notag\,, \\ 
( {\cal C}^{-1} \, \Ga^{\mu} )^{AB} \ \ &= \ \ \ccb (C^{-1} \, \gab^\mu )^{\al \be} &0 \\ 0 & (C^{-1} \, \ga^\mu)_{\dal \dbe} \cce\,.
\label{gaccinv}
\end{align}

\subsection{Symmetry properties}
\label{appA.2}

The transposed gamma matrices $\Ga^t$ satisfy the same Clifford algebra as the original ones. Arguments from
representation theory state that $\Ga$ and $\Ga^t$ must be related by a similarity transformation. This transformation
is given by the charge conjugation matrix and its inverse\footnote{In even dimensions $D \in 2\NN$, the signs in
  (\ref{gammat}) and (\ref{ct}) are a matter of convention due to the freedom to redefine ${\cal C} \mapsto \Ga_{D+1}
  {\cal C}$ with the chirality matrix. In odd dimensions $D \in 2\NN-1$, however, the non-existence of a chirality
  matrix leads to a unique choice.}
\begin{equation} {\cal C}^{-1} \, \Ga^\mu \, {\cal
  C} \eq - \, (\Ga^\mu)^t\,.
\label{gammat}
\end{equation}
Reading this equation on the level of chiral blocks leads to two different scenarios whether $D=0 \ \te{mod} \ 4$ or
$D=2 \ \te{mod} \ 4$. In the former case, transposition (\ref{gammat}) intertwines the two classes of
matrices $\ga^\mu, \gab^\mu$:
\begin{equation}
\ga^\mu_{\be \dal} \eq - \, (C^{-1})_{\dal \dka} \, \gab^{\mu \dka \ka} \, C_{\ka \be} \co \gab^{\mu \dbe \al} \eq - \, (C^{-1})^{\al \ka} \, \ga^{\mu}_{\ka \dka} \, C^{\dka \dbe}\,.
\label{cgac4}
\end{equation}
For $D=2 \ \te{mod} \ 4$, on the other hand, a consistency condition is obtained for $\ga^\mu$ and $\gab^\mu$:
\begin{equation}
\ga^\mu_{\be \dal} \eq - \, (C^{-1})_{\dal } \, \! ^{\ka} \, \ga^{\mu}_{ \ka \dka } \, C^{\dka} \, \! _{\be} \co \gab^{\mu \dbe \al} \eq - \, (C^{-1})^{\al} \, \!_{ \dka} \, \gab^{\mu \dka \ka} \, C_{\ka} \! \,^{\dbe}\,.
\label{cgac6}
\end{equation}
To give a unified way of understanding these conditions: The symmetry property of $(\Ga^\mu {\cal C})$ is opposite to
that of the charge conjugation matrix
\begin{equation}
{\cal C}^t \eq \wp_D \, {\cal C} \ \ \ \Rightarrow \ \ \ (\Ga^\mu \, {\cal C})^t \eq - \,  \wp_D \, (\Ga^\mu \, {\cal C})\,,
\label{ct}
\end{equation}
where $\wp_D=\pm 1$. By iterating this argument, one can infer the symmetry properties of higher order gamma products $(\Ga^{\mu_1} ... \Ga^{\mu_p} {\cal C})^t = \wp_D (-1)^p (\Ga^{\mu_p} ... \Ga^{\mu_1} {\cal C})$ such that antisymmetric chains of $\Ga$'s satisfy
\begin{equation}
(\Ga^{\mu_1 ... \mu_p} \, {\cal C})^t \eq  \wp_D \, (-1)^{\frac{p}{2}(p+1)} \,  (\Ga^{\mu_1 ... \mu_p} \, {\cal C}) \eq \left\{ \begin{array}{rl} + \, \wp_D \, (\Ga^{\mu_1 ... \mu_p} \, {\cal C}) &: \ p \ = \  0,3 \ \te{mod} \ 4 \\
- \, \wp_D \, (\Ga^{\mu_1 ... \mu_p} \, {\cal C}) &: \ p \ = \  1,2 \ \te{mod} \ 4 \end{array} \right. \ .
\label{ct2}
\end{equation}
In order to fix the dimension dependent sign $\wp_D=\pm 1$ in \eqref{ct} one has to make use of the fact that the set $\bigl\{ (\Ga^{\mu_1 ... \mu_p}
{\cal C}) \, : \ 0 \leq p \leq D \}$ forms a basis of the $2^{D/2} \times 2^{D/2}$ matrices. In particular, there must
be $\frac{1}{2} 2^{D/2} (2^{D/2}-1)$ antisymmetric matrices, and this fixes
\begin{equation}
\wp_D  \eq (-1)^{\frac{D}{4} \, \left(\frac{D}{2}+1 \right)} \ \ \ \Rightarrow \ \ \ {\cal C}^t \eq \left\{ \begin{array}{rl} + \, {\cal C} &: \  D \ = \ 0,6 \ \te{mod} \ 8 \\ - \, {\cal C} &: \ D \ = \ 2,4 \ \te{mod} \ 8 \end{array} \right. \,.
\label{ct3}
\end{equation}
Let us explicitly write down all the relevant cases on the level of chiral blocks:
\begin{equation}
\begin{array}{|c | c|}
\hline
D \, = \, 4 & D \, = \, 6 \\  \hline
C_{\al \be} \ \ = \ \ - \, C_{\be \al}  &C_\al \, \!^{\dbe} \ \ = \ \ + \, C^{\dbe} \, \!_\al \\
(\ga^\mu \, C)_\al \, \!^{\dbe} \ \ = \ \  + \, (\gab^\mu \, C)^{\dbe} \, \!_\al  &(\ga^\mu \, C)_{\al \be} \ \ = \ \ - \, (\ga^\mu \, C)_{\be \al}  \\
 (\ga^{\mu \nu} \, C)_{\al \be} \ \ = \ \ + \, (\ga^{\mu \nu} \, C)_{ \be \al } &(\ga^{\mu \nu} \, C)_\al \, \!^{\dbe} \ \ = \ \ - \, (\gab^{\mu \nu } \, C)^{\dbe} \, \!_\al \\
 &(\ga^{\mu \nu \la} \, C)_{\al \be} \ \ = \ \ + \, (\ga^{\mu \nu \la} \, C)_{ \be \al } \\ \hline
D \, = \, 8 & D \, = \, 10  \\ \hline
C_{\al \be} \ \ = \ \ + \, C_{\be \al} &C_\al \, \!^{\dbe} \ \ = \ \ - \, C^{\dbe} \, \!_\al  \\
(\ga^\mu \, C)_\al \, \!^{\dbe} \ \ = \ \  - \, (\gab^\mu \, C)^{\dbe} \, \!_\al &(\ga^\mu \, C)_{\al \be} \ \ = \ \ + \, (\ga^\mu \, C)_{\be \al}   \\
(\ga^{\mu \nu} \, C)_{\al \be} \ \ = \ \ - \, (\ga^{\mu \nu} \, C)_{ \be \al }
&(\ga^{\mu \nu} \, C)_\al \, \!^{\dbe} \ \ = \ \ + \, (\gab^{\mu \nu } \, C)^{\dbe} \, \!_\al  
 \\
(\ga^{\mu \nu \la} \, C)_\al \, \!^{\dbe} \ \ = \ \  + \, (\gab^{\mu \nu \la} \, C)^{\dbe} \, \!_\al
&(\ga^{\mu \nu \la} \, C)_{\al \be} \ \ = \ \ - \, (\ga^{\mu \nu \la} \, C)_{ \be \al } \\
(\ga^{\mu \nu \la \rho} \, C)_{\al \be} \ \ = \ \ + \, (\ga^{\mu \nu \la \rho} \, C)_{ \be \al } &
(\ga^{\mu \nu \la \rho} \, C)_\al \, \!^{\dbe} \ \ = \ \ - \, (\gab^{\mu \nu \la \rho} \, C)^{\dbe} \, \!_\al  
  \\
&
(\ga^{\mu \nu \la \rho \tau} \, C)_{\al \be} \ \ = \ \ + \, (\ga^{\mu \nu \la \rho \tau} \, C)_{ \be \al }   \\\hline 
\end{array}
\label{ct6}
\end{equation}
To avoid over-counting of the independent symmetric matrices, one should be aware of the self-dualities of $D/2$-fold products (with $\phi_D$ denoting a dimension dependent phase):
\begin{align}
(\ga^{\mu_1 ... \mu_{D/2}} \, C)_{\al \be} \eq \pm \; \frac{e^{i \phi_D}}{(D/2) !} \; \vep^{ \mu_1 ... \mu_{D/2} \nu_1 ... \nu_{D/2} } \, ( \ga_{\nu_1  ... \nu_{D/2}} \, C)_{\al \be}\,, \notag \\
(\gab^{\mu_1 ... \mu_{D/2}} \, C)^{\dal \dbe} \eq \mp \; \frac{e^{i \phi_D}}{(D/2) !} \; \vep^{ \mu_1 ... \mu_{D/2} \nu_1 ... \nu_{D/2} } \, ( \gab_{\nu_1  ... \nu_{D/2}} \, C)^{\dal \dbe}\,.
\label{ct8}
\end{align}

\subsection{Fierz identities}

Antisymmetrized gamma products $\Ga^{\mu_1 ... \mu_p}$ with $0 \leq p \leq D$ form a complete set of $2^{D/2} \times 2^{D/2}$ matrices. Therefore, it is possible to expand any bi-spinor in terms of forms. The expansion prescriptions are referred to as Fierz identities.

\medskip
Within the chiral blocks $\ga^\mu, \gab^\mu$, it is sufficient to consider forms up to degree $D/2$ since any $p \leq D$
fold product $\ga^{\mu_1 ... \mu_p}$ is related to $D-p$ products via Hodge duality. Weyl bi-spinors are therefore
expanded as follows:
\begin{itemize}
\item $D \ = \ 0 \ \te{mod} \ 4$:
\begin{align}
\psi_\al \, \chi_\be \ \ &= \ \ 2^{-D/2} \sum_{p \ \te{even}}^{D/2-2} \frac{1}{p!} \; (\ga^{\mu_1 ... \mu_p} \, C)_{\be \al} \, (\psi \, C^{-1} \, \ga_{\mu_p ... \mu_1} \, \chi) \notag \\
& \ \ \ \ \ \ \ \ \ \ + \  \frac{2^{-D/2}}{2 \, (D/2)!} \; (\ga^{\mu_1 ... \mu_{D/2}} \, C)_{\be \al} \, (\psi \, C^{-1} \, \ga_{\mu_{D/2} ... \mu_1} \, \chi)\,,
\label{fierza} \\
\psi_\al \, \bar \chi^{\dbe} \ \ &= \ \ - \, 2^{-D/2} \sum_{p \ \te{odd}}^{D/2-1} \frac{1}{p!} \; (\gab^{\mu_1 ... \mu_p} \, C)^{\dbe} \, \!_{ \al} \, (\psi \, C^{-1} \, \ga_{\mu_p ... \mu_1} \, \bar \chi)\,,
\label{fierzb}
\end{align}
\item $D \ = \ 2 \ \te{mod} \ 4$:
\begin{align}
\psi_\al \, \chi_\be \ \ &= \ \ - \, 2^{-D/2} \sum_{p \ \te{odd}}^{D/2-2} \frac{1}{p!} \; (\ga^{\mu_1 ... \mu_p} \, C)_{\be \al} \, (\psi \, C^{-1} \, \gab_{\mu_p ... \mu_1} \, \chi) \notag \\
& \ \ \ \ \ \ \ \ \ \ - \ \frac{2^{-D/2}}{2 \, (D/2)!} \; (\ga^{\mu_1 ... \mu_{D/2}} \, C)_{\be \al} \, (\psi \, C^{-1} \, \gab_{\mu_{D/2} ... \mu_1} \, \chi)\,,
\label{fierzc} \\
\psi_\al \, \bar \chi^{\dbe} \ \ &= \ \ 2^{-D/2} \sum_{p \ \te{even}}^{D/2-1} \frac{1}{p!} \; (\gab^{\mu_1 ... \mu_p} \, C)^{\dbe} \, \!_{ \al} \, (\psi \, C^{-1} \, \gab_{\mu_p ... \mu_1} \, \bar \chi)\,.
\label{fierzd}
\end{align}
\end{itemize}
Their proof basically relies on the fact that the $\ga^{\mu_1 ... \mu_p}$ are orthonormal with respect to the trace (up to the subtlety that the trace $\te{Tr}\{ \ga^{\mu_1... \mu_{D/2}} \stackrel{_{(-)}}{\ga}_{\nu_{1}...\nu_{D/2}} \}$ of $D$ gamma matrices additionally involves the $\vep^{\mu_1 ... \mu_{D/2}} \, \!_{ \nu_1 ... \nu_{D/2}}$ symbol).

\medskip
Let us display the Fierz identities in $D=4,6,8,10$ dimensions explicitly:
\begin{itemize}
\item $D=4$:
\begin{align}
\psi_\al \, \chi_\be \ \ &= \ \ \frac{1}{2} \; C_{\be \al} \, (\psi  \, C^{-1} \,  \chi) \ + \ \frac{1}{8} \; (\ga^{\mu
  \nu } \, C)_{\be \al} \, \bigl(\psi  \, C^{-1} \, \ga_{ \nu \mu} \, \chi \bigr)\label{F4.1}\,, \\
\psi_\al \, \bar{\chi}^{\dbe} \ \ &= \ \ - \; \frac{1}{2} \; (\gab^\mu \, C)^{\dbe} \, \! _\al \, \bigl(\psi \, C^{-1} \, \ga_\mu \, \bar \chi  \bigr)\,,
\label{F4.2} 
\end{align}
\item $D=6$:
\begin{align}
\psi_\al \, \chi_\be \ \ &= \ \ - \, \frac{1}{4} \; (\ga^\mu \, C)_{\be \al} \, (\psi \, C^{-1} \, \gab_\mu \, \chi) \ - \ \frac{1}{48} \; (\ga^{\mu \nu \la} \, C)_{\be \al} \, (\psi \, C^{-1} \, \gab_{\la \nu \mu} \, \chi)\label{F6.1}\,, \\
\psi_\al \, \bar{\chi}^{\dbe} \ \ &= \ \ \frac{1}{4} \; C^{\dbe} \! \, _{ \al} \, (\psi \, C^{-1} \, \bar \chi) \ + \ \frac{1}{8} \; (\gab^{\mu \nu } \, C)^{\dbe} \, \! _{ \al} \, (\psi \, C^{-1} \, \gab_{ \nu \mu} \, \bar \chi)\label{F6.2} \,,
\end{align}
\item $D=8$:
\begin{align}
\psi_\al \, \chi_\be \ \ &= \ \ \frac{1}{8}\; C_{\be \al} \, ( \psi \, C^{-1} \, \chi) \ + \ \frac{1}{16}\; (\ga^{\mu \nu} \, C)_{\be \al} \, ( \psi \, C^{-1}  \, \ga_{\nu \mu} \, \chi) \ + \ \frac{1}{384}\; (\ga^{\mu \nu \la \rho} \, C)_{\be \al} \, ( \psi \, C^{-1}  \, \ga_{\rho \la \nu \mu} \, \chi)\label{F8.1}\,, \\
\psi_\al \, \bar{\chi}^{\dbe} \ \ &= \ \ - \, \frac{1}{8} \; (\gab^\mu \, C)^{\dbe} \, \! _\al \, (\psi \, C^{-1} \, \ga_\mu \, \bar \chi) \ - \ \frac{1}{48} \; (\gab^{\mu \nu \la} \, C)^{\dbe} \, \! _\al \, (\psi \, C^{-1} \, \ga_{\la \nu \mu} \, \bar \chi)\label{F8.2}\,, 
\end{align}
\item $D=10$:
\begin{align}
\psi_\al \, \chi_\be \ \ = \ \ &-\, \frac{1}{16} \; (\ga^\mu \, C)_{\be \al} \, (\psi \, C^{-1} \, \gab_\mu \, \chi) \ - \ \frac{1}{96} \; (\ga^{\mu \nu \la} \, C)_{\be \al} \, (\psi \, C^{-1} \, \gab_{\la \nu \mu} \, \chi) \notag \\
&-\, \frac{1}{3840} \; (\ga^{\mu \nu \la \rho \tau} \, C)_{\be \al} \, (\psi \, C^{-1} \, \gab_{\tau \rho \la \nu \mu} \, \chi)\label{F10.1}\,,\\
\psi_\al \, \bar{\chi}^{\dbe} \ \ = \ \ &\frac{1}{16} \; C^{\dbe} \! \, _{ \al} \, (\psi \, C^{-1} \, \bar \chi) \ + \ \frac{1}{32} \; (\gab^{\mu \nu } \, C)^{\dbe} \, \! _{ \al} \, (\psi \, C^{-1} \, \gab_{ \nu \mu} \, \bar \chi) \notag \\
 &+\,\frac{1}{384} \; (\gab^{\mu \nu \la \rho } \, C)^{\dbe} \, \! _{ \al} \, (\psi \, C^{-1} \, \gab_{ \rho \la \nu \mu} \, \bar \chi)\label{F10.2}\,.
\end{align}
\end{itemize}
Fierz identities allow to derive relations between various $SO(1,D-1)$ Clebsch--Gordan coefficients by making clever
choices for $\psi_\al, \chi_\be, \bar \chi^{\dbe}$. For $D=4$ (\ref{F4.2}) immediately implies that
\begin{equation}
  (\ga^\mu)_{\al \dbe} \, (\ga_\mu)_{\ga \dde} \eq -2\,C_{\al \ga}\,C_{\dbe \dde}\,,
\end{equation}
whereas (\ref{F4.1}) with $\psi_\al =C_{\al \ga}$ and $ \chi_{\be} = C_{\be \de}$ yields
\begin{equation}
  (\ga^{\mu \nu})_\al \, \! ^\be \, (\ga_{\mu \nu})_\ga \, \!^{\de}  \eq 4\,\de_\al^\be\,\de_\ga^\de \  - \  8\,\de_\al^\de\,\de_\ga^\be\,. 
\end{equation}
Analogous equations in higher dimensions will be given in the subsections of appendix \ref{appB}.

\section{Gamma matrix identities in $\bm{D=6,8,10}$ dimensions}
\label{appB}
Each correlation function $\langle \psi^{\mu_1}\dots\psi^{\mu_n} \, S_{\al_1}\dots S_{\al_r} \, S^{\dbe_1}\dots
S^{\dbe_s} \rangle$ will be expressed in terms of Clebsch--Gordan coefficients taking the the tensor product of $n$
vectors, $r$ left-handed spinors and $s$ right-handed spinors into the scalar representation of $SO(1,D-1)$. The number
of linearly independent tensors $T^{\mu_1...\mu_n}\, \!_{\al_1 ... \al_r} \, \! ^{\dbe_1... \dbe_s}$ is equal to the
number of scalar representations in the corresponding tensor product. These are summarized in Table \ref{scalars} for all correlators with fixed number of fields given in
this paper.

\begin{table}
  \small
  \centering
  \begin{tabular}{|r|c|c| | r |c|c|}\hline $\otimes$
    & $D=4$ & $D=8$ &$\otimes$ &$D=6$ &$D=10$ 
    \\ \hline \hline 
    $\langle S_\al S_\be \rangle$ & 1 & 1 &$\langle S_\al S^{\dbe} \rangle$ & 1 & 1 \\ \hline
    $\langle \psi^\mu S_\al S^{\dbe} \rangle$ & 1 & 1 & $\langle \psi^\mu S_\al S_{\be} \rangle$ & 1 & 1 \\ \hline
    $\langle \psi^\mu \psi^\nu S_\al S_\be \rangle$ &2 &2 &$\langle \psi^\mu \psi^\nu S_\al S^{\dbe} \rangle$ &2 &2  \\ \hline
    $\langle \psi^\mu \psi^\nu \psi^\la S_\al S^{\dbe} \rangle$ &4 &4 
    &$\langle \psi^\mu \psi^\nu \psi^\la S_\al S_{\be} \rangle$ &4 &4 \\ \hline
    $\langle \psi^\mu \psi^\nu \psi^\la \psi^\rho S_\al S_\be \rangle$ &10 &10 
    &$\langle \psi^\mu \psi^\nu \psi^\la \psi^\rho S_\al S^{\dbe} \rangle$ &10 &10  \\ \hline
    $\langle \psi^\mu \psi^\nu \psi^\la \psi^\rho \psi^\tau S_\al S^{\dbe} \rangle$ &25 &26 &$\langle \psi^\mu \psi^\nu \psi^\la \psi^\rho \psi^\tau S_\al S_{\be} \rangle$ &26 &26 \\ \hline
    $\langle \psi^\mu \psi^\nu \psi^\la \psi^\rho \psi^{\tau} \psi^{\xi} S_\al S_\be \rangle$ &70 &76
    &$\langle \psi^\mu \psi^\nu \psi^\la \psi^\rho \psi^{\tau} \psi^{\xi} S_\al S^{\dbe} \rangle$ &76 &76  \\ \hline \hline
$\langle S_\al S_{\be} S^{\dga} S^{\dde}  \rangle$ &1 &2  
&$\langle S_\al S_\be S_\ga S_\de  \rangle$ &1 &2
\\ \hline
$\langle S_\al S_\be S_\ga S_\de  \rangle$ &2 &3 
&$\langle S_\al S_{\be} S^{\dga} S^{\dde}  \rangle$ &2 &3
\\ \hline
$\langle \psi^\mu S_\al S_\be S_\ga S^{\dde}  \rangle$ &2 &4 
&$\langle \psi^\mu S_\al S_\be S_\ga S^{\dde}  \rangle$ &3 &5  \\ \hline 
$\langle \psi^\mu \psi^\nu S_\al S_{\be} S^{\dga} S^{\dde}  \rangle$ &4 &9 
&$\langle \psi^\mu \psi^\nu S_\al S_\be S_\ga S_\de  \rangle$ &6 &11
\\ \hline
$\langle \psi^\mu \psi^\nu S_\al S_\be S_\ga S_\de  \rangle$ &5 &10
&$\langle \psi^\mu \psi^\nu S_\al S_{\be} S^{\dga} S^{\dde}  \rangle$ &7 &12 \\ \hline
$\langle \psi^\mu \psi^\nu \psi^\la S_\al S_\be S_\ga S^{\dde}  \rangle$ &10 &24 
&$\langle \psi^\mu \psi^\nu \psi^\la S_\al S_\be S_\ga S^{\dde}  \rangle$ &17 &31 \\ \hline 
$\langle \psi^\mu \psi^\nu \psi^\la \psi^\rho S_\al S_{\be} S^{\dga} S^{\dde}  \rangle$ &25 &68 
&$\langle \psi^\mu \psi^\nu \psi^\la \psi^\rho S_\al S_\be S_\ga S_\de  \rangle$ &45 &88
\\ \hline
$\langle \psi^\mu \psi^\nu \psi^\la \psi^\rho S_\al S_\be S_\ga S_\de  \rangle$ &28 &71
&$\langle \psi^\mu \psi^\nu \psi^\la \psi^\rho S_\al S_{\be} S^{\dga} S^{\dde}  \rangle$ &48 &91 
\\ \hline \hline
$\langle S_\al S_{\be} S_\ga S_{\de} S^{\dep} S^{\di} \rangle$ &2 &10 
&$\langle S_\al S_\be S_\ga S_\de S_\ep S^{\di} \rangle$ &4 &16
\\ \hline
$\langle S_\al S_\be S_\ga S_\de S_\ep S_\io \rangle$ &5 &15 
&$\langle S_\al S_{\be} S_\ga S^{\dde} S^{\dep} S^{\di} \rangle$ &6 &19   \\ \hline 
$\langle \psi^\mu S_\al S_\be S_\ga S^{\dde} S^{\dep} S^{\di} \rangle$ &4 &24 
&$\langle \psi^\mu S_\al S_{\be} S_\ga S_{\de} S_{\ep} S_{\io} \rangle$ &9 &40 
\\ \hline
$\langle \psi^\mu S_\al S_{\be} S_\ga S_{\de} S_{\ep} S^{\di} \rangle$ &5 &26 
&$\langle \psi^\mu S_\al S_\be S_\ga S_\de S^{\dep} S^{\di} \rangle$ &12 &45  \\ \hline
$\langle \psi^\mu \psi^\nu S_\al S_{\be} S_\ga S_{\de} S^{\dep} S^{\di} \rangle$ &10 &68 
&$\langle \psi^\mu \psi^\nu S_\al S_\be S_\ga S_\de S_\ep S^{\di} \rangle$ &29 &125   \\ \hline
$\langle \psi^\mu \psi^\nu S_\al S_\be S_\ga S_\de S_\ep S_\io \rangle$ &14 &76 
&$\langle \psi^\mu \psi^\nu S_\al S_{\be} S_\ga S^{\dde} S^{\dep} S^{\di} \rangle$ &32 &130
\\ \hline \hline
$\langle S_\al S_{\be} S_\ga S_{\de} S^{\dep} S^{\di} S^{\dot \ka} S^{\dot \la} \rangle$ &4 &71 
&$\langle S_\al S_\be S_\ga S_\de S_\ep S_\io S_{\ka} S_{ \la} \rangle$ &14 &175
\\ \hline
$\langle S_\al S_\be S_\ga S_\de S_\ep S_\io S^{\dot \ka} S^{\dot \la} \rangle$ &5 &76 
&$\langle S_\al S_\be S_\ga S_\de S_\ep S_\io S^{\dot \ka} S^{\dot \la} \rangle$ &19 &196   \\ \hline
$\langle S_\al S_\be S_\ga S_\de S_\ep S_\io S_{\ka} S_{ \la} \rangle$ &14 &106  
&$\langle S_\al S_{\be} S_\ga S_{\de} S^{\dep} S^{\di} S^{\dot \ka} S^{\dot \la} \rangle$ &24 &210  \\ \hline
\end{tabular}
\caption{\small Number of linearly independent Clebsch--Gordan coefficients in various tensor products. Since the chiral structure differs for $D=4,8$ and $D=6,10$, these two cases are separated into
  different sets of columns}
\label{scalars}
\end{table}

In the following subsections we will give identities necessary to express correlation functions in terms of a minimal
basis.

\subsection[$D=6$ dimensions]{$\bm{D=6}$ dimensions}

In six dimensions, tensors with four spinor indices are severely constrained by Fierz identities.

\bigskip
\noindent
\paragraph{\small\underline{Correlator $\langle S_{\al} S_{\be} S_{\ga} S_{\de} \rangle$}}\hfill\\
The most important relation between $(\ga^\mu C)_{\al \be} (\ga_\mu C)_{\ga \de}$ and permutations in the spinor indices arises from (\ref{F6.1}) with $\psi_\al = (\ga^\mu C)_{\al \ga}$ and $\chi_\be = (\ga_\mu C)_{\be \de}$. After
some manipulations, one arrives at\footnote{Useful tools for deriving relations like this are
\begin{subequations}
\begin{align}
\Ga^\mu \, \Ga^{\nu_1 ... \nu_m} \, \Ga_\mu \ \ &= \ \ (-1)^{m-1} \, (D-2m) \, \Ga^{\nu_1 ... \nu_m}  \label{F6.3a} \\
\Ga^{\mu \nu} \, \Ga^{\la_1 ... \la_m} \, \Ga_{\mu \nu} \ \ &= \ \ \bigl( D \, - \, (D-2m)^2 \bigr) \, \Ga^{\la_1 ... \la_m}  \label{F6.3b}
\end{align}
\end{subequations}}
\begin{equation}
(\ga^\mu \, C)_{\al \ga} \, (\ga_\mu \, C)_{\be \de} \eq (\ga^\mu \, C)_{\be \al } \, (\ga_\mu \, C)_{\ga \de} \ .
\label{F6.3}
\end{equation}
Together with the antisymmetry of $(\ga^\mu C)_{\al \be} = (\ga^\mu C)_{[\al \be]}$, this implies that the contraction
$(\ga^\mu C)_{\al \be} (\ga_\mu C)_{\ga \de}$ is totally antisymmetric in four indices and therefore proportional to the
$\vep$ tensor in the four-dimensional chiral spinor representations. Normalizing $\vep_{1234} = 1$, we get
\begin{equation} 
(\ga^\mu \, C)_{\al \be} \, (\ga_\mu \, C)_{\ga \de} \eq -\,2 \, \vep_{\al \be \ga \de} \ .
\label{F6.4}
\end{equation}

\paragraph{\small\underline{Correlator $\langle S_{\al} S_{\be} S^{\dga} S^{\dde} \rangle$}}\hfill\\
The choices $\psi_\al = C_\al \, \! ^{\dga}, \ \chi_\be = C_\be \, \! ^{\dde}$ in (\ref{F6.1}) and $\psi_\al = (\ga^\mu C)_{\al \ga}, \ \bar \chi^{\dbe} = (\gab_\mu C)^{\dbe \dde}$ in (\ref{F6.2}) yield the set of equations:
\begin{align}
(\ga^\mu \, C)_{\al \be} \, (\gab_\mu \, C)^{\dga \dde} \ \ &= \ \ 2 \, \bigl( C_\al \, \!^{\dga} \, C_\be \, \!^{\dde} \ - \ C_\al \, \!^{\dde} \, C_\be \, \!^{\dga} \bigr)\,, \label{F6.5} \\
(\ga^{\mu \nu} \, C)_{\al} \, \! ^{\dga} \, (\ga_{\mu \nu} \, C)_{\be} \, \! ^{\dde} \ \ &= \ \ 2 \, C_\al \, \!^{\dga} \, C_\be \, \!^{\dde} \ - \ 8 \, C_\al \, \!^{\dde} \, C_\be \, \!^{\dga}\,, \label{F6.6} \\
(\ga^{\mu \nu \la} \, C)_{\al \be} \, (\gab_{\mu \nu \la} \, C)^{\dga \dde} \ \ &= \ \ 24 \, \bigl( C_\al \, \!^{\dde} \, C_\be \, \!^{\dga} \ + \ C_\al \, \!^{\dga} \, C_\be \, \!^{\dde} \bigr)\,. \label{F6.7}
\end{align}
Hence, only $C_\al \, \!^{\dde}  C_\be \, \!^{\dga}$ and $C_\al \, \!^{\dga} \, C_\be \, \!^{\dde}$ remain as independent Clebsch--Gordan coefficients.

\paragraph{\small\underline{Correlator $\langle \psi^\mu S_{\al} S_{\be} S_{\ga} S^{\dde} \rangle$}}\hfill\\
For this correlator equations of the type
\begin{equation}
  \label{F6.8}
  (\ga^\mu\,\gab^\nu\,C)_\ga{}^{\dde}\,(\ga_\nu\,C)_{\al\be}\,=\,2\,(\ga^\mu\,C)_{\be\ga}\,\cc[\al]{\dde}-2\,(\ga^\mu\,C)_{\al\ga}\,\cc[\be]{\dde}
\end{equation}
prove to be useful, which can be derived by multiplying \eqref{F6.5} with $\ga^\mu_{\ga\dga}$.

\paragraph{\small\underline{Correlator $\langle S_{\al} S_{\be} S_{\ga} S_\de S_\ep S^{\di} \rangle$}}\hfill\\
As Weyl spinors in six dimensions only have $2^{6/2-1}=4$ independent components, the trivial relation
\begin{equation}
  \label{F6.9}
  (\ga^\mu\,C)_{[\al\be}\,(\ga_\mu\,C)_{\ga\de}\,C_{\ep]}{}^{\di}\,=\,0
\end{equation}
holds for this correlator.

\paragraph{\small\underline{Correlator $\langle \psi^\mu \psi^\nu S_{\al} S_{\be} S_{\ga} S_\de \rangle$}}\hfill\\
This correlation function can be expressed in terms of $\eta^{\mu\nu}\,(\ga^\la\,C)_{\al\be}\,(\ga_\la\,C)_{\ga\de}$,
$(\ga^\mu\,C)_{\al\be}\,(\ga^\nu\,C)_{\ga\de}$ and permutations in $\al,\be,\ga,\de$ thereof. These are related by the
equation
\begin{equation}
  \label{F6.10}
  \eta^{\mu\nu}\,(\ga^\la\,C)_{\al\be}\,(\ga_\la\,C)_{\ga\de}\,=\,-(\ga^\mu\,C)_{[\al\be}\,(\ga^\nu\,C)_{\ga\de]}\,.
\end{equation}

\paragraph{\small\underline{Correlator $\langle \psi^\mu \psi^\nu S_{\al} S_{\be} S^{\dga} S^{\dde} \rangle$}}\hfill\\
In this case the ten index terms $\eta^{\mu\nu}\,\cc[\al]{\dga}\,\cc[\be]{\dde}$,
$(\ga^\mu\,C)_{\al\be}\,(\gab^\nu\,C)^{\dga\dde}$, $(\ga^\nu\,C)_{\al\be}\,(\gab^\mu\,C)^{\dga\dde}$,
$(\ga^\mu\,\gab^\nu\,C)_\al{}^{\dga}\,\cc[\be]{\dde}$,
$(\ga^\mu\,\gab^\nu\,\ga^\la\,C)_{\al\be}\,(\gab_\la\,C)^{\dga\dde}$ and
$(\gab^\mu\,\ga^\nu\,\gab^\la\,C)^{\dga\dde}\,(\ga_\la\,C)_{\al\be}$ including permutations in $\al,\be,\dga,\dde$
appear in the calculations. Applying (\ref{F6.5}) to $(\ga^\la\,C)(\ga_\la\,C)$ in the terms above yields
\begin{align}
  \label{F6.11}
  (\ga^\mu\,\gab^\nu\,\ga^\la\,C)_{\al\be}\,(\gab_\la\,C)^{\dga\dde}\, = \, &
  2\,(\ga^\mu\,\gab^\nu\,C)_\al{}^{\dga}\,\cc[\be]{\dde}-2\,(\ga^\mu\,\gab^\nu\,C)_\al{}^{\dde}\,\cc[\be]{\dga}\,,\notag\\
  (\gab^\mu\,\ga^\nu\,\gab^\la\,C)^{\dga\dde}\,(\ga_\la\,C)_{\al\be}\,=\,&
  2\,(\ga^\mu\,\gab^\nu\,C)_\be{}^{\dga}\,\cc[\al]{\dde}-2\,(\ga^\mu\,\gab^\nu\,C)_\al{}^{\dga}\,\cc[\be]{\dde}+4\,\eta^{\mu\nu}\,\cc[\al]{\dde}\,\cc[\be]{\dga}-4\,\eta^{\mu\nu}\,\cc[\al]{\dga}\,\cc[\be]{\dde}\,.
\end{align}
A further relation is obtained by anti-symmetrizing the previous result in the spinor indices $\al,\be$:
\begin{align}
  \label{F6.12}
  2\,\eta^{\mu\nu}\,\cc[\al]{\dde}\,&\cc[\be]{\dga}-2\,\eta^{\mu\nu}\,\cc[\al]{\dga}\,\cc[\be]{\dde}
  -(\ga^\mu\,C)_{\al\be}\,(\gab^\nu\,C)^{\dga\dde}+(\ga^\nu\,C)_{\al\be}\,(\gab^\mu\,C)^{\dga\dde}\, = \,\notag\\
  &(\ga^\mu\,\gab^\nu\,C)_\al{}^{\dga}\,\cc[\be]{\dde}-(\ga^\mu\,\gab^\nu\,C)_\al{}^{\dde}\,\cc[\be]{\dga}
  -(\ga^\mu\,\gab^\nu\,C)_\be{}^{\dga}\,\cc[\al]{\dde}+(\ga^\mu\,\gab^\nu\,C)_\be{}^{\dde}\,\cc[\al]{\dga}\,.
\end{align}

\paragraph{\small\underline{Correlator $\langle \psi^\mu S_{\al} S_{\be} S_{\ga} S_\de S^{\dep} S^{\di} \rangle$}}\hfill\\
The relevant index terms for this correlator are $(\gab^\mu\,C)^{\dep\di}\,\vep_{\al\be\ga\de}$,
$(\ga^\mu\,C)_{\al\be}\cc[\ga]{\dep}\,\cc[\de]{\di}$ and permutations in $\al,\be,\ga,\de$. By replacing $\dga,\dde$
with $\dep,\di$ in \eqref{F6.12} and multiplying with $(\ga_\nu\,C)_{\ga\de}$ they turn out to be related:
\begin{equation}
  \label{F6.13}
  2 (\gab^\mu\,C)^{\dep\di}\,\vep_{\al\be\ga\de}\,=\,-(\ga^\mu\,C)_{[\al\be}\cc[\ga]{\dep}\,C_{\de]}^{\di}\,.
\end{equation}

\subsection[$D=8$ dimensions]{$\bm{D=8}$ dimensions}

A lot of tensor equations in eight dimensions can be related by $SO(8)$ triality. In section \ref{sec:triality}, their most general form is given for some cases of interest. In the following, we will list the specializations needed for correlation functions with four or more spin fields:

\paragraph{\small\underline{Correlator $\langle S_{\al} S_{\be} S^{\dga} S^{\dde} \rangle$}}\hfill\\
Here, the choices $\psi_\al = (\ga^\mu C)_\al \, \! ^{\dga}, \ \chi_\be = (\ga_\mu C)_\be \, \! ^{\dde}$ and $\psi_\al = C_{\al \ga} , \ \bar \chi ^{\dbe} = C^{\dbe \dde}$ in the eight-dimensional Fierz identities (\ref{F8.1}) and (\ref{F8.2}) yield
\begin{align}
C_{\al \be} \, C^{\dga \dde} \ \ &= \ \  \frac{1}{2} \; \bigl(   (\ga^\mu \, C)_\al \, \! ^{\dga} \, (\ga_\mu \, C)_\be \, \! ^{\dde} \ + \ (\ga^\mu \, C)_\al \, \! ^{\dde} \, (\ga_\mu \, C)_\be \, \! ^{\dga} \bigr)\,, \label{F8.3} \\
(\ga^{\mu \nu} \, C)_{\al \be} \, (\gab_{\mu \nu} \, C)^{\dga \dde} \ \ &= \ \  2 \, \bigl(   (\ga^\mu \, C)_\al \, \! ^{\dga} \, (\ga_\mu \, C)_\be \, \! ^{\dde} \ - \ (\ga^\mu \, C)_\al \, \! ^{\dde} \, (\ga_\mu \, C)_\be \, \! ^{\dga} \bigr)\,, \label{F8.4} \\
(\ga^{\mu \nu \la} \, C)_\al \, \! ^{\dga} \, (\ga_{\mu \nu \la} \, C)_\be \, \! ^{\dde} \ \ &= \ \  18 \,  (\ga^\mu \, C)_\al \, \! ^{\dga} \, (\ga_\mu \, C)_\be \, \! ^{\dde} \ + \ 24 \, (\ga^\mu \, C)_\al \, \! ^{\dde} \, (\ga_\mu \, C)_\be \, \! ^{\dga}\,, \label{F8.5}
\end{align}
such that it suffices to express this correlator in terms of $(\ga^\mu  C)_\al \! ^{\dga} \, (\ga_\mu C)_\be \, \! ^{\dde}$ and $(\ga^\mu C)_\al \, \! ^{\dde} (\ga_\mu C)_\be \, \! ^{\dga}$.

\paragraph{\small\underline{Correlator $\langle S_{\al} S_{\be} S_{\ga} S_{\de} \rangle$}}\hfill\\
Fierz identities with $\psi_\al = C_{\al \ga}$ and $\chi_\be = C_{\be \de}$ lead us to
\begin{align}
(\ga^{\mu \nu} \, C)_{\al \be} \, (\ga_{\mu \nu} \, C)_{\ga \de} \ \ &= \ \ 8 \, \bigl( C_{\al \ga} \, C_{\be \de} \ - \
C_{\al \de} \, C_{\be \ga} \bigr)\,, \label{F8.6} \\
(\ga^{\mu \nu \la \rho} \, C)_{\al \be} \, (\ga_{\mu \nu \la \rho} \, C)_{\ga \de} \ \ &= \ \ 192 \, \bigl( C_{\al \ga} \, C_{\be \de} \ + \ C_{\al \de} \, C_{\be \ga} \bigr) \ - \ 48 \, C_{\al \be} \, C_{\ga \de} \ . \label{F8.7}
\end{align}

\paragraph{\small\underline{Correlator $\langle \psi^{\mu} S_{\al} S_{\be} S_{\ga} S^{\dde} \rangle$}}\hfill\\
Multiplying (\ref{F8.3}) by $\ga^\mu_{\ga \dga}$ gives
\begin{equation}
C_{\al \be} \, (\ga^\mu \, C)_{\ga}{}^{\dde} \eq - \, \frac{1}{2} \; \Bigl[ \, (\ga^\la \, \gab^\mu \, C)_{\al \ga} \, (\ga_\la \, C)_{\be}{}^{\dde} \ + \ (\ga^\la \, \gab^\mu \, C)_{\be \ga} \, (\ga_\la \, C)_{\al}{}^{\dde} \, \Bigr]\,.
\label{F8.8}
\end{equation} 
Hence, it is possible to eliminate two out of the six tensors $C_{\al \be} (\ga^\mu \, C)_{\ga}{}^{\dde}$, $(\ga^\la \,
\gab^\mu \, C)_{\al \ga} \, (\ga_\la \, C)_{\be}{}^{\dde}$ and permutations in $\al,\be,\ga$
thereof.

\paragraph{\small\underline{Correlator $\langle \psi^{\mu} \psi^{\nu} S_{\al} S_{\be} S^{\dga} S^{\dde} \rangle$}}\hfill\\
This is a triality invariant index structure for which the tensor identities are particularly interesting. In our nine
dimensional basis of (\ref{6pt,tri}), only the antisymmetric part in $\mu,\nu$ of the tensor $(\ga^{\mu} {}_\la C)_{\al
  \be} (\gab^{\nu \la } C)^{\dga \dde}$ is kept because the symmetric piece can be reduced to
\begin{align}
(\ga^{(\mu} {}_\la \, C)_{\al \be} \, (\gab^{\nu ) \la } \, C)^{\dga \dde}  \ \ &= \ \ \eta^{\mu \nu} \, (\ga_\la \, C)_{[\al}{}^{\dga} \, (\ga^\la \, C)_{\be]} {}^{\dde} \ - \ (\ga^\mu \, C)_{\al}{}^{\dga} \, (\ga^\nu \, C)_{\be}{}^{\dde} \ + \ (\ga^\mu \, C)_{\al}{}^{\dde} \, (\ga^\nu \, C)_{\be}{}^{\dga} \notag \\
& \ \ \ \ - \ (\ga^\mu \, C)_{\be}{}^{\dde} \, (\ga^\nu \, C)_{\al}{}^{\dga} \ + \ (\ga^\mu \, C)_{\be}{}^{\dga} \, (\ga^\nu \, C)_{\al}{}^{\dde} \ .
\label{F8.11}
\end{align}
This can be derived from (\ref{F8.8}) multiplied by $(\gab^{\nu})^{ \dga \ga}$. Then, triple products of $\ga$ matrices
can be eliminated as follows:
\begin{align}
(\ga^\mu \, &\gab^\nu \, \ga^\la \, C)_\al{}^{\dga} \, (\ga_\la \,  C)_\be{}^{\dde} \ \ = \ \ - \, (\ga^{[\mu} {}_\la \, C)_{\al \be} \, (\gab^{\nu ] \la } \, C)^{\dga \dde} \ - \ \eta^{\mu \nu} \, (\ga_\la \, C)_{ [\al} {}^{\dga} \, (\ga^\la \, C)_{ \be]}{}^{\dde} \ - \ \eta^{\mu \nu} \, C_{\al \be} \, C^{\dga \dde} \notag \\
&\ \ \ \ \ \ \ \ \ \ \ \  - \ (\ga^\mu \, C)_\al {}^{\dga} \, (\ga^\nu \, C)_{\be}{}^{\dde} \ + \ (\ga^\mu \, C)_\al {}^{\dde} \, (\ga^\nu \, C)_{\be}{}^{\dga} \ + \ (\ga^\mu \, C)_\be {}^{\dde} \, (\ga^\nu \, C)_{\al}{}^{\dga} \notag \\
&\ \ \ \ \ \ \ \ \ \ \ \  - \ (\ga^\mu \, C)_\be {}^{\dga} \, (\ga^\nu \, C)_{\al}{}^{\dde} \ + \ (\ga^{\mu \nu} \, C)_{\al \be} \, C^{\dga \dde} \ - \ C_{\al \be} \, (\gab^{\mu \nu} \, C)^{\dga \dde}\ .
\label{F8.12a} 
\end{align}
Observe that, upon adding $\al \leftrightarrow \be, \dga \leftrightarrow \dde$ images and anti-symmetrizing $\ga^\mu
\gab^\nu \ga^\la$, one finds
\begin{equation}
(\ga^{\mu \nu \la} \, C)_\al {}^{\dga} \, (\ga_\la \, C)_\be {}^{\dde} \ + \ (\ga^{\mu \nu \la} \, C)_\be {}^{\dga} \, (\ga_\la \, C)_\al {}^{\dde} \ + \ (\ga^{\mu \nu \la} \, C)_\al {}^{\dde} \, (\ga_\la \, C)_\be {}^{\dga} \ + \ (\ga^{\mu \nu \la} \, C)_\be {}^{\dde} \, (\ga_\la \, C)_\al {}^{\dga} \eq 0 \ .
\label{F8.13}
\end{equation}

\paragraph{\small\underline{Correlator $\langle \psi^{\mu} \psi^{\nu} S_{\al} S_{\be} S_{\ga} S_{\de} \rangle$}}\hfill\\
Possible Clebsch--Gordan coefficients for this correlator are permutations in $\al,\be,\ga,\de$ of the tensor $(\ga^\mu
\gab^\la C)_{\al \be} (\ga_\la \gab^\nu C)_{\ga \de}$. However these are not independent as multiplication of
\eqref{F8.8} with $\ga^\nu_{\de \dde}$ shows:
\begin{equation}
(\ga^\mu \, \gab^\la \, C)_{\al \be} \, (\ga_\la \, \gab^\nu \, C)_{\ga \de} \eq - \, (\ga^\mu \, \gab^\la \, C)_{\al \ga} \, (\ga_\la \, \gab^\nu \, C)_{\be \de} \ - \ 2 \, (\ga^\mu \, \gab^\nu \, C)_{\al \de} \, C_{\be \ga} \ .
\label{F8.9}
\end{equation}
By repeating this index shift in the last three spinor indices and anti-symmetrizing in the vector indices, one finds the
following relation for the antisymmetric part of $(\ga^{\mu \la} \, C)_{\al \be} \, (\ga_\la {} ^\nu \, C)_{\ga \de}$ in $\mu,\nu$:
\begin{equation}
(\ga^{\la [\mu} \, C)_{\al \be} \, (\ga^{\nu ] }{}_\la \, C)_{\ga \de} \eq C_{\al \de} \, (\ga^{\mu \nu} \, C)_{\ga \be} \ - \ C_{\al \ga} \, (\ga^{\mu \nu} \, C)_{\de \be} \ - \ C_{\be \de} \, (\ga^{\mu \nu} \, C)_{\ga \al} \ + \ C_{\ga \be} \, (\ga^{\mu \nu} \, C)_{\de \al}\,.
\label{F8.10}
\end{equation}

\paragraph{\small\underline{Correlator $\langle S_{\al} S_{\be} S_{\ga} S_{\de} S^{\dep} S^{\di} \rangle$}}\hfill\\
The novel tensors here are $(\ga^\la C)_\al{}^{\dep} \, (\ga_{\la \rho} C)_{\be \ga} \, (\ga^{\rho} C)_\de {}^{\di}$
and permutations in the spinor indices thereof. The antisymmetric piece in the indices $\dep, \di$ can be expressed in
terms of simpler $\ga$ matrix combinations:
\begin{align}
(\ga^\la \, C)_\al{}^{ [ \dep} \, &(\ga_{\la \rho} \, C)_{\be \ga} \, (\ga^{|\rho|} \, C)_\de {}^{\di ]} \ \ = \ \ C_{\al \ga} \, (\ga_\la \, C)_{[\be}{}^{\dep} \, (\ga^{\la} \, C)_{\de ]} {}^{\di} \ + \ C_{\de \ga} \, (\ga_\la \, C)_{[\be}{}^{\dep} \, (\ga^{\la} \, C)_{\al ]} {}^{\di} \notag \\
&- \ C_{\al \be} \, (\ga_\la \, C)_{[\ga}{}^{\dep} \, (\ga^{\la} \, C)_{\de ]} {}^{\di} \ - \ C_{\de \be} \, (\ga_\la \, C)_{[\ga}{}^{\dep} \, (\ga^{\la} \, C)_{\al ]} {}^{\di} \ - \ C_{\al \de} \, (\ga_\la \, C)_{[\be}{}^{\dep} \, (\ga^{\la} \, C)_{\ga ]} {}^{\di}\,.
\label{F8.14}
\end{align}

\subsection[$D=10$ dimensions]{$\bm{D=10}$ dimensions}

Most of the following tensor identities can be traced back to the fundamental relation
\begin{equation}
(\ga^\mu \, C)_{\al \be} \, (\ga_\mu \, C)_{\ga \de} \ + \ (\ga^\mu \, C)_{\be \ga} \, (\ga_\mu \, C)_{\al \de}  \ + \
(\ga^\mu \, C)_{\ga \al} \, (\ga_\mu \, C)_{\be \de} \eq 0 \,,
\label{symm}
\end{equation}
due to the fact that $(S)^{\otimes_s 3} \otimes (S)$ does not contain any scalars. Here $(S)^{\otimes_s 3}$ denotes a
totally symmetric threefold tensor product of the left-handed $SO(1,9)$ spinor representation $(S)$.

In general, correlators in $D=10$ dimensions will involve more independent Lorentz tensors, which enter more difficult
relations compared to their $D=6$ relatives. Observe, for instance, that no direct analogue of the relations
\eqref{F6.4} and \eqref{F6.5} holds.

\paragraph{\small\underline{Correlator $\langle S_{\al} S_{\be} S_{\ga} S_{\de}  \rangle$}}\hfill\\
The Fierz identity (\ref{F10.1}) with $\psi_\al = (\ga^\mu C)_{\al \ga}$ and $\chi_\be =(\ga_\mu C)_{\be \de}$ admits to eliminate
\begin{equation}
(\ga^{\mu \nu \la} \, C)_{\al \be} \, (\ga_{\mu \nu \la} \, C)_{\ga \de} \eq 12 \, \bigl( 
(\ga^{\mu  } \, C)_{\al \de} \, (\ga_{\mu } \, C)_{\be \ga} \ - \ (\ga^{\mu  } \, C)_{\al \ga} \, (\ga_{\mu } \, C)_{\be \de} \bigr) \,,
\label{F10.3}
\end{equation}
and $(\ga^\mu C)_{\al \ga} (\ga_\mu C)_{\be \de}$ is redundant by (\ref{symm}).

\paragraph{\small\underline{Correlator $\langle S_{\al} S_{\be} S^{\dga} S^{\dde} \rangle$}}\hfill\\
Setting $\psi_\al = C_\al \, \! ^{\dga}, \ \chi_\be = C_\be \, \! ^{\dde}$ and $\psi_\al = (\ga^{\mu \nu}C)_\al \, \! ^{\dga}, \ \chi_\be = (\ga_{\mu \nu}C)_\be \, \! ^{\dde}$ in (\ref{F10.1}) as well as $\psi_\al = C_\al \, \! ^{\dga}, \ \bar \chi^{\dbe} = C_\de \, \! ^{\dbe}$ and $\psi_\al = (\ga^\mu C)_{\al \ga}, \ \bar \chi^{\dbe} = (\gab_\mu C)^{\dbe \dde}$ in (\ref{F10.2}) gives rise to the following identities:
\begin{align}
(\ga^{\mu \nu} \, C)_{\al } \, \! ^{\dga} \, (\ga_{\mu \nu} \, C)_\be \, \! ^{\dde} \ \ &= \ \ - \, 2 \, C_{\al } \, \! ^{\dga} \,  C_\be \,  \! ^{\dde} \ - \ 8 \, C_{\al } \, \! ^{\dde} \,  C_\be \,  \! ^{\dga} \ + \ 4 \, (\ga^{\mu} \, C)_{\al \be} \, (\gab_\mu \, C)^{\dga \dde}\,,
\label{F10.4} \\
(\ga^{\mu \nu \la} \, C)_{\al \be} \, (\gab_{\mu \nu \la} \, C)^{\dga \dde} \ \ &= \ \  48 \, \bigl( C_{\al } \, \! ^{\dga} \,  C_\be \,  \! ^{\dde} \ - \ C_{\al } \, \! ^{\dde} \,  C_\be \,  \! ^{\dga}  \bigr)\,,
\label{F10.5} \\
(\ga^{\mu \nu \la \rho} \, C)_{\al } \, \! ^{\dga} \, (\ga_{\mu \nu \la \rho} \, C)_\be \, \! ^{\dde} \ \ &= \ \ - \, 48 \, C_{\al } \, \! ^{\dga} \,  C_\be \,  \! ^{\dde} \ + \ 288 \, C_{\al } \, \! ^{\dde} \,  C_\be \,  \! ^{\dga} \ + \ 48 \, (\ga^{\mu} \, C)_{\al \be} \, (\gab_\mu \, C)^{\dga \dde}\,,
\label{F10.6} \\
(\ga^{\mu \nu \la \rho \tau} \, C)_{\al \be} \, (\gab_{\mu \nu \la \rho \tau} \, C)^{\dga \dde}  \ \ &= \ \ 1920 \,  \bigl( C_{\al } \, \! ^{\dga} \,  C_\be \,  \! ^{\dde} \ + \ C_{\al } \, \! ^{\dde} \,  C_\be \,  \! ^{\dga} \bigr) \ - \ 240 \, (\ga^{\mu} \, C)_{\al \be} \, (\gab_\mu \, C)^{\dga \dde}\,.
\label{F10.7} 
\end{align}
Hence, the three tensors on the right hand side are sufficient to express $\langle S_\al S_\be S^{\dga} S^{\dde} \rangle$.

\paragraph{\small\underline{Correlator $\langle \psi^\mu S_{\al} S_{\be} S_{\ga} S^{\dde} \rangle$}}\hfill\\
The six tensors $C_\al {}^{\dde} (\ga^\mu C)_{\be \ga}$, $(\ga^{\nu} \gab^{\mu} C)_\al {}^{\dde} (\ga_\nu C)_{\be \ga}$
and permutations in $\al,\be,\ga$ can be used to express this correlation function. However, (\ref{symm}) multiplied by
$(\gab^\mu)^{\dde \de}$ admits to eliminate one of them:
\begin{equation}
   (\ga^{\nu} \, \gab^{\mu} \, C)_\al {}^{\dde} \, (\ga_\nu \, C)_{\be \ga} \ + \ (\ga^{\nu} \, \gab^{\mu} \, C)_\be
   {}^{\dde} \, (\ga_\nu \, C)_{ \ga \al} \ + \ (\ga^{\nu} \, \gab^{\mu} \, C)_\ga {}^{\dde} \, (\ga_\nu \, C)_{\al \be}\,
   =\, 0\,.
   \label{F10.8}
\end{equation}

\paragraph{\small\underline{Correlator $\langle \psi^\mu \psi^\nu S_{\al} S_{\be} S_{\ga} S_{\de} \rangle$}}\hfill\\
Among the fifteen tensors obtained from $\eta^{\mu \nu} (\ga^\la C)_{\al \be} (\ga_{\la} C)_{\ga \de}, \ (\ga^\mu
C)_{\al \be} (\ga^\nu C)_{\ga \de}$ and $(\ga^{\mu \nu \la} C)_{\al \be} (\ga_\la C)_{\ga \de}$ and permutations in
$\al,\be,\ga,\de$, there are four relations. We choose to work with antisymmetric $\ga$-products because then the
tensors involving $\eta^{\mu\nu}$ decouple from the others in the relations. Equation (\ref{symm}) directly implies
\begin{align}
&\eta^{\mu \nu} \, (\ga^\la \, C)_{\al \be} \, (\ga_{\la} \, C)_{\ga \de} \ +  \ \eta^{\mu \nu} \, (\ga^\la \, C)_{\al \ga} \, (\ga_{\la} \, C)_{\be \de} \ + \ \eta^{\mu \nu} \, (\ga^\la \, C)_{\al \de} \, (\ga_{\la} \, C)_{\ga \be} \eq 0\,,
\label{F10.9}
\end{align}
and from (\ref{F10.8}) we derive:
\begin{align}
0 \eq &(\ga^{\mu \nu \la} \, C)_{\al \be} \, (\ga_\la \, C)_{\ga \de} \ + \ (\ga^{\mu \nu \la} \, C)_{\al \ga} \, (\ga_\la \, C)_{\de \be} \ + \ (\ga^{\mu \nu \la} \, C)_{\al \de} \, (\ga_\la \, C)_{\be \ga} \notag \\
& \ \ \ + \ (\ga^\mu \, C)_{\ga \de} \, (\ga^\nu \, C)_{\al \be} \ - \ (\ga^\mu \, C)_{\al \be} \, (\ga^\nu \, C)_{\ga \de} \ + \ (\ga^\mu \, C)_{\be \de} \, (\ga^\nu \, C)_{\al \ga} \notag \\
& \ \ \ - \ (\ga^\mu \, C)_{\al \ga} \, (\ga^\nu \, C)_{\be \de} \ + \ (\ga^\mu \, C)_{\be \ga} \, (\ga^\nu \, C)_{\al \de} \ - \ (\ga^\mu \, C)_{\al \de} \, (\ga^\nu \, C)_{\be \ga}
\label{F10.10a} \ , 
\end{align}
In the result (\ref{sen}) for this correlator, we have used only two out of the three independent permutations of relation (\ref{F10.10a}) to eliminate $(\ga^{\mu \nu \la}  C)_{\al \ga}  (\ga_\la C)_{\be \de}$ and $(\ga^{\mu \nu \la}  C)_{\be \de}  (\ga_\la C)_{\al \ga}$. The missing third identity can be recast as
\begin{align}
(\ga^{\mu \nu \la} \, C)_{\al \be} \, (\ga_\la \, C)_{\ga \de} \ + \ & (\ga^{\mu \nu \la} \, C)_{\al \de} \, (\ga_\la \, C)_{\ga \be} \ + \ (\ga^{\mu \nu \la} \, C)_{\ga \be} \, (\ga_\la \, C)_{\al \de} \notag \\ 
 + \ &(\ga^{\mu \nu \la} \, C)_{\ga \de} \, (\ga_\la \, C)_{\al \be} \ - \ 4 \, (\ga^{[\mu} \, C)_{\al \ga} \, (\ga^{\nu]} \, C)_{\be \de} \eq 0 \ .
\label{F10.11}
\end{align}

\paragraph{\small\underline{Correlator $\langle \psi^\mu \psi^\nu S_{\al} S_{\be} S^{\dga} S^{\dde} \rangle$}}\hfill\\
This correlator can be expressed in terms of $\eta^{\mu\nu}\,\cc[\al]{\dga}\,\cc[\be]{\dde}$,
$\eta^{\mu\nu}\,(\ga^\la\,C)_{\al\be}\,(\gab_\la\,C)^{\dga\dde}$, $(\ga^\mu\,C)_{\al\be}\,(\gab^\nu\,C)^{\dga\dde}$,
$(\ga^\nu\,C)_{\al\be}\,(\gab^\mu\,C)^{\dga\dde}$, $(\ga^\mu\,\gab^\nu\,C)_\al{}^{\dga}\,\cc[\be]{\dde}$,
$(\ga^\mu\,\gab^\la\,C)_\al{}^{\dga}\,(\ga^\nu\,\gab_\la\,C)_\be{}^{\dde}$ and permutations in $\al,\be,\dga,\dde$. In
addition there are $(\ga^\mu\,\gab^\nu\,\ga^\la\,C)_{\al\be}\,(\gab_\la\,C)^{\dga\dde}$ and
$(\gab^\mu\,\ga^\nu\,\gab^\la\,C)^{\dga\dde}\,(\ga_\la\,C)_{\al\be}$. However only twelve of these fifteen index terms are
independent. One relation is found be replacing $\nu$ with $\la$ and $\mu$ with $\nu$ in \eqref{F10.8} and multiplying
with $\gab^\mu{}^{\dga\ga}$, another by treating the complex conjugate of \eqref{F10.8} in the same manner:
\begin{align}
  (\ga^\mu\,\gab^\nu\,\ga^\la\,C)_{\al\be}\,(\gab_\la\,C)^{\dga\dde}\,=\,-2\,(\ga^\mu\,C)_{\al\be}\,(\gab^\nu\,C)^{\dga\dde}
  -(\ga^\mu\,\gab^\la\,C)_\al{}^{\dga}\,(\ga^\nu\,\gab_\la\,C)_\be{}^{\dde}-(\ga^\mu\,\gab^\la\,C)_\al{}^{\dde}\,(\ga^\nu\,\gab_\la\,&C)_\be{}^{\dga}\,,\notag\\
  (\gab^\mu\,\ga^\nu\,\gab^\la\,C)^{\dga\dde}\,(\ga_\la\,C)_{\al\be}\,=\,-2\,(\ga^\nu\,C)_{\al\be}\,(\gab^\mu\,C)^{\dga\dde}
  -(\ga^\mu\,\gab^\la\,C)_\al{}^{\dga}\,(\ga^\nu\,\gab_\la\,C)_\be{}^{\dde}-(\ga^\mu\,\gab^\la\,C)_\be{}^{\dga}\,(\ga^\nu\,\gab_\la\,&C)_\al{}^{\dde}\notag\\
  -2\,(\ga^\mu\,\gab^\nu\,C)_\al{}^{\dga}\,\cc[\be]{\dde}+2\,(\ga^\mu\,\gab^\nu\,C)_\al{}^{\dga}\,\cc[\be]{\dga}
  -2\,(\ga^\mu\,\gab^\nu\,C)_\be{}^{\dga}\,\cc[\al]{\dde}+2\,(\ga^\mu\,\gab^\nu\,C)_\be{}^{\dde}\,&\cc[\al]{\dga}\,.
  \label{F10.13}
\end{align}
A third equation is found by symmetrizing the previous result in the vector indices $\mu$ and $\nu$:
\begin{align}
  (\ga^\mu\,\gab^\la\,C)_\be{}^{\dde}\,(\ga^\nu\,\gab_\la\,C)_\al{}^{\dga}\,=\,
   2\,\eta^{\mu\nu}\,(\ga^\la\,C)_{\al\be}\,(\gab_\la\,C)^{\dga\dde}
  -2\,(\ga^\mu\,C)_{\al\be}\,(\gab^\nu\,C)^{\dga\dde}
  -2\,(\ga^\nu\,C)_{\al\be}\,(\gab^\mu\,&C)^{\dga\dde}\notag\\
 -(\ga^\mu\,\gab^\la\,C)_\al{}^{\dga}\,(\ga^\nu\,\gab_\la\,C)_\be{}^{\dde}
  -(\ga^\mu\,\gab^\la\,C)_\al{}^{\dde}\,(\ga^\nu\,\gab_\la\,C)_\be{}^{\dga}
  -(\ga^\mu\,\gab^\la\,C)_\be{}^{\dga}\,(\ga^\nu\,\gab_\la\,&C)_\al{}^{\dde}\,.
  \label{F10.14}
\end{align}

\paragraph{\small\underline{Correlator $\langle S_{\al} S_{\be} S_{\ga} S_{\de} S_{\ep} S^{\di} \rangle$}}\hfill\\
The relevant index terms for this correlator are $(\ga^\mu\,C)_{\al\be}\,(\ga_\mu\,C)_{\ga\de}\,\cc[\ep]{\di}$,
$(\ga_\mu\,C)_{\al\be}\,(\ga_\nu\,C)_{\ga\de}\,(\ga^\mu\,\gab^\nu\,C)_\ep{}^{\di}$ and permutation in the spinor indices
thereof. Using \eqref{symm} one can eliminate six out of the fifteen tensors of the first type. Changing the index $\dde$ in
\eqref{F10.8} to $\di$ and multiplying with $(\ga_\nu\,C)_{\de\ep}$ gives rise to the relation:
\begin{equation}
  (\ga^\mu\,\gab^\nu\,C)_\al{}^{\di}\,(\ga_\mu\,C)_{\be\ga}\,(\ga_\nu\,C)_{\de\ep}+
  (\ga^\mu\,\gab^\nu\,C)_\be{}^{\di}\,(\ga_\mu\,C)_{\al\ga}\,(\ga_\nu\,C)_{\de\ep}+
  (\ga^\mu\,\gab^\nu\,C)_\ga{}^{\di}\,(\ga_\mu\,C)_{\al\be}\,(\ga_\nu\,C)_{\de\ep}\,=\,0\,.
  \label{F10.15}
\end{equation}
By permuting the spinor indices in this equation one obtains eight further independent relations, which can be used to
eliminate in total nine tensors of the second type.

\paragraph{\small\underline{Correlator $\langle S_{\al} S_{\be} S_{\ga} S^{\dde} S^{\dep} S^{\di} \rangle$}}\hfill\\
\noindent This correlations function can be expressed in terms of the 24 tensors $\cc[\al]{\dde}\,\cc[\be]{\dep}\,\cc[\ga]{\di}$,
$(\ga^\mu\,C)_{\al \be}(\gab_\mu\,C)^{\dde\dep}\,\cc[\ga]{\di}$, $(\ga^\mu\, \gab^\nu\,C)_\al
{}^{\dde}\,(\ga_\mu\,C)_{\be\ga}\,(\gab_\nu\,C)^{\dep\di}$ and permutations in the spinor indices. However only 16 of
these are independent. By multiplying \eqref{symm} with $\gab^{\nu\,\dde\de}\,(\gab_\nu\,C)^{\dde\di}$ and proceeding in
the same way with the complex conjugate of \eqref{symm} one obtains the equations
\begin{align}
  \label{F10.12}
  (\ga^\mu\,\gab^\nu\,C)_\al {}^{\dde}\,(\ga_\mu\,C)_{\be\ga}\,(\gab_\nu\,C)^{\dep\di}+
  (\ga^\mu\,\gab^\nu\,C)_\be {}^{\dde}\,(\ga_\mu\,C)_{\al\ga}\,(\gab_\nu\,C)^{\dep\di}+
  (\ga^\mu\,\gab^\nu\,C)_\ga {}^{\dde}\,(\ga_\mu\,C)_{\al\be}\,(\gab_\nu\,C)^{\dep\di}&\,=\,0\,,\notag\\
  (\ga^\mu\,\gab^\nu\,C)_\al {}^{\dde}\,(\ga_\mu\,C)_{\be\ga}\,(\gab_\nu\,C)^{\dde\di}+
  (\ga^\mu\,\gab^\nu\,C)_\al {}^{\dep}\,(\ga_\mu\,C)_{\be\ga}\,(\gab_\nu\,C)^{\dde\di}+
  (\ga^\mu\,\gab^\nu\,C)_\al {}^{\di}\,(\ga_\mu\,C)_{\be\ga}\,(\gab_\nu\,C)^{\dde\dep}&\,=\,0\,.
\end{align}
Upon permutation in the spinor indices these yield five independent equations, which are sufficient to reduce the number
of index terms to 16.

\break

\end{document}